\documentclass[12pt]{iopart}

\usepackage[capitalize]{cleveref} 
\usepackage{graphicx} 
\usepackage{subcaption} 
\usepackage{bm} 
\usepackage{braket} 
\usepackage{qcircuit} 
\usepackage{algorithm,setspace} 
\usepackage{algpseudocode} 
\usepackage{amssymb} 
\usepackage{xcolor}

\DeclareMathAlphabet{\bbold}{U}{bbold}{m}{n}
\newcommand{\id}{\ensuremath{\bbold{1}}}  
\crefname{appendix}{}{} 

\makeatletter
\DeclareRobustCommand{\text}{%
  \ifmmode\expandafter\text@\else\expandafter\mbox\fi}
\let\nfss@text\text
\def\text@#1{{\mathchoice
  {\textdef@\displaystyle\f@size{#1}}%
  {\textdef@\textstyle\f@size{#1}}%
  {\textdef@\textstyle\sf@size{#1}}%
  {\textdef@\textstyle \ssf@size{#1}}%
  \check@mathfonts
  }%
}
\def\textdef@#1#2#3{\hbox{{%
                    \everymath{#1}%
                    \let\f@size#2\selectfont
                    #3}}}
\makeatother

\begin{document}

\title[Analysis of the non-variational QWOA]{Analysis of the non-variational quantum walk-based optimisation algorithm}

\author{Tavis Bennett, Lyle Noakes and Jingbo B Wang}

\address{Department of Physics, The University of Western Australia, Perth, Australia}
\ead{\mailto{tavis.bennett@research.uwa.edu.au}, \mailto{lyle.noakes@uwa.edu.au}, \mailto{jingbo.wang@uwa.edu.au}}

\begin{abstract}
This paper introduces in detail a non-variational quantum algorithm designed to solve a wide range of combinatorial optimisation problems, including constrained problems and problems with non-binary variables. The algorithm returns optimal and near-optimal solutions from repeated preparation and measurement of an amplified state. The amplified state is prepared via repeated application of two unitaries; one which phase-shifts solution states dependent on objective function values, and the other which mixes phase-shifted probability amplitudes via a continuous-time quantum walk (CTQW) on a problem-specific mixing graph. The general interference process responsible for amplifying optimal solutions is derived in part from statistical analysis of objective function values as distributed over the mixing graph. The algorithm's versatility is demonstrated through its application to various problems: weighted maxcut, k-means clustering, quadratic assignment, maximum independent set and capacitated facility location. In all cases, efficient circuit implementations of the CTQWs are discussed. A penalty function approach for constrained problems is also introduced, including a method for optimising the penalty function. For each of the considered problems, the algorithm's performance is simulated for a randomly generated problem instance, and in each case, the amplified state produces a globally optimal solution within a small number of iterations.

\end{abstract}


\section{Introduction}
Combinatorial optimisation problems, such as the practically important routing and assignment problems encountered in logistics, pose a significant and often insurmountable challenge to classical computers. Problems of this kind are intrinsically difficult to solve in large part because the number of feasible solutions scales exponentially with the size of a problem instance. Exact algorithms, designed to find a globally optimal solution, even those capable of pruning large sections of the solution space, as in branch and bound \cite{land2010automatic}, must necessarily exhaust the entire exponentially-large space of solutions. These exact algorithms therefore have exponential time-complexity (they require an amount of time which scales exponentially in the size of the problem instance). On the other hand, heuristic algorithms, such as local search methods \cite{crama2005local} or simulated annealing \cite{kirkpatrick1983optimization}, designed to run in polynomial time, are not allowed sufficient time to explore the entire optimisation landscape, and therefore are not generally expected to find a globally optimal solution, or may do so but with a probability which necessarily decays exponentially with increasing problem size. Typically, heuristic algorithms produce ``good enough'' solutions, efficiently, but the improved margins delivered by globally optimal solutions may in many cases provide considerable economic impact. The ability to find globally optimal solutions in polynomial time, even just in the average case, to practically important and currently intractable combinatorial optimisation problems would prove a significant application of quantum computation and an instance of genuine quantum advantage.

Typical variational approaches for quantum combinatorial optimisation, such as the quantum approximate optimisation algorithm (QAOA) \cite{Farhi2014QAOA} and the variational quantum eigensolver \cite{peruzzo2014variational}, rely on the use of parameterised quantum circuits, and their effective performance relies on the computationally expensive tuning of an often prohibitively large number of variational parameters. The performance of these variational algorithms is highly dependent on the initialisation of variational parameters, for which various strategies have been explored, including, for the QAOA, a machine learning approach \cite{alam2020accelerating} and the use of annealing inspired parameters \cite{sack2021quantum}. In addition, for some binary problems, the ``concentration of parameters" has been observed, enabling the transfer of an effective set of parameters between different problem instances \cite{streif2020training, brandao2018fixed, akshay2021parameter, galda2021transferability}, though these effective parameter sets often resemble those of an annealing-inspired approach. The use of annealing-inspired parameter sets may also be described as a Trotterized implementation of the quantum adiabatic algorithm \cite{farhi2001quantum} (Trotterized QAA), as in \cite{abbas2023quantum}. The (digital) Trotterized QAA may have some advantages over the analogue QAA, as the gate-based model, with its greater expressibility, enables approaches such as the parameterised warm-starting employed in \cite{sachdeva2024quantum}, as well as the potential for implementation of more complex problem and mixing Hamiltonians. 

In any case, the QAA, or quantum annealing in general, currently sees successful application to binary problems, or at least those problems which have been embedded as a quadratic unconstrained binary optimisation (QUBO) problem. So a significant weakness of (Trotterized) QAA, or more generally, methods of parameter transfer, is a lack of generaliseability to constrained and/or non-binary problems. There are QAOA variants which attempt to deal with other problem structures, as in \cite{hadfield2017quantum,hadfield2019quantum,wang2020xy}, however they face the challenge of effective parameter initialisation and more generally the computational expense of the variational procedure. 

A recently introduced algorithm, the non-variational quantum walk-based optimisation algorithm (non-variational QWOA) \cite{bennett2024non}, in contrast to the aforementioned algorithms, generalises to non-binary and constrained problems, while simultaneously solving the challenges related to the variational approach. For binary, unconstrained problems, the non-variational QWOA may resemble Trotterized QAA, however, it is motivated independently of the quantum adiabatic theorem. The algorithm exploits problem structures via continuous-time quantum walks on problem-specific mixing graphs. This theoretical framework provides the algorithm its key strength, which is the clearly motivated generalisation to non-binary and constrained problem structures, and the design of suitable penalty-functions for constrained problems, where necessary. The purpose of this paper is to explore this algorithm in detail, motivating it clearly from first principles and presenting a detailed analysis of the general interference process at its heart. It is our hope that this detailed analysis clarifies the inner workings of the non-variational QWOA, better enabling its generalisation to practically important problems.

\section{Motivation from first principles}
\label{sec:Motivation}
Quantum computing has at its disposal the quantum phenomena of superposition, entanglement and interference, which when methodically combined, give quantum computers the capability to efficiently solve certain problems which would be intractable to classical computers. Quantum superposition is quite naturally related to the exponential scaling of combinatorial problems; an isolated quantum system (the system of qubits at the heart of a quantum computer) has the capability of existing in a superposition of orthogonal states, the number of which grows exponentially with respect to the size of the system (the number of qubits). Where a travelling salesman visiting 60 cities has as many routes to choose from as there are atoms in the observable universe, a quantum computer would require only 360 qubits to contain a superposition of states directly encoding each of these routes. 

The first component of a quantum algorithm for solving a combinatorial optimisation problem is the equal superposition state,
\begin{equation}
    \label{eq:equal_superposition}
    \ket{s} = \frac{1}{\sqrt{N}} \sum_{\bm{x} \in S} \ket{\bm{x}},
\end{equation}
where $S$ is the set of $N$ feasible solutions to the problem, and $\ket{\bm{x}}$ are computational basis states which directly encode the solutions in $S$, referred to as solution states. Within this equal superposition, each solution state $\ket{\bm{x}}$ is occupied with the same probability amplitude. The set of qubits used to represent solutions, those that occupy states $\ket{\bm{x}}$, are referred to as the input register. 

Consider a scalar objective function $f(\bm{x})$ whose output quantifies the quality or cost associated with a solution $\bm{x} \in S$. Solving the combinatorial optimisation problem naturally involves finding the solution or solutions which either minimise or maximise the objective function, depending on the nature of the problem. So the value of $f(\bm{x})$ across all solutions $\bm{x}$ defines the optimisation landscape for a particular problem. Now consider a set of quantum logic operations which computes and stores this objective function. More specifically, introducing a set of ancilla qubits (the output register), the objective function implemented within the quantum computing hardware can be understood as a unitary operator $U_f$ acting on both the input and output registers such that,
\begin{equation}
    U_f \ket{\bm{x}}\otimes\ket{0} = \ket{\bm{x}}\otimes\ket{f(\bm{x})}.
\end{equation}
As shown here, the objective function evaluation for a single arbitrary solution $\bm{x} \in S$ is not particularly impressive, and akin to the same objective function computed on a regular computer. However, implementing the same set of quantum logic operations or executing the same objective function evaluation in the case where the input register has first been initialised in the equal superposition, produces a rather remarkable outcome,
\begin{equation}
    U_f \ket{s}\otimes\ket{0} = \frac{1}{\sqrt{N}} \sum_{\bm{x} \in S}  \ket{\bm{x}}\otimes\ket{f(\bm{x})}.
\end{equation}
The input and output registers have become entangled; evaluating the objective function, or implementing the unitary $U_f$, has resulted in the simultaneous computation and storage of the objective function value for every single solution in the solution space. Initialising the input register in the equal superposition effectively opens multiple computational paths, one for each solution in the solution space. The subsequent computation of the objective function is executed within each of these computational paths in parallel, and each path sees its associated objective function value stored in the output register, such that the state of the output register is dependent on the state of the input register. Where a classical computational processing unit is restricted to inspection of just a single point in the optimisation landscape at a time, the quantum computer is able to occupy and assess the entire landscape at once. 

This property may seem like it should be extremely powerful; if the quantum system occupies the entire space of solutions, and contains all the associated objective function values, there must be some way to reliably extract the optimal solution. However, the catch is that each computational path is occupied with an equal, and exponentially small probability amplitude $\frac{1}{\sqrt{N}}$ and when measured, the probability that we find the quantum computer to occupy any particular computational path is determined by the modulus squared of the probability amplitude contained within that path. In other words, measuring the system at this point would be akin to taking a single random sample from a uniform probability distribution over all solutions (exceedingly unlikely to return an optimal or even near-optimal solution). Rather than measuring right away, it is by making use of quantum interference between the computational paths that we are able to extract the optimal solution(s) from this entangled superposition. In fact, it is often the case that quantum interference holds the key to extracting meaningful quantum advantage. Shor's algorithm \cite{shor1994algorithms,shor1999polynomial}, for example, which is able to efficiently find prime factors of a large integer, and hence crack the RSA encryption scheme, relies upon the quantum Fourier transform \cite{coppersmith2002approximate} to extract the period of a periodic function over an exponentially large discrete domain. The quantum Fourier transform is efficiently implemented via quantum interference between computational paths, initially assigned, one each, to the discrete points in the function domain.

In order to similarly make use of quantum interference, we first apply a phase shift to each computational path proportional to the objective function value stored in the output register, with proportionality constant $\gamma$. We must also subsequently disentangle the input and output registers, by uncomputing the objective function (applying ${U_f}^\dagger$), such that the computational paths are left free to interfere. The net effect of these operations (discarding the ancilla qubits in the output register) is the phase separation unitary $U_Q(\gamma)$, controlled by a classically controlled parameter $\gamma$,
\begin{equation}
    \label{eq:action_of_UQ}
    U_Q(\gamma) \ket{\bm{x}} = e^{-\text{i} \gamma f(\bm{x})} \ket{\bm{x}},
\end{equation}
which acts on the equal superposition to produce the state,
\begin{equation}
    U_Q(\gamma) \ket{s} = \frac{1}{\sqrt{N}} \sum_{\bm{x} \in S} e^{-\text{i} \gamma f(\bm{x})} \ket{\bm{x}}.
\end{equation}

At this stage, the quantum system uniformly occupies the entire optimisation landscape, one computational path assigned to each feasible solution, with equal-sized probability amplitudes phase shifted proportional to objective function values of assigned solutions. Interference occurs when multiple phase-shifted probability amplitudes combine within a single computational path, and in order for interference to take effect globally, probability amplitudes should mix across all computational paths. 

The primary motivation of the non-variational QWOA is to provide a method by which to mix these probability amplitudes in order to reliably produce significant constructive interference within the computational path(s) assigned to the globally optimal solution(s). The mixing process is achieved via continuous-time quantum walks over graphs connecting computational paths (solution states), as in the quantum walk optimisation algorithm (QWOA) introduced by Marsh and Wang \cite{Marsh2019,marsh2020combinatorial}. The present method differs due mainly to the use of problem specific graphs and a non-variational amplification process.

\section{The algorithm (non-variational QWOA)}
For clarity and ease of reference, this section contains, in part, content duplicated from \cite{bennett2024non}.

\subsection{Definitions}
\label{sec:definitions}
Consider a combinatorial optimisation problem with a solution space $S$ containing $N$ feasible solutions ${\bm{x} = (x_1, x_2, ..., x_n)}$ each composed of $n$ decision variables, ${x_j \in \{0,1,...,k-1\}}$. The solution space may contain all possible solutions of this form, of which there are $N=k^n$. Alternatively, it may be restricted to solutions which satisfy some constraint, such as in the case of permutation-based problems where $k=n$ and $N=n!$. The solutions to the problem can be encoded in the computational basis states of a quantum computer by allocating to each variable a sub-register such that $\bm{x}$ is represented by the solution state,
\begin{equation}
    \ket{\bm{x}} = \prod_{j=1}^n \ket{x_j},
\end{equation}
where $\ket{x_j}$ is the computational basis state of the $j^{\text{th}}$ sub-register which directly encodes the decision variable $x_j$. If using a binary encoding, each sub-register will be assigned $\lceil \log_2 k \rceil$ qubits, whereas a one-hot encoding will use $k$ qubits. These encodings are described in more detail in \ref{sec:kmeans}. The binary encoding is more space efficient and lends itself to a more efficient implementation of the mixing unitary. However, the one-hot encoding may enable a more efficient implementation of the phase-separation unitary.

The state of the qubits in the input register is initialised in the equal superposition state, with equal probability amplitude assigned to each solution state, as in \cref{eq:equal_superposition}. The preparation of this equal superposition state is discussed for several problems in \cref{sec:maxcut,sec:kmeans,sec:QAP,sec:MIS,sec:CFLP}.

The non-variational QWOA is designed to find optimal or near-optimal solutions to the combinatorial optimisation problem, with respect to maximisation or minimisation of an objective function $f(\bm{x})$. The interference process responsible for amplifying the measurement probability of optimal and near-optimal solution states is driven by repeated applications of two unitary operations. The first of these is the phase-separation unitary, which applies a phase shift to each solution state, proportional to the associated objective function value, as described in \cref{sec:Motivation} and shown in \cref{eq:action_of_UQ}.

The second unitary performs a continuous-time quantum walk for time $t$ on the mixing graph which connects feasible solution states, 
\begin{equation}
    U_M(t) =  e^{-\text{i} t A},    
\end{equation}
where $A$ is the adjacency matrix which defines the mixing graph's structure. This unitary is referred to as the mixing unitary, or mixer for short, since it acts to distribute probability amplitudes between the solution states. Different versions of the mixer, appropriate for different problem structures, and the efficient implementation of each, is discussed in \cref{sec:maxcut,sec:kmeans,sec:QAP}.

\subsection{The mixing graph}
\label{sec:mixing_graph}
The adjacency matrix $A$ is formed by connecting solution states associated with solutions that are nearest neighbours. For the mixing graphs discussed in this paper, nearest neighbour solutions are those separated by a minimum non-zero Hamming distance, where the Hamming distance between any two solutions counts the number of mismatched decision variables between them. This generally corresponds with a Hamming distance of $1$, though not necessarily, as in the case of permutations, where the smallest possible Hamming distance between two different permutations is $2$.

More generally, we specify a set of $d$ moves (polynomial in problem size), each of which modifies any feasible solution to return a different feasible solution with similar configuration. Given a particular solution, its nearest neighbours are defined as those that can be generated with a single move from the set of available moves. Distance on the mixing graph therefore acts as a measure of similarity, with the distance between any two vertices counting the minimum number of moves required to transform between the associated solutions. In addition, since the nearest neighbours of each solution are generated by the same set of $d$ moves, the graph is vertex transitive with degree $d$.

\begin{figure}[htbp]
    \centering
    \captionsetup{margin=1cm, font=small}
    \hspace{0.4cm}
    \includegraphics[width=0.7\columnwidth]{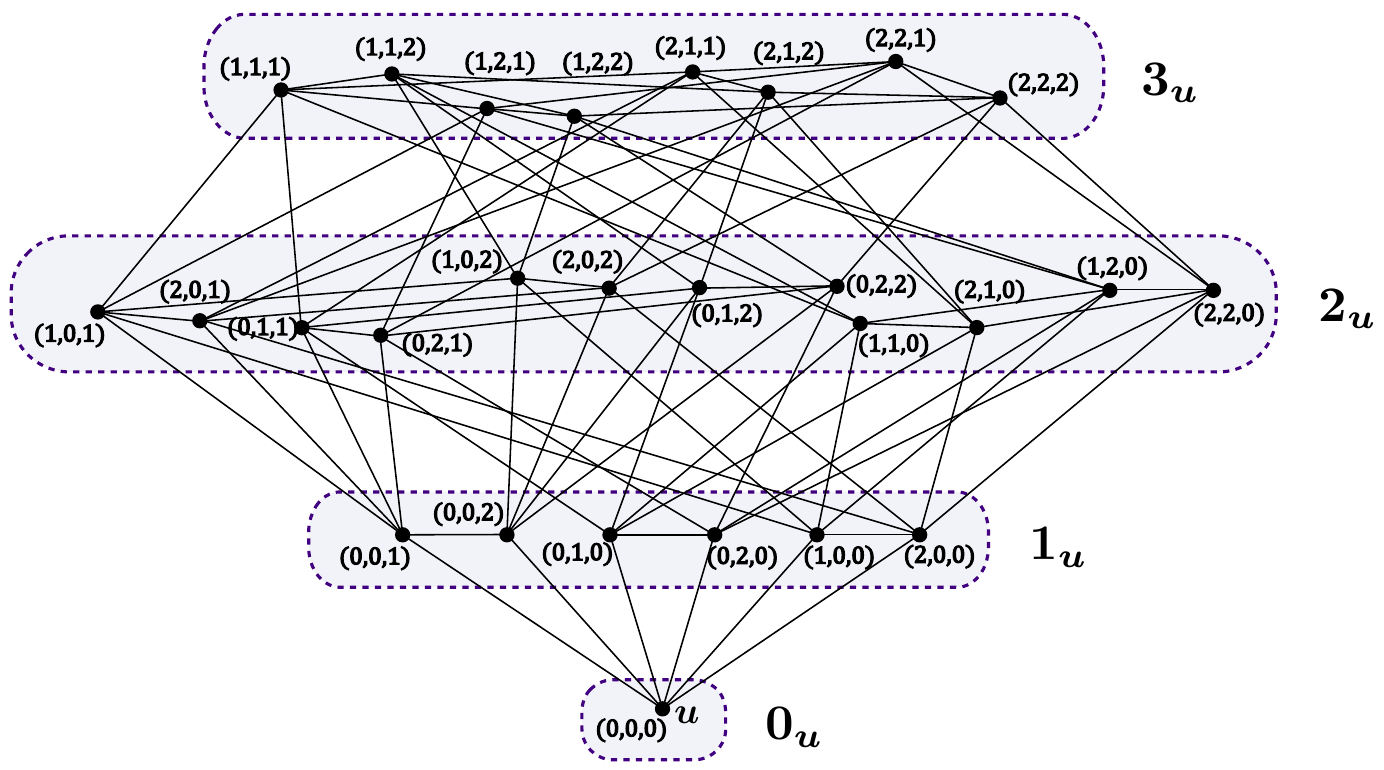}
    \caption{Example mixing graph connecting solutions to an integer variable problem, for $n=3$, $k=3$. Solution $\bm{u}=(0,0,0)$ is selected, and the graph is partitioned into subsets $h_{\bm{u}}$ for $h=0,1,2,3$.}
    \label{fig:vertex_subsets}
\end{figure}

The mixing graph defined by the adjacency matrix should have a diameter $D$ which scales linearly in the size of the problem instance. For effective performance the mixer should also satisfy two important conditions. Before describing these, it is useful to first define distance-based subsets of vertices on the graph. For any two vertices $\bm{u}$ and $\bm{v}$ we define ${\text{dist}(\bm{u},\bm{v})}$ as the distance on the graph between these vertices. For any vertex $\bm{u}$ and distance $h$, we define a subset of vertices,
\begin{equation}
    h_{\bm{u}} = \{\bm{v} : \text{dist}(\bm{u},\bm{v}) = h\},
\end{equation}
in other words, $h_{\bm{u}}$ is the subset of vertices which are a distance $h$ from $\bm{u}$ on the graph.  \cref{fig:vertex_subsets} provides an illustration of the partitioning of a mixing graph into these distance based subsets for a particular choice of vertex $\bm{u}$.

It may often be the case (as in the mixer corresponding with Fig.~\ref{fig:vertex_subsets} where the Hilbert space of each sub-register has dimension larger than $k=3$). In this case, the adjacency matrix defines a disconnected graph with several components (isolated subgraphs). The relevant subgraph connecting just the feasible solution states, which we refer to as the mixing graph, is just one of the components, where the others can be ignored. This is because the continuous-time quantum walk distributes probability amplitude only between connected vertices/states, such that the feasible solution states form an invariant subspace of the mixer.

\subsection{Necessary conditions}
Here we impose two conditions under which the algorithm is expected to work well. These conditions are satisfied within a surprisingly wide range of problems, as demonstrated in \cref{sec:maxcut,sec:kmeans,sec:QAP,sec:MIS,sec:CFLP}. The first condition, which relates to selecting a neighbourhood and designing a mixing graph, is that the action of the resulting mixer on an arbitrary solution state $\ket{\bm{u}}$, must be as follows:
\begin{equation}
    U_M(t) \ket{\bm{u}} = \sum_{h=0}^D \left( e^{-\text{i} h \phi(t)} \sum_{\bm{x} \in h_{\bm{u}}} r_{\bm{x}}(t) \ket{\bm{x}} \right),
    \label{eq:mixer_requirement_1}
\end{equation}
where $D$ is the diameter of the graph and (for sufficiently small $t$) $r_{\bm{x}}$ and $\phi$ are positive real-valued functions, also $0 < \phi(t) < \pi$. In other words, probability amplitude from an initial vertex is distributed to other vertices such that the complex phase of the distributed probability amplitudes is proportional to their distance from the initial vertex.

The second condition relates to the tendency for solutions with similar configurations to possess similar objective function values, and manifests due to the mixing graph clustering solutions with similar configurations. Firstly, the following relationship should be at least approximately satisfied,
\begin{equation}
    \left( \mu_{h\bm{x}} - f(\bm{x}) \right) \approx - \alpha_h \left( f(\bm{x}) - \mu \right),
    \label{eq:mixer_requirement_2}
\end{equation}
where $\mu_{h\bm{x}}$ is defined as the mean objective function value of solutions contained in $h_{\bm{x}}$, $\mu$ is the mean objective function value of all solutions in $S$, and $\alpha_h$ is a positive constant. Secondly, the constant of proportionality $\alpha_h$ should increase monotonically with increasing $h$ (up to a distance which is a considerable fraction of the graph's diameter $D$). Examples of this relationship are plotted in \cref{fig:maxcut_subset_means,fig:k_means_subset_means_transformed,fig:QAP_subset_means,fig:MIS_subset_means,fig:CFLP_tuned_pen_subset_means}.

\subsection{The amplified state}
Having established the phase-separation and mixing unitaries, and given a pre-selected number $p$ of iterations (which should be polynomial in the size of the problem instance), we define the amplified state,
\begin{equation}
    \ket{\gamma,t,\beta} = \left[ \prod_{i=0}^{p-1} U_M\left(t_i\right) U_Q\left(\frac{\pm \gamma_i}{\sigma}\right) \right] \ket{s},
    \label{eq:Amplified_state}
\end{equation}
where $t_i$ and $\gamma_i$ are obtained from specific formulae in terms of $\gamma$, $t$ and $\beta$ and where $\gamma>0$, $t>0$ and $0<\beta<1$. In other words, these three user-defined parameters determine the applied phase separations and the mixing times for each iteration. Specifically, the values of $\gamma_i$ increase linearly over the domain $[\beta\gamma,\gamma]$ and are given by the formula,
\begin{equation}
    \gamma_i = \left( \beta + (1-\beta) \frac{i}{p-1} \right) \gamma.
\end{equation}
Similarly the mixing times $t_i$ decrease linearly over the domain $[\beta t, t]$ and are given by the formula,
\begin{equation}
    t_i = \left( 1 - (1-\beta) \frac{i}{p-1} \right) t.
\end{equation}

These increasing and decreasing profiles are reminiscent of the annealing protocols for the adiabatic algorithm, though they are independently motivated, as discussed in more detail in \cref{sec:general_interference_process}. The $\pm$ in \cref{eq:Amplified_state} accounts for whether the goal is maximisation ($+$) or minimisation ($-$), and $\sigma$ is the standard deviation of objective function values, controlling for variation across problems and problem instances such that appropriate values of $\gamma$ are consistently of order $\gamma \approx 1$. An approximate value for $\sigma$ is sufficient (acquired through random sampling, for instance). An appropriate value of $\beta$ accounts for the nature of the problem, and the number of iterations, approaching zero for large iterations.

\subsection{Repeated state preparation and measurement}

Subject to the conditions described above, and allowing a sufficient number of iterations (polynomial in problem size), a wide range of appropriate values for $\gamma$, $t$ and $\beta$ significantly increase the measurement probability of globally optimal and near-optimal solutions within the amplified state $\ket{\gamma,t,\beta}$, such that a globally optimal or near-optimal solution is exceedingly likely to be measured following repeated state preparation and measurement of $\ket{\gamma,t,\beta}$.  

The total number of state preparations should be fixed, regardless of iteration count $p$ or problem size. 
The best solution measured during the process of repeated state preparation and measurement is taken as the solution, either exact or approximate, to the optimisation problem. Optionally, at regular intervals during the repeated state preparation and measurement, the values of $\gamma$, $t$ and $\beta$ can be updated in order to improve the amplification of optimal and near-optimal solutions, as indicated by improvement in approximated values of either the expectation value of the objective function, or a related measure, such as the Conditional Value at Risk (CVaR) \cite{barkoutsos2020improving}. 

The reason that the algorithm remains effective while constrained to a fixed total number of state-preparations and measurements, and the reason it is best characterised as non-variational, is because the optimal set of parameters can be determined via a fixed complexity 3-dimensional optimisation which seeks only a local extremum (that closest to the origin). In any case, even sub-optimal parameter values significantly amplify optimal and near-optimal solutions, such that the improvement/optimisation of parameters is optional.

\subsection{Constrained problems}
\label{sec:constrained_problems}
There are two ways that problem constraints can be dealt with in this framework. The first approach is to restrict the space of feasible solutions $S$ to include just those solutions which are valid (do not violate the constraints). This approach is only viable if the space of valid solutions can be characterised independently of the problem instance, if it is possible to efficiently implement a mixer that connects only valid solutions and if the mixer satisfies the necessary condition in \cref{eq:mixer_requirement_1}. An example of this approach is demonstrated in \cref{sec:QAP} for the quadratic assignment problem.

The other approach is to include both valid and invalid solutions and for the constraints to manifest as additional terms in the objective function, which act to penalise invalid solutions. This approach is often necessary, as it is frequently the case that the space of valid solutions cannot be predetermined. This penalty function approach is discussed and demonstrated in \cref{sec:pen_func,sec:MIS,sec:CFLP}. 

Note that while it may, in principle, be possible to use amplitude amplification and the Grover mixer \cite{Grover_mixer} to prepare the initial equal superposition over only the valid states and to subsequently mix between them, this is not an effective approach. Interpreted as a CTQW, the Grover mixer mixes between valid states connected via a complete graph such that it cannot exploit problem structure and additionally can be demonstrated to produce an upper bound quadratic speedup, certainly not sufficient to solve NP-hard problems. 

\section{The interference process}
\label{sec:general_interference_process}

Given an initial arbitrary superposition over solution states,
\begin{equation}
    \ket{\psi} = \sum_{\bm{x} \in S} c_{\bm{x}} \ket{\bm{x}},
    \label{eq:arb_state}
\end{equation}
a corollary to \cref{eq:mixer_requirement_1} is that the probability amplitude arriving at an arbitrary solution $\bm{u}$ following application of the mixer is given by,
\begin{equation}
    \bra{\bm{u}}U_M(t)\ket{\psi} = \sum_{h=0}^D \left( e^{-\text{i} h \phi(t)} \sum_{\bm{x} \in h_{\bm{u}}} r_{\bm{x}}(t) c_{\bm{x}} \right).
\end{equation}
Recall that $c_{\bm{x}}$ in \cref{eq:arb_state}, and the probability amplitudes in general, are complex valued, and their modulus squared determines measurement probability.

As the mixing graph is vertex transitive with degree $d$, the equal superposition state is an eigenstate of the mixer with eigenvalue $e^{-\text{i} d t}$. Therefore, for the equal superposition, the probability amplitude at an arbitrary solution $\bm{u}$ evolves under the mixer as follows,
\begin{equation}
    \bra{\bm{u}}U_M(t)\ket{s} = \frac{1}{\sqrt{N}} \sum_{h=0}^D \left( e^{-\text{i} h \phi(t)} \sum_{\bm{x} \in h_{\bm{u}}} r_{\bm{x}}(t) \right) = \frac{e^{-\text{i} d t}}{\sqrt{N}}.
    \label{eq:Um_on_s}
\end{equation}
Isolating just the following equality,
\begin{equation}
    \sum_{h=0}^D \left( e^{-\text{i} h \phi(t)} \sum_{\bm{x} \in h_{\bm{u}}} r_{\bm{x}}(t) \right) = e^{-\text{i} d t},
    \label{eq:rotating_unit_vector}
\end{equation}
we have the sum of exponentially many individually contributing complex terms, each of magnitude $r_{\bm{x}}(t)$, and these contributions sum to a resultant of unit magnitude. 

Consider now the action of the mixer on solution $\bm{u}$ as per \cref{eq:mixer_requirement_1}. At time $t=0$, probability amplitude is contained just at $\bm{u}$. Due to the nature of a continuous-time quantum walk, and the fact that the mixing graph contains exponentially many vertices within a small distance of $\bm{u}$, for sufficient walk times $t$, probability amplitude is distributed across exponentially many vertices. Though we do not prove it here, we make the reasonable assumption that this results in exponentially many not insignificant terms, $r_{\bm{x}}(t)$. While the sum of their squares is constrained to 1 (due to normalisation), these terms represent a total probability amplitude much greater than $1$ (${\sum_{\bm{x} \in S} r_{\bm{x}}(t) >> 1}$). 

Referring back to \cref{eq:rotating_unit_vector}, the exponentially many individually contributing complex terms arriving at $\bm{u}$ represent a total probability amplitude which is significantly larger than one (for sufficient time), and yet, the magnitude of their resultant remains equal to one, with a phase that rotates in time $t$. In other words, the individual terms largely destructively interfere. This simple dynamic emerges from the orchestrated arrival of probability amplitudes from other solutions, each of which is phase-shifted depending on distance from the solution $\bm{u}$, with each increment in distance rotating phase negatively/anti-clockwise by an additional $\phi(t)$ radians. If there were a way to reduce this offset between the complex phase of subsequent subsets, for example, by introducing some offset phase proportional to $h$, then it would be possible to bring the individually contributing probability amplitudes closer in phase with each other, allowing for significant constructive interference and hence amplification of a particular solution state. This offset in phase can be partly achieved with the phase-separation unitary, but before demonstrating this, we establish a useful approximation.

Consider the sum of complex numbers which are approximately in-phase,
\begin{equation}
    \sum_m r_m e^{\text{i} \phi_m},
\end{equation}
for which we want to approximate the resultant. A detailed derivation and demonstration is presented in \cref{sec:deriving_the_approximation}, where by defining the weighted mean $E$ and variance $V$ of $\phi_m$,
\begin{equation}
    E = \frac{\sum_m r_m \phi_m}{\sum_m r_m}, \hspace{2cm}  V = \frac{\sum_m r_m \left( \phi_m - E \right)^2}{\sum_m r_m},
\end{equation}
the resultant can be approximated as,
\begin{equation}
    \sum_m r_m e^{\text{i} \phi_m} \approx \left( \sum_m r_m \right) e^{-\frac{V}{2}} e^{\text{i} E}.
    \label{eq:approximation}
\end{equation}
Each term in the expression can be understood intuitively, $e^{\text{i} E}$ corresponds with the fact that the resultant has a phase centred within that of the individual contributions, and $e^{\frac{-V}{2}}$ corresponds with the tendency of the resultant to shrink as the phases of individual contributions become more dispersed. This result, as detailed in \cref{sec:deriving_the_approximation}, is significant, as it allows us to analyse the sum of an arbitrarily large number of complex terms, simply by studying their statistics, given of course that the terms are approximately in-phase.

The probability amplitude at an arbitrary solution state $\ket{\bm{u}}$, following both phase-separation and mixing applied to the equal superposition, can be expressed as,
\begin{equation}
    \bra{\bm{u}}U_M(t)U_Q(\gamma)\ket{s} = \frac{1}{\sqrt{N}} \sum_{h=0}^D \left( e^{-\text{i} h \phi(t)} \sum_{\bm{x} \in h_{\bm{u}}} r_{\bm{x}}(t) e^{-\text{i} \gamma f(\bm{x})} \right).
    \label{eq:single_iteration}
\end{equation}

We can now use the established approximation (\cref{eq:approximation}) to significantly simplify this expression, by taking advantage of the statistics of the objective function values contained within each subset. First we define the weighted mean $E_{h\bm{u}}$ and variance $V_{h\bm{u}}$ of $f(\bm{x})$ within each subset $h_{\bm{u}}$, where we take $r_{\bm{x}}(t)$ as the weights. Now the sum of complex terms from \cref{eq:single_iteration} can be simplified,
\begin{equation}
    \sum_{\bm{x} \in h_{\bm{u}}} r_{\bm{x}}(t) e^{-\text{i} \gamma f(\bm{x})} \approx \left(\sum_{\bm{x} \in h_{\bm{u}}} r_{\bm{x}}(t)\right) e^{\frac{-\gamma^2 V_{h\bm{u}}}{2}} e^{-\text{i} \gamma E_{h\bm{u}}}.
\end{equation}
This approximation should be accurate if $\gamma$ is sufficiently small. Note that the algorithm takes an initial value of $\gamma<\frac{1}{\sigma}$ where $\sigma$ is the standard deviation of $f(\bm{x})$ across all solutions. This ensures that dispersion within the resulting phases should be within the regime where this approximation is appropriate. Making the substitution, \cref{eq:single_iteration} simplifies,
\begin{equation}
    \bra{\bm{u}}U_M(t)U_Q(\gamma)\ket{s} \approx \frac{1}{\sqrt{N}} \sum_{h=0}^D \left( e^{-\text{i} h \phi(t)} e^{-\text{i} \gamma E_{h\bm{u}}} e^{\frac{-\gamma^2 V_{h\bm{u}}}{2}} \sum_{\bm{x} \in h_{\bm{u}}} r_{\bm{x}}(t) \right).
\end{equation}

It is expected that the values of $r_{\bm{x}}(t)$ for each subset $h_{\bm{u}}$ would be uncorrelated with objective function value $f(\bm{x})$, so the weighted mean can be approximated as the unweighted mean, $E_{h\bm{u}} \approx \mu_{h\bm{u}}$. Since it's assumed that the necessary conditions have been satisfied, \cref{eq:mixer_requirement_2} can be substituted, simplifying the expression further,
\begin{equation}
    \fl \hspace{1cm} \bra{\bm{u}}U_M(t)U_Q(\gamma)\ket{s} \approx \frac{e^{-\text{i} \gamma f(\bm{u})}}{\sqrt{N}} \sum_{h=0}^D \left( e^{-\text{i} h \phi(t)} e^{\text{i} \gamma \alpha_h \left( f(\bm{u}) - \mu \right)} e^{\frac{-\gamma^2 V_{h\bm{u}}}{2}} \sum_{\bm{x} \in h_{\bm{u}}} r_{\bm{x}}(t) \right).
    \label{eq:simplified_single_iteration}
\end{equation}
Since $\alpha_h$ increases monotonically with increasing distance $h$, the term $e^{\text{i} \gamma \alpha_h \left( f(\bm{u}) - \mu \right)}$, depending on the sign of $\gamma (f(\bm{u}) - \mu)$, acts to either counteract or increase the phase offsets produced by $e^{-\text{i} h \phi(t)}$, causing either constructive or destructive interference, respectively. This means for positive $\gamma$, solutions with $f(\bm{x})>\mu$ are amplified, and those for which $f(\bm{x})<\mu$ are suppressed. The opposite is true for negative $\gamma$, hence the role of the $\pm$ in \cref{eq:Amplified_state}. Moreover, the size of $\left( f(\bm{u}) - \mu \right)$ also determines the amount of constructive/destructive interference, with solutions of more extreme objective function values experiencing larger interference effects. The effect of this is that the resulting amplification of solutions either increases or decreases monotonically with objective function value, depending whether we are dealing with a maximisation or minimisation problem, respectively. \emph{For the remainder of this section, and for the sake of clarity, we discuss only the maximisation case.} This involves no loss of generality, since the $\pm$ in \cref{eq:Amplified_state} ensures functional equivalence.

Rather than maximising the interference effects for a single iteration, a sufficiently small value of $\gamma$ produces less amplification of optimal solutions but better preserves coherence in the probability amplitudes assigned to each solution state (following the first iteration). Where coherence is (approximately) preserved (probability amplitudes remain approximately in-phase), subsequent iterations can continue to produce additional constructive interference in much the same manner as the first iteration. Consider the state of the system after $i$ iterations,
\begin{equation}
    \ket{\psi_i} = \frac{1}{\sqrt{N}} \sum_{\bm{x} \in S} a_{\bm{x}} e^{\text{i} \varphi_{\bm{x}}} \ket{\bm{x}},
\end{equation}
where $a_{\bm{x}}$ is a real value accounting for the amplification (constructive and destructive interference) produced by the prior iterations, and $\varphi_{\bm{x}}$ is also a real value which accounts for the rotation (and dispersion) in complex phase induced by the previous iterations. The probability amplitude at an arbitrary solution state $\bm{u}$ following an additional iteration can be expressed as,
\begin{equation}
    \bra{\bm{u}}U_M(t)U_Q(\gamma)\ket{\psi_i} = \frac{1}{\sqrt{N}} \sum_{h=0}^D \left( e^{-\text{i} h \phi(t)} \sum_{\bm{x} \in h_{\bm{u}}} a_{\bm{x}} r_{\bm{x}}(t) e^{-\text{i} \gamma f(\bm{x})}  e^{\text{i} \varphi_{\bm{x}}} \right).
\end{equation}
The expression can be simplified by using the approximation from \cref{eq:approximation} and substituting the weighted means and variances, 
\numparts
\begin{eqnarray}
    E_{f_{h\bm{u}}} =  \frac{\sum_{\bm{x} \in h_{\bm{u}}} a_{\bm{x}} r_{\bm{x}}(t) f(\bm{x})}{\sum_{\bm{x} \in h_{\bm{u}}} a_{\bm{x}} r_{\bm{x}}(t)}\\
    E_{\varphi_{h\bm{u}}} =  \frac{\sum_{\bm{x} \in h_{\bm{u}}} a_{\bm{x}} r_{\bm{x}}(t) \varphi_{\bm{x}}}{\sum_{\bm{x} \in h_{\bm{u}}} a_{\bm{x}} r_{\bm{x}}(t)} \\
    E_{h\bm{u}} = -\gamma E_{f_{h\bm{u}}} + E_{\varphi_{h\bm{u}}} \\
    V_{f_{h\bm{u}}} = \frac{\sum_{\bm{x} \in h_{\bm{u}}} a_{\bm{x}} r_{\bm{x}}(t) \left( f(\bm{x}) - E_{f_{h\bm{u}}} \right)^2 }{\sum_{\bm{x} \in h_{\bm{u}}} a_{\bm{x}} r_{\bm{x}}(t)} \\
    V_{\varphi_{h\bm{u}}} = \frac{\sum_{\bm{x} \in h_{\bm{u}}} a_{\bm{x}} r_{\bm{x}}(t) \left( \varphi_{\bm{x}} - E_{\varphi_{h\bm{u}}} \right)^2 }{\sum_{\bm{x} \in h_{\bm{u}}} a_{\bm{x}} r_{\bm{x}}(t)} \\
    C_{h\bm{u}} = \frac{\sum_{\bm{x} \in h_{\bm{u}}} a_{\bm{x}} r_{\bm{x}}(t) \left[ f(\bm{x}) \left( \varphi_{\bm{x}} - E_{\varphi_{h\bm{u}}}\right) + E_{f_{h\bm{u}}} \left( E_{\varphi_{h\bm{u}}} - \varphi_{\bm{x}} \right)   \right] }{\sum_{\bm{x} \in h_{\bm{u}}} a_{\bm{x}} r_{\bm{x}}(t)} \\
    V_{h\bm{u}} = (\gamma)^2 V_{f_{h\bm{u}}} + V_{\varphi_{h\bm{u}}} - 2\gamma C_{h\bm{u}}
\end{eqnarray}
\endnumparts
such that, 
\begin{equation}
    \bra{\bm{u}}U_M(t)U_Q(\gamma)\ket{\psi_i} \approx \frac{1}{\sqrt{N}} \sum_{h=0}^D \left( e^{-\text{i} h \phi(t)} e^{\text{i} E_{h\bm{u}}} e^{\frac{-V_{h\bm{u}}}{2}} \sum_{\bm{x} \in h_{\bm{u}}} a_{\bm{x}} r_{\bm{x}}(t) \right).
\end{equation}

\begin{figure}[htbp]
    \centering
    \captionsetup{margin=1cm, font=small}
    \begin{subfigure}{0.48\columnwidth}
        \centering
        \includegraphics[width=1\columnwidth]{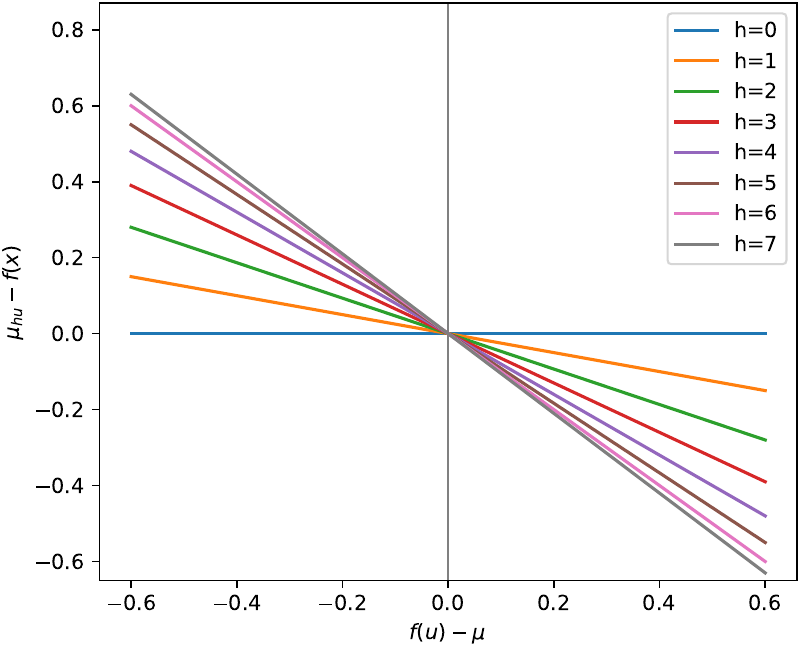}
        \caption{}
        \label{fig:unamplified_subset_means}
    \end{subfigure}
    \vspace{-0.25cm}
    \begin{subfigure}{0.48\columnwidth}
        \centering
        \includegraphics[width=1\columnwidth]{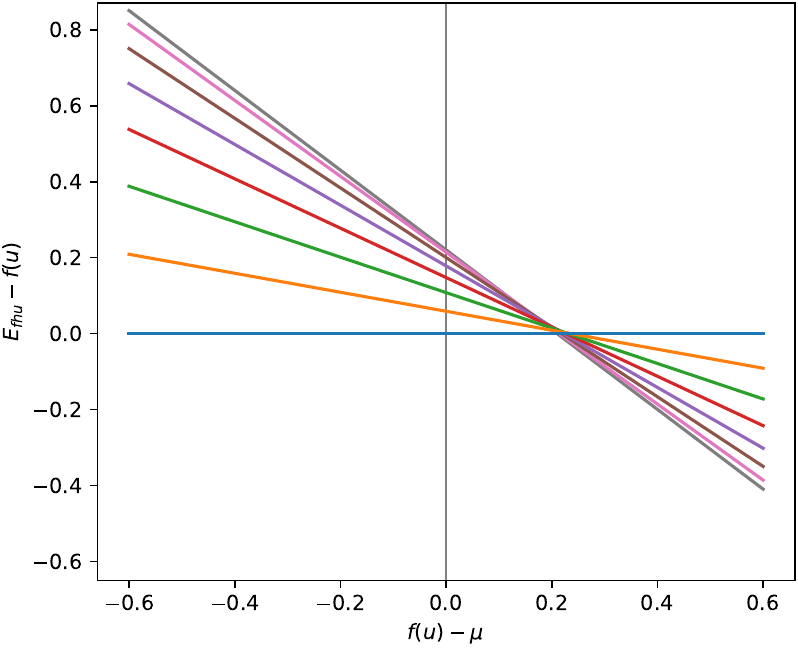}
        \caption{}
        \label{fig:maximisation_subset_means}
    \end{subfigure}
    \caption{(a) Example subset means, $\mu_{h\bm{u}}$ distributed as per \cref{eq:mixer_requirement_2}. (b) Weighted subset means $E_{f_{h\bm{u}}}$ as they may be expected to appear under an amplification profile which increases monotonically with increasing objective function value $f(\bm{u})$.}
    \label{fig:evolving_subset_means}
\end{figure}

As with the first iteration, constructive interference in subsequent iterations is produced by offsetting the phase rotation induced by the term $e^{-\text{i} h \phi(t)}$. This is achieved by the phase-separation unitary through the term, $e^{\text{i} E_{h\bm{u}}}$. Since ${E_{h\bm{u}} = -\gamma E_{f_{h\bm{u}}} + E_{\varphi_{h\bm{u}}}}$ this should be effective if the weighted-mean values $E_{f_{h\bm{u}}}$ increase/decrease monotonically with distance $h$.
We expect this to be the case when amplification values $a_{\bm{x}}$ increase  monotonically with objective function value $f(\bm{x})$, as they do after the first iteration. Without amplification, the weighted subset means take on values as per \cref{eq:mixer_requirement_2}, an example illustration of which is shown in \cref{fig:unamplified_subset_means}, clearly displaying the necessary increasing/decreasing weighted subset means. Under the monotonically increasing amplification profile, we expect the distribution of weighted subset means to be modified as per the example illustration in \cref{fig:maximisation_subset_means}. This can be explained by the simultaneous amplification of high objective function solutions and suppression of low objective function solutions, which causes $E_{f_{h\bm{u}}}$ to increase for all solutions $\bm{u}$. Due to the fact that solutions in $h_{\bm{u}}$ are composed by $h$ moves applied to solution $\bm{u}$, more distant subsets explore a wider range of solution configurations and therefore $E_{f_{h\bm{u}}}$ should increase more in more distant subsets. This kind of modified distribution of weighted subset means is also demonstrated via simulations in \cref{sec:maxcut_multiple_iterations}.

\begin{figure}[htbp]
    \centering
    \captionsetup{margin=1cm, font=small}
    \begin{subfigure}{0.48\columnwidth}
        \centering
        \includegraphics[width=1\columnwidth]{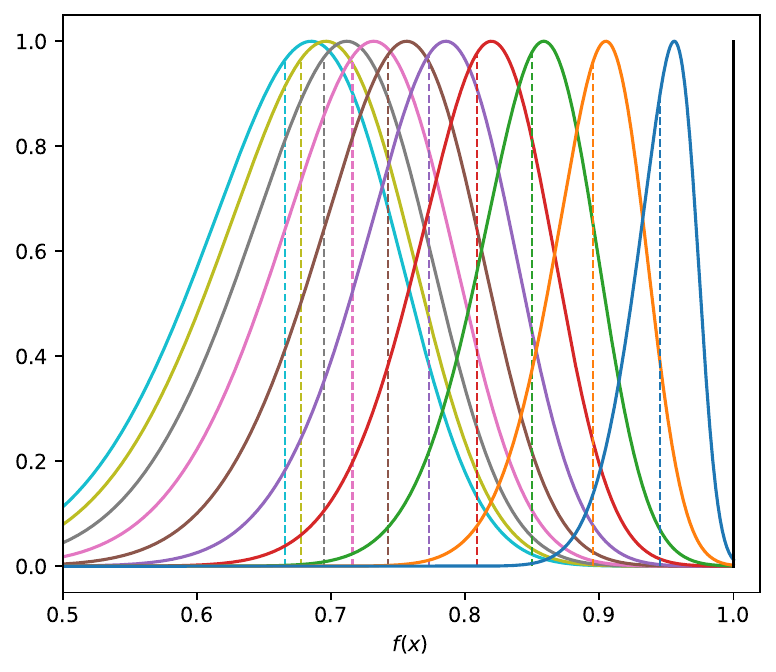}
        \caption{}
    \end{subfigure}
    \vspace{-0.25cm}
    \begin{subfigure}{0.48\columnwidth}
        \centering
        \includegraphics[width=1\columnwidth]{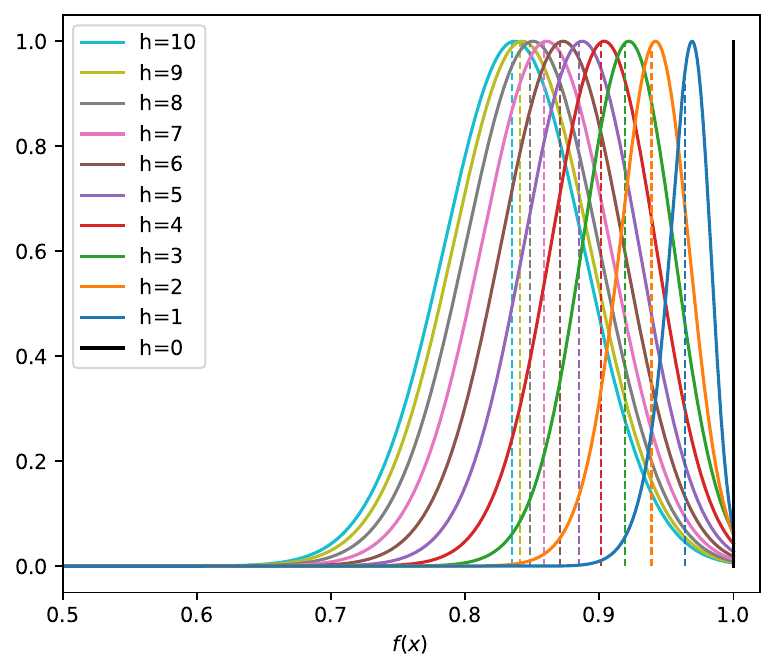}
        \caption{}
    \end{subfigure}
    \caption{(a) Illustrative example of how objective function values are likely to be distributed in each subset $h_{\bm{u}}$ for an optimal solution $\bm{u}$ to a maximisation problem. Skew-normal distributions are skewed away from the optimal value, as may be expected for a space of solutions with initially normally distributed objective function values and for which mean values in $h_{\bm{u}}$ are monotonically decreasing. (b) Weighted distributions with an amplification profile, $a_{\bm{x}} = 10^{20(f(\bm{x})-0.8)}$. In all cases, the weighted-mean location is shown with the dashed vertical lines. It is clear that after the amplification process, weighted-mean values preserve the monotonic relationship. All distributions are scaled relative to their maximum value, for clarity.}
    \label{fig:evolving_weighted_means}
\end{figure}

Several key features are displayed in \cref{fig:maximisation_subset_means}, namely, the relationship from \cref{fig:unamplified_subset_means} is largely preserved. The transition from monotonically increasing to monotonically decreasing $E_{f_{h\bm{u}}}$ has shifted to the right, isolating a smaller number of higher objective function solutions within the constructive interference regime. Note the optimal (and nearest optimal) solutions are expected to retain their monotonically decreasing subset means, even with significant amplification. An illustration of this is included in \cref{fig:evolving_weighted_means}, using skew normal distributions to model the distribution of objective function values in $h_{\bm{u}}$, where for illustrative purpose it is assumed that the solution space cardinality is sufficiently large, such that the distribution of objective function values within each subset can be approximated by a continuous and smooth curve. 

With larger amounts of amplification, a reducing number of solutions continue to be amplified in subsequent iterations. In addition, the monotonic decreases/increases in values of $E_{f_{h\bm{u}}}$ increase in size for solutions with more extreme values of $f(\bm{u})$. All of this together results in an amplification profile that we expect to continue to maintain its monotonically increasing property, and for which we expect amplification of the optimal and near-optimal solutions to experience compounding growth with subsequent iterations. Note that numerical evidence supporting these expectations is provided in \cref{sec:maxcut_multiple_iterations}.

As the amplification process concentrates probability amplitude in a decreasing number of optimal and near-optimal solutions, the resulting decreasing differences between values of $E_{f_{h\bm{u}}}$ for optimal and near-optimal solutions, illustrated in both \cref{fig:evolving_subset_means} and \cref{fig:evolving_weighted_means}, necessitates an increasing value of $\gamma$ to achieve an equivalent (or increasing) amount of relative phase rotation each iteration between probability amplitudes contributed from different subsets $h_{\bm{u}}$. In addition, the concentration of probability amplitudes reduces weighted variances $V_{f_{h\bm{u}}}$, reducing subsequent dispersion of phases and better enabling the increased values of $\gamma$.

An additional effect of a monotonic amplification profile which increases probability amplitude in near-optimal solutions and decreases probability amplitude for sub-optimal solutions is that it affects the distribution of probability amplitudes within subsets $h_{\bm{u}}$ depending on objective function value $f(\bm{u})$. Optimal and near-optimal solutions are expected to see an increasing probability amplitude in nearby subsets (small $h$) and a decreasing probability amplitude in more distant subsets. This is simply because nearby subsets are expected to contain a larger fraction of near-optimal solutions, which are preferentially amplified. On the other hand, sub-optimal solutions, those with $f(\bm{x})<\mu$, are expected to see a decreasing probability amplitude in nearby subsets and an increasing probability amplitude in distant subsets. This is because the nearby subsets contain a larger fraction of suppressed solutions and more distant subsets typically contain a larger fraction of amplified and near-optimal solutions. It is for this reason that a decreasing mixing time is effective. As probability amplitude converges closer to optimal solutions and further from sub-optimal solutions, the smaller mixing times minimise the distribution of probability amplitudes back into the suppressed sub-optimal solutions. In addition, the constructive interference process at optimal and near-optimal solutions becomes more effective with smaller mixing times, simply because the arrival of probability amplitudes from nearby subsets is favoured by smaller mixing times.

\section{Demonstrating the interference process for weighted maxcut}
\label{sec:maxcut}

\subsection{Defining the weighted maxcut problem and establishing an appropriate mixer}
A weighted maxcut problem of size $n$ is characterised by a weighted graph $G(V,E,W)$ containing $n$ vertices, where the aim is to find a partitioning of the graph's vertices into two subsets such that the total weight of edges passing between the two subsets is maximised. Each solution to an instance of the maxcut problem, or equivalently, each partitioning of the $n$ vertices, can be identified with a bit-string, $\bm{x} = (x_1, x_2, ..., x_{n})$ of length $n$, where each entry, or decision variable, $x_j \in \{0,1\}$, corresponds to placement of vertex $j$ into the first or second subset respectively. The objective function can then be expressed,
\begin{equation}
    f(\bm{x}) = \sum_{(i,j) \in E} w_{ij} (x_i - x_j)^2.
\end{equation}

Upon careful consideration, enumerating all solutions as all possible bit-strings of length $n$ introduces a mirror symmetry and hence a two-fold degeneracy in the solution space. This degeneracy or mirror symmetry is not detrimental to the performance of the algorithm, since solutions are connected via a vertex-transitive graph, and each pair of mirror-symmetric solutions behaves identically within the graph due to functionally identical connectivity with neighbouring solutions. There is also no preference for measuring one solution or the other from a mirror-symmetric pair, since they are both equivalent. In addition, if the degeneracy is removed by freezing one vertex of the maxcut graph into one of the two subsets, one of the available moves is removed when defining the mixing-graph, which may (slightly) negatively affect algorithm performance.

Consider now an $n$ qubit quantum system with a Hilbert space of dimensionality, $N=2^n$. A natural embedding of the solutions can be achieved by allowing computational basis states of each qubit to encode each of the binary variables $x_j$. In this way, each of the computational basis states of the $n$ qubit system, is a solution state which directly encodes the solution with a matching bit-string. This corresponds with the solution state definition in \cref{sec:definitions} where each sub-register contains only a single qubit. Preparing the initial equal superposition over all solution states is trivial, as every computational basis state of the input register is also a solution state,
\begin{equation}
    \ket{s} = H^{\otimes n} \ket{0}^{\otimes n} = \frac{1}{\sqrt{2^n}} \sum_{x=0}^{2^n-1} \ket{x} = \frac{1}{\sqrt{N}} \sum_{\bm{x} \in S} \ket{\bm{x}}.   
\end{equation}

The smallest and most natural perturbation in the configuration of a maxcut solution would be achieved by moving one of the vertices of the maxcut graph to a subset different to the one to which it is already assigned, which would correspond with a single bit-flip in the solution's bit-string. A natural choice for the nearest neighbours of any given solution is the set of solutions which differ by a single bit-flip. So we define the set of moves for a size $n$ maxcut problem as the $n$ available bit-flips. Assigning vertices to each solution $\bm{x}$ and generating edges via bit-flips, results in a hypercube graph. The adjacency matrix for this hypercube can also be expressed in terms of the Pauli X matrix $\sigma_x$ which generates these bit-flips,
\begin{equation}
    A = \sum_{j=1}^{n} \id^{\otimes j-1} \otimes \sigma_x \otimes \id^{\otimes n-j}.
    \label{eq:Adjacency_binary_mixer}
\end{equation}
Note that this adjacency matrix is identical to the mixing Hamiltonian (transverse-field Hamiltonian) used within the originally conceived QAOA by Farhi \emph{et al.} \cite{Farhi2014QAOA}. Time evolution under this Hamiltonian for time $t$ performs a CTQW over the hypercube defined by $A$, so the mixing unitary is given by,
\numparts
\begin{eqnarray}
    \fl \hspace{0.5cm} U_M(t) &= e^{-\text{i} t A} \\
    \fl       &= \prod_{j=1}^n e^{-\text{i} t \sigma_x} \\
    \fl       &= \prod_{j=1}^n \left( \left(\cos{t}\right) \id - \text{i} \left(\sin{t}\right) \sigma_x \right) \\
    \fl       &= \sum_{h=0}^n \left( (-\text{i})^h(\cos{t})^{n-h}(\sin{t})^{h} \sum_{\text{h-comb}} \left( \prod_{j=1}^n \cases{\sigma_x & \mbox{if $ j \in $ h-comb} \\ \id & \mbox{otherwise} \\} \right) \right),
\end{eqnarray}
\endnumparts
where the second sum is over all h-combinations, i.e. summing over each of the ways in which $h$ bits can be selected for bit-flips. Note that all of the solutions which are $h$ bit-flips away from a solution $\bm{u}$ are those in set $h_{\bm{u}}$, so the action of the mixer on an arbitrary solution state can be expressed,
\begin{equation}
    U_M(t) \ket{\bm{u}} = \sum_{h=0}^{n} \left( (-\text{i})^h (\cos{t})^{n-h} (\sin{t})^h \sum_{\bm{x} \in h_{\bm{u}}}  \ket{\bm{x}} \right).
    \label{eq:dispersing_amplitude}
\end{equation}

This expression satisfies the first condition for an appropriate mixer, because the complex phase of the distributed probability amplitudes is proportional to distance, $e^{-\text{i} h \phi}$ where $\phi = \frac{\pi}{2}$. Also the expression $(\cos{t})^{n-h} (\sin{t})^h$ is real and positive valued for $t<\frac{\pi}{2}$. The diameter of the graph is linear in the problem size ($D=n$), which makes intuitive sense, since any two solutions are separated by at most $n$ bit-flips. Note that this mixer also satisfies a stronger condition, that the distributed probability amplitudes are exactly equal within each subset $h_{\bm{u}}$, though we shall see with the quadratic assignment problem in \cref{sec:QAP} (and in general any permutation based problem) that this is not necessary for effective performance.

For many problems, the second condition for the mixer, as in \cref{eq:mixer_requirement_2}, can be verified numerically but for weighted maxcut, it can be demonstrated analytically. Consider a single edge on the maxcut graph. Whether the edge is included in the cut or not, and hence whether its weight is added to the objective function value, depends only on the placement of the two vertices to which it connects. With respect to the bit-string encoding of solutions, the two bits which determine the placement of these vertices must have opposite values in order for the connecting edge to be included in the cut. Considering all possible bit-strings or solutions, we expect any two bits to have opposite values in exactly half of the solutions. Therefore, out of $N=2^n$ possible solutions, each edge is included in $N/2$ solutions. Therefore, the mean objective function value of all solutions is determined solely by the sum of all edge weights in the parent graph,
\begin{equation}
    \mu = \frac{1}{N} \sum_{\bm{x} \in S} f(\bm{x}) = \frac{1}{2} \sum w.
    \label{eq:mu}
\end{equation}

In general, when considering the subset of solutions $h_{\bm{u}}$, relative to solution $\bm{u}$, a single edge is either removed from the cut if it was already included, or added to the cut if it was not initially included. This happens only when exactly one of the edge's connecting vertices has its bit value flipped. Of the ${n \choose h}$ solutions in $h_{\bm{u}}$, this occurs in exactly $2{n-2 \choose h-1}$ solutions. The total weight of edges already included in the cut is given by $f(\bm{u})$, and the total weight of edges not included in the cut is therefore $\left(\sum\omega\right)-f(\bm{u})$. So the mean objective function value of solutions in $h_{\bm{u}}$ can be determined as,
\begin{equation}
    \fl \hspace{1.5cm} \mu_{h\bm{u}} = \frac{1}{{n \choose h}} \sum_{\bm{x} \in h_{\bm{u}}} f(\bm{x}) = \frac{{n \choose h} f(\bm{u}) + 2{n-2 \choose h-1}\left[ \left(\sum\omega\right)-f(\bm{u}) \right] - 2{n-2 \choose h-1}f(\bm{u})}{{n \choose h}}.
\end{equation}
By substituting $\sum\omega=2\mu$, and expressing the binomial coefficients in terms of factorials, the expression simplifies to
\begin{equation}
    \mu_{h\bm{u}} =  f(\bm{u}) - 4  \frac{h(n-h)}{n(n-1)} \left( f(\bm{u}) - \mu \right),
    \label{eq:maxcut_2nd_condition}
\end{equation}
which can be rearranged, such that it is exactly equivalent to \cref{eq:mixer_requirement_2} by making the substitution,
\begin{equation}
    \alpha_h = 4  \frac{h(n-h)}{n(n-1)}.
\end{equation}

The $h(n-h)$ in the numerator is the only part of the expression for $\alpha_h$ which varies with distance $h$, and it increases monotonically with increasing distance $h$ for half of the diameter ${0 < h \leq \frac{n}{2}}$, which satisfies the necessary condition. The relationship in \cref{eq:maxcut_2nd_condition} is illustrated for a particular weighted maxcut problem instance in \cref{fig:subset_means}. The reason that $\alpha_h$ increases over only half the graph's diameter is due to the mirror symmetry, as two different subsets, $h_{\bm{u}}$ and $h_{\bm{u}}'$ consist of mirror-image solutions when $h+h'=n$, in which case,  $\mu_{h\bm{u}} = \mu_{h'\bm{u}}$. 

\subsection{Interference effects: Single iteration}

Building on \cref{eq:dispersing_amplitude}, we can show that for an arbitrary state,
\begin{equation}
    \ket{\psi} = \sum_{\bm{x} \in S} c_{\bm{x}} \ket{\bm{x}},
\end{equation}
the action of the hypercube mixer transforms the probability amplitude of arbitrary solution state $\ket{\bm{u}}$ as follows:
\begin{equation}
    \bra{\bm{u}}U_M(t)\ket{\psi} =  \sum_{h=0}^{n} \left( (-\text{i})^h (\cos{t})^{n-h} (\sin{t})^h \sum_{\bm{x} \in h_{\bm{u}}}  c_{\bm{x}} \right).
    \label{eq:collecting_amplitude}
\end{equation}

Since the hypercube is a vertex-transitive graph with degree $d=n$, the equal superposition over all vertices is the maximal eigenvector of the adjacency matrix with eigenvalue $n$. Therefore, the action of the mixer on the equal superposition is,
\begin{equation}
    U_M(t) \ket{s} = e^{-\text{i} n t} \ket{s}.
\end{equation}

Consider just the probability amplitude at a particular vertex (solution state) during the mixing process for the equal superposition,
\begin{equation}
    \bra{\bm{u}}U_M(t)\ket{s} = \frac{1}{\sqrt{N}} \sum_{h=0}^{n} {n \choose h} e^{-\text{i}\frac{\pi}{2}h} (\cos{t})^{n-h} (\sin{t})^h =  \frac{1}{\sqrt{N}} e^{-\text{i} n t}.
\end{equation}

The probability amplitude at each solution state remains at a constant magnitude, but rotates in phase. As described in \cref{sec:general_interference_process}, this simple dynamic emerges from the orchestrated arrival of an exponentially large number of individual probability amplitudes, one from each of the solution states, where the probability amplitudes largely destructively interfere. The probability amplitudes arriving from other solutions-states are phase-shifted depending on distance, with each increment in distance rotating the phase (clockwise/negatively) by an additional $\frac{\pi}{2}$ radians.

Consider modifying the initial equal superposition by introducing a phase offset proportional to distance from $\bm{u}$,
\begin{equation}
    \ket{s'} = \frac{1}{\sqrt{N}} \sum_{h=0}^{n} e^{\text{i} \theta h} \sum_{\bm{x} \in h_{\bm{u}}} \ket{\bm{x}},
\end{equation}
then under the mixer, the probability amplitude at $\bm{u}$ evolves as,
\begin{equation}
    \bra{\bm{u}}U_M(t)\ket{s'} = \frac{1}{\sqrt{N}} \sum_{h=0}^{n} {n \choose h} e^{-\text{i}\frac{\pi}{2}h} e^{\text{i} \theta h} (\cos{t})^{n-h} (\sin{t})^h.
    \label{eq:with_phase_offset}
\end{equation}

\begin{figure}[htbp]
    \centering
    \captionsetup{margin=1cm, font=small}
    \includegraphics[width=1\columnwidth]{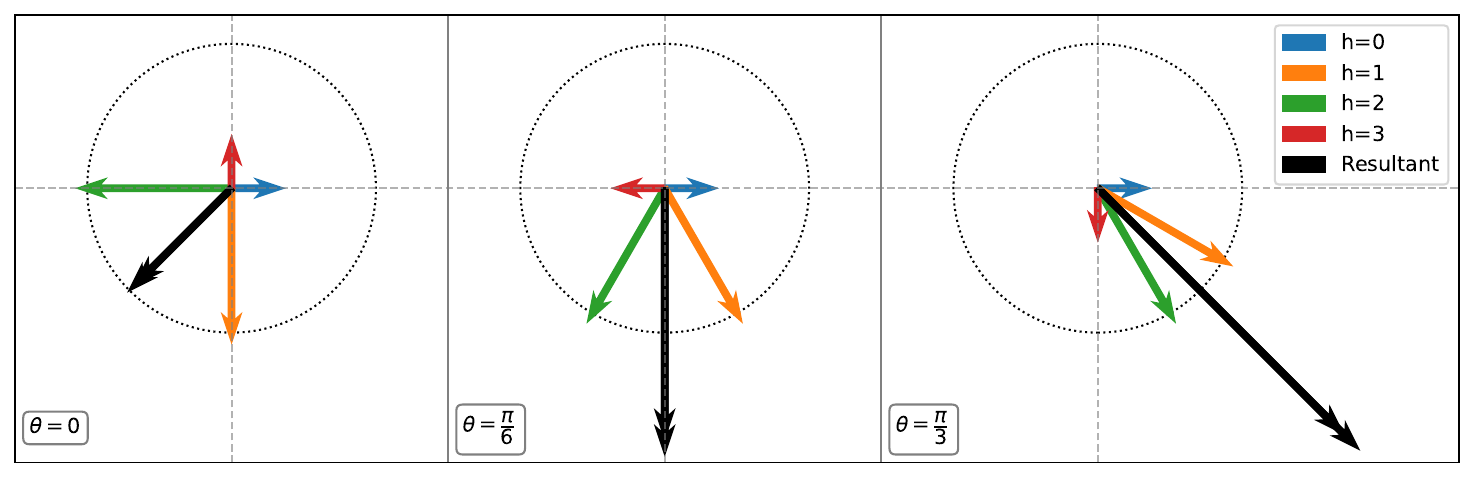}
    \caption{Vector diagrams in the complex plane depicting probability amplitude contributions from $h_{\bm{u}}$ with the additional phase rotation defined by $\theta$, all for the case of $n=3$ and $t=\frac{\pi}{4}$, where constructive interference is clear in the increased magnitude/length of the resultant vectors in black. The inscribed circles define the scale with radius equal to the unamplified magnitude $\frac{1}{\sqrt{N}}$.}
    \label{fig:n=3_vectors}
\end{figure}

To illustrate how this produces constructive interference, \cref{fig:n=3_vectors} shows how the contributing probability amplitudes from each subset $h_{\bm{u}}$ combine for various $\theta$ and for the simple $n=3$ case.

We can also rewrite \cref{eq:with_phase_offset} as,
\numparts
\begin{eqnarray}
    \bra{\bm{u}}U_M(t)\ket{s'} &= \frac{1}{\sqrt{N}} \left(\cos{t} - \text{i} e^{\text{i} \theta} \sin{t} \right)^n \\ &= \frac{1}{\sqrt{N}} \left( \left( \cos{t} + \sin{t} \sin{\theta} \right) - \text{i} \left( \sin{t} \cos{\theta} \right) \right)^n,
\end{eqnarray}
    
\endnumparts
from which the amplification of $\bm{u}$ can be determined,
\begin{equation}
    a_u^2 = \frac{{|\bra{\bm{u}}U_M(t)\ket{s'}|}^2}{{|\braket{u|s}|}^2} = \left( 1 + \sin{2t} \sin{\theta} \right)^n,
    \label{eq:idealised_amplification}
\end{equation}
where amplification defines the increase in measurement probability relative to the initial equal superposition. The significance of this result is that the amount of amplification that the mixer is intrinsically capable of producing for just a single iteration increases exponentially in the problem size. This is certainly a good sign for an algorithm which we hope to have polynomial time-complexity despite an exponentially growing space of solutions. 

This impressive performance is clearly subject to a rather contrived ability to phase-shift each solution state exactly proportional to distance from some target state (a target state of which we have no knowledge in the case of combinatorial optimisation). However, a similarly impressive result can be achieved by ensuring the mean phase-shift applied within each subset, $h_{\bm{u}}$, increases monotonically with distance (in this case up to a distance of half the diameter). In order to demonstrate this, consider the probability amplitude at $\bm{u}$ following a single application of the phase-separation unitary followed by the mixing unitary,
\begin{equation}
    \fl \hspace{1.6cm} \bra{\bm{u}}U_M(t)U_Q(\gamma)\ket{s} = \frac{1}{\sqrt{N}} \sum_{h=0}^{n} \left( (-\text{i})^h (\cos{t})^{n-h} (\sin{t})^h \sum_{\bm{x} \in h_{\bm{u}}}  e^{-\text{i} \gamma f(\bm{x})} \right).
    \label{eq:maxcut_single_iteration}
\end{equation}

Using the approximation from \cref{eq:approximation}, substituting $\mu_{h\bm{u}}$, the mean value of $f(\bm{x})$ within each subset $h_{\bm{u}}$, as well as the variance $V_{h\bm{u}}$, the expression simplifies,
\begin{equation}
    \fl \hspace{1.6cm} \bra{\bm{u}}U_M(t)U_Q(\gamma)\ket{s} = \frac{1}{\sqrt{N}} \sum_{h=0}^{n} e^{-\frac{\gamma^2 V_{h\bm{u}}}{2}} e^{-\text{i}\frac{h \pi}{2}} e^{-\text{i} \gamma \mu_{h\bm{u}}} {n \choose h} (\cos{t})^{n-h} (\sin{t})^h.
    \label{eq:simplified_amplitude}
\end{equation}

\begin{figure}[htbp]
    \centering
    \captionsetup{margin=1cm, font=small}
    \begin{subfigure}{0.5\columnwidth}
        \centering
        \includegraphics[width=1\columnwidth]{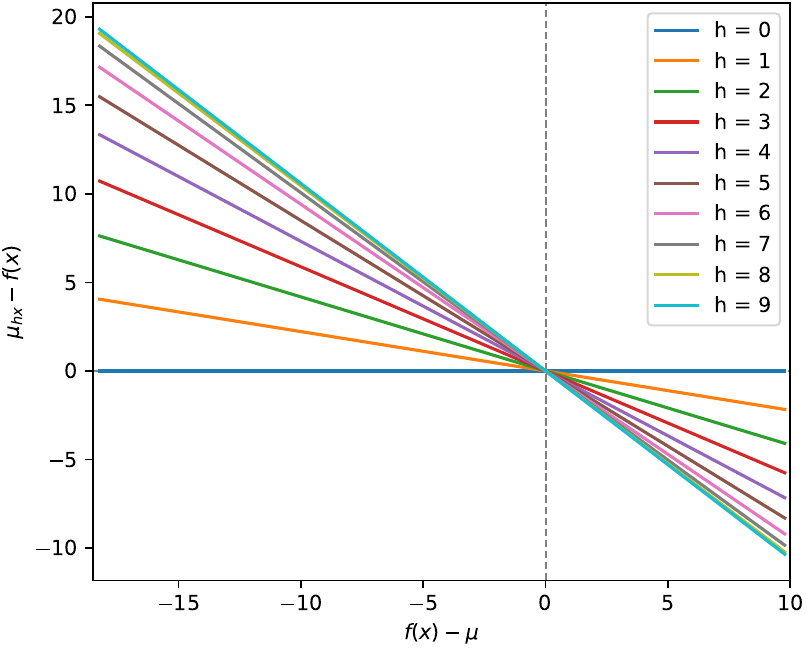}
        \caption{}
        \label{fig:subset_means}
    \end{subfigure}
    \vspace{-0.25cm}
    \begin{subfigure}{0.49\columnwidth}
        \centering
        \includegraphics[width=1\columnwidth]{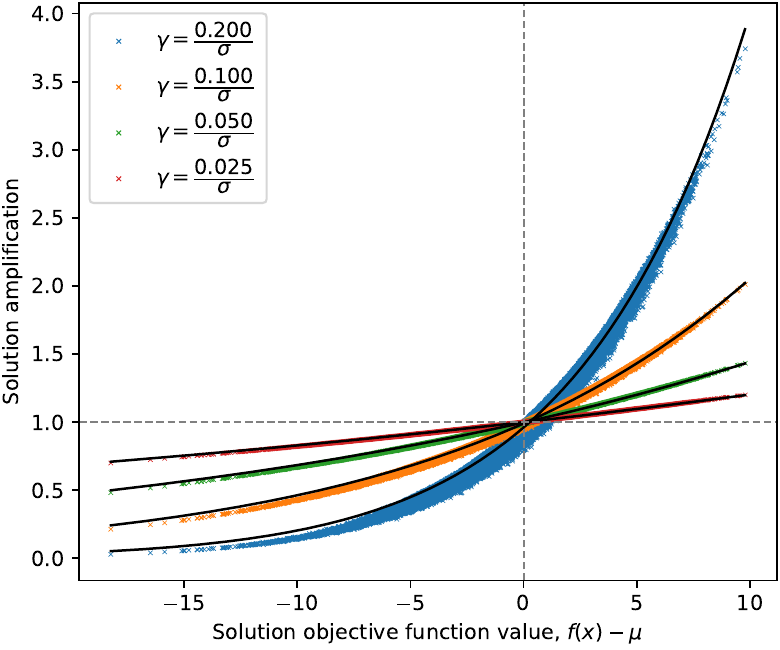}
        \caption{}
        \label{fig:amplification_profile}
    \end{subfigure}
    \caption{These figures relate to the random $n=18$ weighted maxcut problem instance. (a) This figure illustrates how subset means $\mu_{h\bm{x}}$ vary both with distance $h$ from a solution $\bm{x}$, and with objective function value, $f(\bm{x})$. (b) Amplification profile ($a_{\bm{x}}^2$) for the state $U_M(0.3)U_Q(\gamma)\ket{s}$ and for all solutions as a function of their objective function values. The black lines show expected curves according to \cref{eq:expected_amp_profiles} with $\delta = 0.18$.}
    \label{fig:maxcut_subset_means}
\end{figure}

Comparing the readily achievable \cref{eq:simplified_amplitude} to the more contrived \cref{eq:with_phase_offset}, the phase rotation ${e^{-\text{i} \gamma \mu_{h\bm{u}}}}$ plays the role of the artificially introduced phase-shift ${e^{\text{i} \theta h}}$, while the additional phase dispersion term ${e^{-\frac{\gamma^2 V_{h\bm{u}}}{2}}}$ limits the achievable amplification. The size of ${\gamma\left(\mu_{h\bm{u}}-\mu_{h'u}\right)}$ controls the relative phase rotation between the resultant probability amplitudes contributed from $h_{\bm{u}}$ and $h_{\bm{u}}'$. Referring to \cref{eq:maxcut_2nd_condition}, and more generally the second condition in \cref{eq:mixer_requirement_2}, the size of ${\gamma\left(\mu_{h\bm{u}}-\mu_{h'u}\right)}$, and hence the size of relative phase rotations, should be directly proportional to ${\left(f(\bm{u})-\mu\right)}$ (\cref{fig:subset_means} shows this). The amplification profile for solutions with objective function value $f(\bm{x})$ may be expected to loosely follow the relationship,
\begin{equation}
    a_{\bm{x}}^2 \approx e^{-\gamma^2 \sigma^2} \left[ 1 + \sin \left(2t\right) \sin \left(\gamma  \delta \left(f(\bm{x})-\mu\right) \right) \right]^n,
    \label{eq:expected_amp_profiles}
\end{equation}
which is similar to the derived result from \cref{eq:idealised_amplification}, with the leading exponential term added to account for the dispersion, crudely assuming that the subset variances can be approximated as constant and equal to $\sigma^2$, the population variance. The parameter $\delta$ accounts for the fact that the phase shifts between subsets $h_{\bm{u}}$ are not directly proportional to $h$. In order to provide supporting numerical evidence, we generate a random $n=18$ weighted maxcut problem instance (see \cref{sec:maxcut_graph}). The amplification of individual solution states within a single-iteration amplified state ${U_M(0.3)U_Q(\gamma)\ket{s}}$ are shown in \cref{fig:amplification_profile} for various $\gamma$ and plotted alongside the predicted curves according to \cref{eq:expected_amp_profiles}.

\subsection{Interference effects: Multiple iterations}
\label{sec:maxcut_multiple_iterations}

\begin{figure}[htbp]
    \centering
    \captionsetup{margin=1cm, font=small}
    \includegraphics[width=0.7\columnwidth]{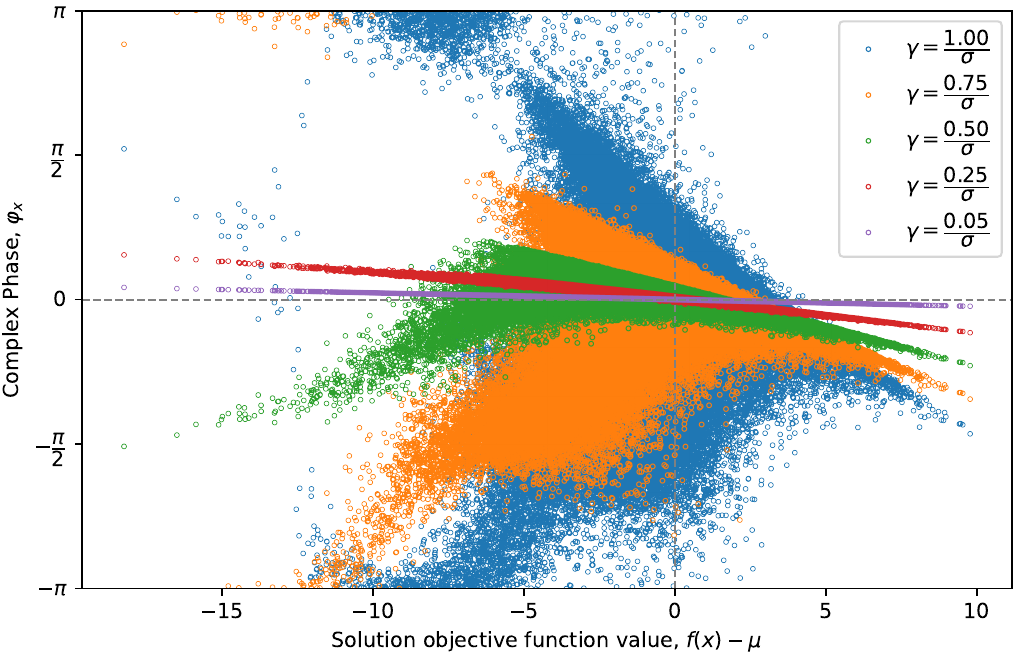}
    \caption{Relative phases $\varphi_{\bm{x}}$ of individual solution states within a single-iteration amplified state ${U_M(0.3)U_Q(\gamma)\ket{s}}$ of the $n=18$ weighted maxcut problem for various $\gamma$.}
    \label{fig:phase_correlations}
\end{figure}

Ideally, we would like to perform this amplification process several times in a row, focusing probability amplitude into the optimal and nearest-optimal solutions such that they can be measured with relatively high probability from the final amplified state. Several key assertions were made in \cref{sec:general_interference_process} with regard to achieving effective interference and amplification over multiple iterations. One of these assertions was that smaller initial values of $\gamma$ better preserve coherence in probability amplitudes assigned to each solution state. Consider the state of the system after a single iteration expressed as,
\begin{equation}
    U_M(t)U_Q(\gamma)\ket{s} = \frac{e^{-\text{i}(nt+\gamma\mu)}}{\sqrt{N}} \sum_{\bm{x} \in S} a_{\bm{x}} e^{\text{i}\varphi_{\bm{x}}} \ket{\bm{x}}
\end{equation}
where $a_{\bm{x}}$ and $\varphi_{\bm{x}}$ are real-valued and describe the magnitude and relative phase of the probability amplitude assigned to each solution. The global phase is included just to correct for differences in global phase due to application times $\gamma$ and $t$. The relative phases $\varphi_{\bm{x}}$ of individual solution states within a single-iteration amplified state ${U_M(0.3)U_Q(\gamma)\ket{s}}$ are shown in \cref{fig:phase_correlations}, which provides numerical evidence of the claim that smaller values of $\gamma$ produce less dispersion in phases. A key requirement for considerable amplification is that the probability amplitudes contained in each subset $h_{\bm{x}}$ should be approximately in-phase. If the phases are too dispersed, the interference process will not be effective. It is for this reason that a smaller value of $\gamma$, while producing less amplification initially, preserves the ability to produce further amplification in subsequent iterations. Maximising total amplification over several iterations, therefore, requires limiting amplification during initial iterations, i.e. a greedy approach is not optimal.

\begin{figure}[htbp]
    \centering
    \captionsetup{margin=1cm, font=small}
    \begin{subfigure}{0.49\columnwidth}
        \centering
        \includegraphics[width=1\columnwidth]{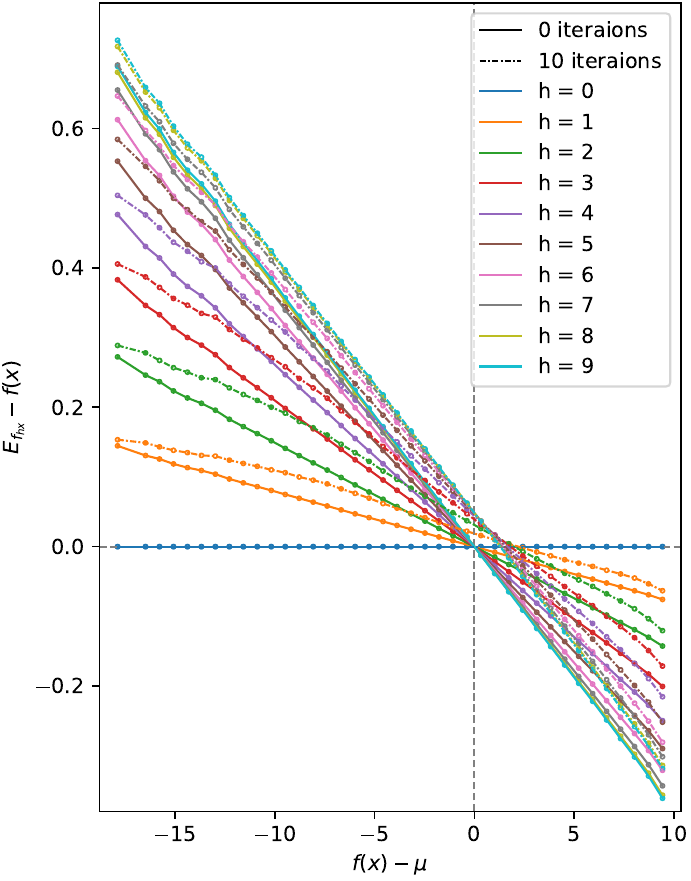}
    \end{subfigure}
    \begin{subfigure}{0.49\columnwidth}
        \centering
        \includegraphics[width=1\columnwidth]{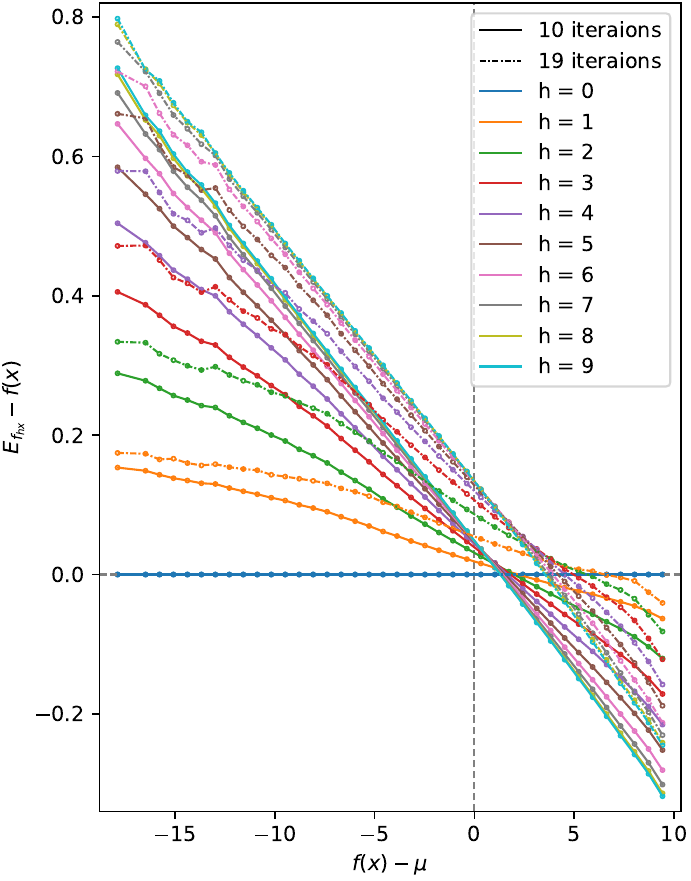}
    \end{subfigure}
    \caption{Distribution of weighted subset means $E_{f_{h\bm{x}}}$ as a function of objective function value $f(\bm{x})$. The left figure compares and contrasts between the initial unamplified state and the partially amplified state after 10 iterations. The right figure compares the same partially amplified state with the partially amplified state after the penultimate 19th iteration.}
    \label{fig:maxcut_n18_evolving_subset_means}
\end{figure}

\begin{figure}[htbp]
    \centering
    \captionsetup{margin=1cm, font=small}
    \includegraphics[width=0.99\columnwidth]{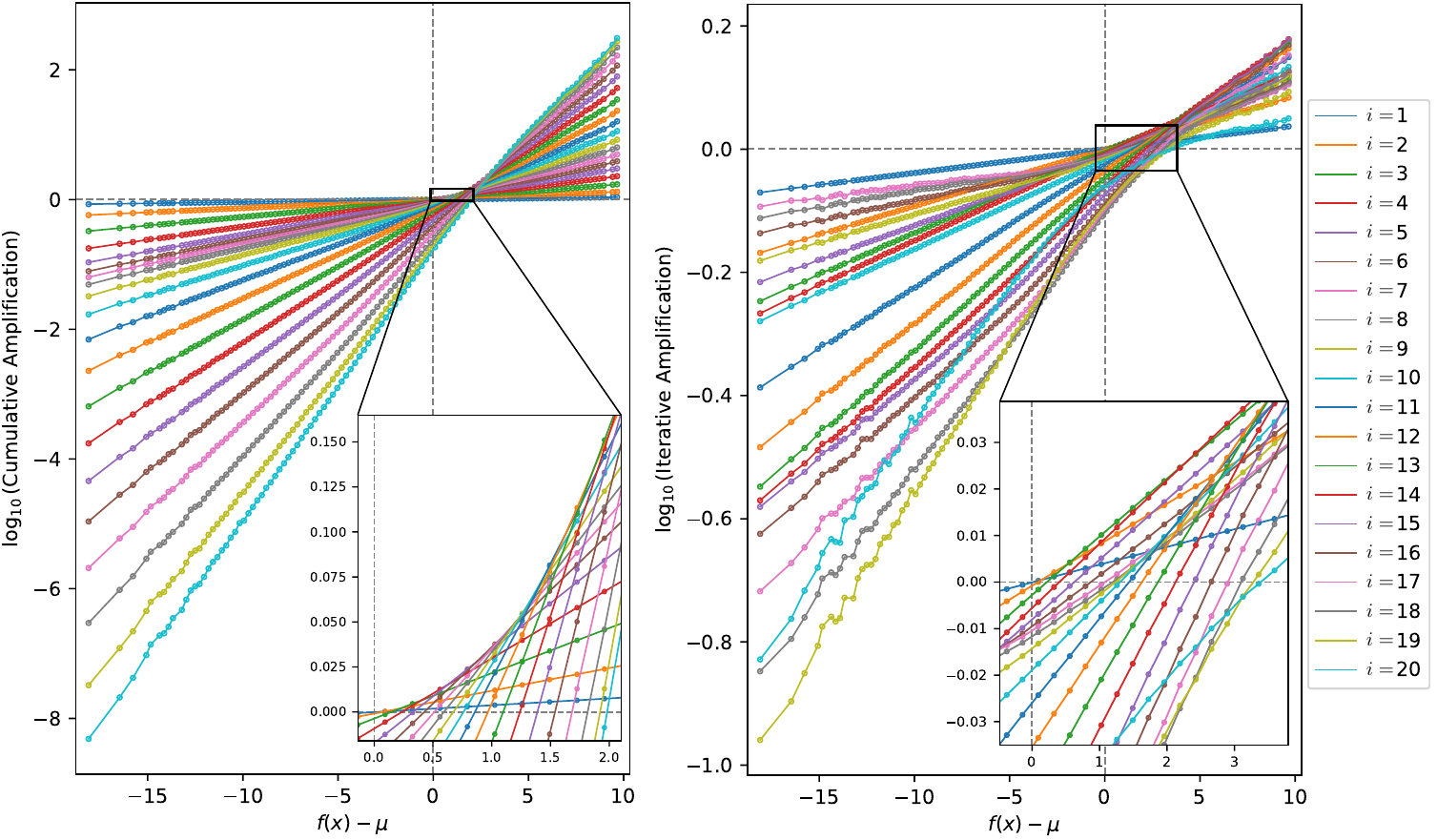}
    \caption{(left) The evolving amplification profile throughout the $20$ iterations. (right) The amplification produced from each individual iteration.}
    \label{fig:maxcut_n18_evolving_amplification_profiles}
\end{figure}

Another key assertion made in \cref{sec:general_interference_process} is how the weighted mean objective function values $E_{f_{h\bm{x}}}$ in each subset $h_{\bm{x}}$ will be distributed as a function of objective function value $f(\bm{x})$, and how this distribution will change under a monotonic amplification profile as the interference process focuses probability amplitude selectively into a reducing number of more optimal solutions. It is also stated that this evolving distribution of $E_{f_{h\bm{x}}}$ affects the amplification profile produced in each iteration. In order to test these assertions, we simulate the amplified state $\ket{\gamma,t,\beta}$ for the $n=18$ maxcut problem, for $p=20$ iterations, with $\gamma=0.3$ and $t=0.2$ and with $\beta = \frac{1}{p}$. Small values for $\beta$ and $\gamma$ are selected because in this case we are interested in preserving coherence and producing ideal behaviour for the full space of solutions, rather than just the near optimal solutions, or rather than simply maximising the final probability amplitude of the two optimal solution states. The results of this analysis are illustrated in \cref{fig:maxcut_n18_evolving_subset_means} and \cref{fig:maxcut_n18_evolving_amplification_profiles}.   

\cref{fig:maxcut_n18_evolving_subset_means} shows that the weighted subset means do vary roughly as anticipated. In general, the values of $E_{f_{h\bm{x}}}$ increase for all solutions $\bm{x}$ and continue to increase throughout the $20$ iterations. The point at which $E_{f_{h\bm{x}}}$ switches from monotonically increasing to monotonically decreasing (with respect to $h$) moves to the right, isolating a decreasing number of more optimal solutions within the constructive interference regime. In addition, the gaps between successive $E_{f_{h\bm{x}}}$ (relative to $h$) are roughly proportional to the offset from this transition point, meaning we should expect amplification in each iteration to be itself monotonically increasing.

\cref{fig:maxcut_n18_evolving_amplification_profiles} provides evidence for both the monotonically increasing amplification profile throughout the $p=20$ iterations, as well as its relationship with the evolving $E_{f_{h\bm{x}}}$ depicted in \cref{fig:maxcut_n18_evolving_subset_means}. The amplification produced in each iteration is also monotonically increasing and the transition between destructive and constructive interference moves towards increasing $f(\bm{x})$ as expected.

\subsection{Simulation Results}
\begin{figure}[htbp]
    \centering
    \captionsetup{margin=1cm, font=small}
    \begin{subfigure}{0.475\columnwidth}
        \centering
        \includegraphics[width=1\columnwidth]{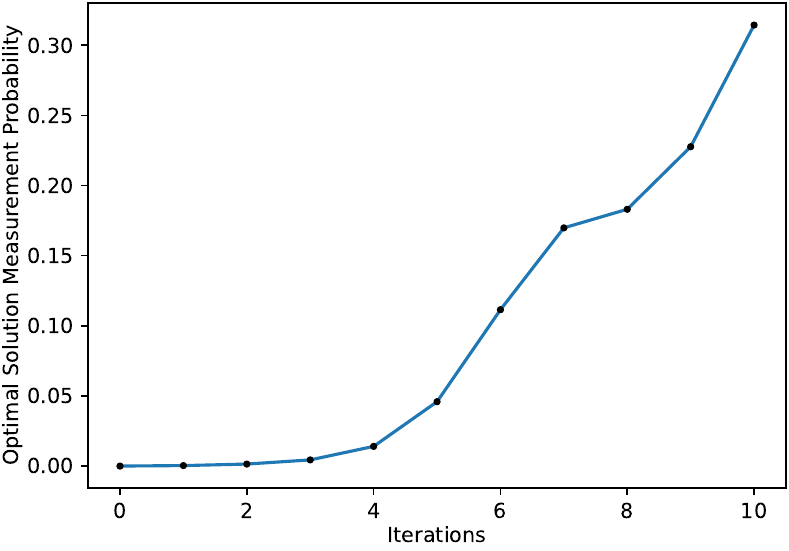}
    \end{subfigure}
    \begin{subfigure}{0.475\columnwidth}
        \centering
        \includegraphics[width=1\columnwidth]{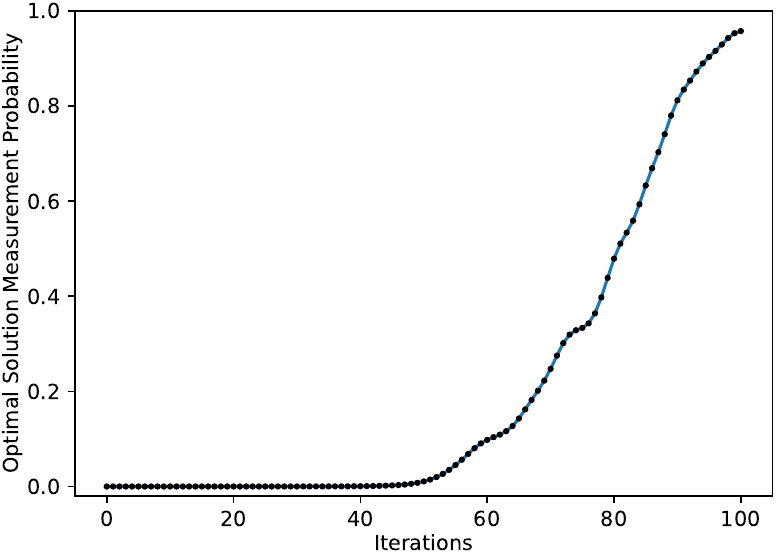}
    \end{subfigure}
        \begin{subfigure}{0.49\columnwidth}
        \centering
        \includegraphics[width=1\columnwidth]{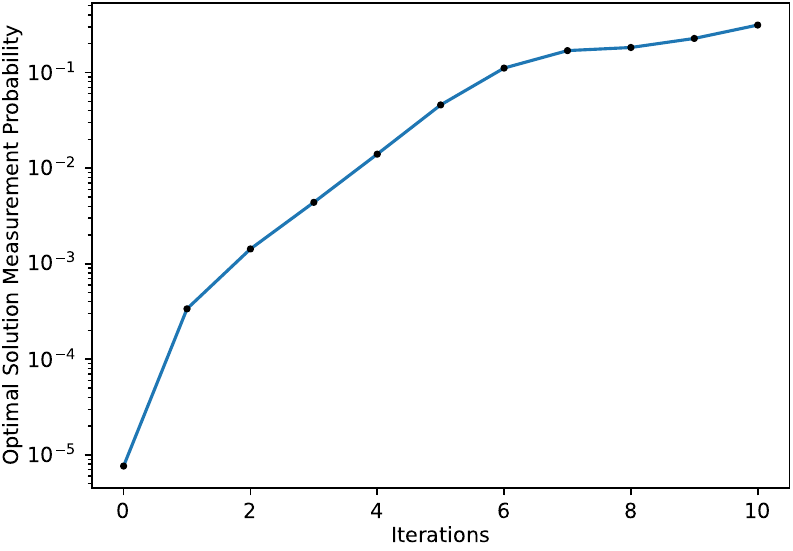}
    \end{subfigure}
    \begin{subfigure}{0.49\columnwidth}
        \centering
        \includegraphics[width=1\columnwidth]{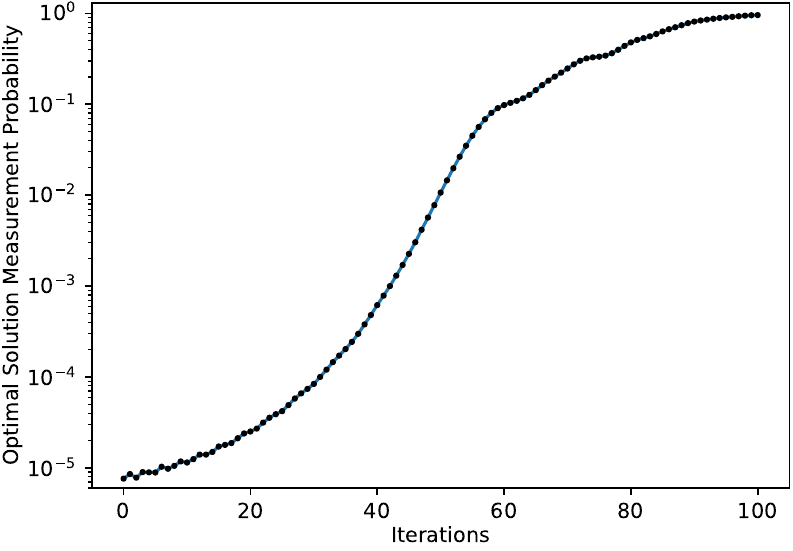}
    \end{subfigure}
    \caption{Simulation results for $p=10$ and $p=100$ non-variational QWOA applied to the $n=18$ weighted maxcut problem. Optimal solution measurement probability at each iteration (top) linear scale and (bottom) logarithmic scale.}
    \label{fig:Maxcut_optimal_prob}
\end{figure}

In this section, the performance of the algorithm is demonstrated by simulation on the randomly generated $n=18$ weighted maxcut problem in \cref{sec:maxcut_graph}. In order to show that the algorithm is robust with regard to the number of iterations, the amplified state $\ket{\gamma,t,\beta}$ is analysed for two cases. In one case, we select an appropriate number of iterations, $p=10$, and in the other case, we select an unreasonably large number of iterations (for a problem of this size), $p=100$. In either case, the optimal parameters are determined from a gradient ascent process initialised with $\gamma = 1$, $t=0.1$ and $\beta=\frac{1}{p}$. For $p=10$ the optimal values are $\gamma=2.4340$, $t=0.4517$ and $\beta=0.2844$. For $p=100$ the optimal values are $\gamma=2.0718$, $t=0.6395$ and $\beta=0.0126$. \cref{fig:Maxcut_optimal_prob} shows how measurement probability for the optimal solution increases steadily throughout the iterations. This provides evidence to suggest that the interference process compounds in each iteration to multiplicatively increase the probability amplitude assigned to the optimal solution(s) (as expected).

\begin{figure}[htbp]
    \centering
    \captionsetup{margin=1cm, font=small}
    \begin{subfigure}{0.49\columnwidth}
        \centering
        \includegraphics[width=1\columnwidth]{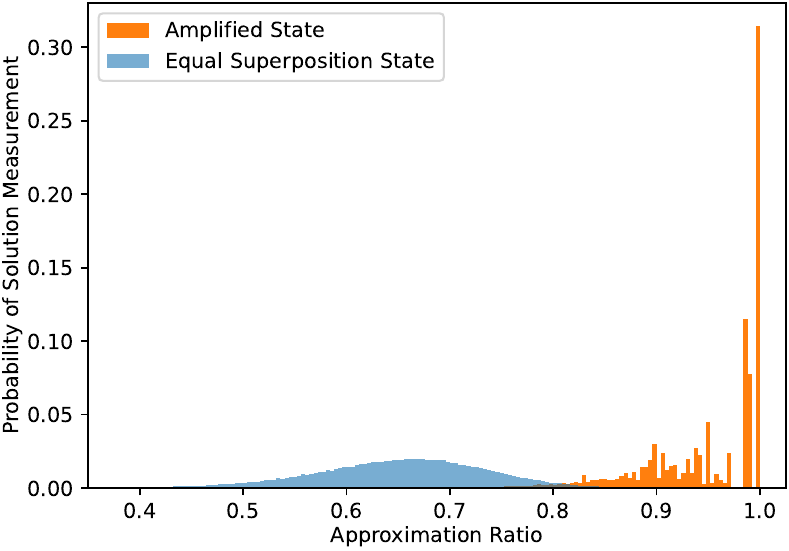}
    \end{subfigure}
    \begin{subfigure}{0.49\columnwidth}
        \centering
        \includegraphics[width=1\columnwidth]{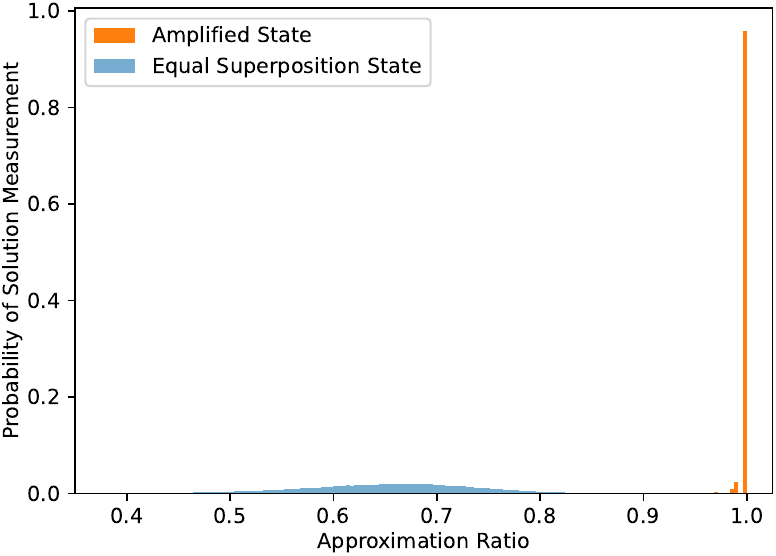}
    \end{subfigure}
    \begin{subfigure}{0.49\columnwidth}
        \centering
        \includegraphics[width=1\columnwidth]{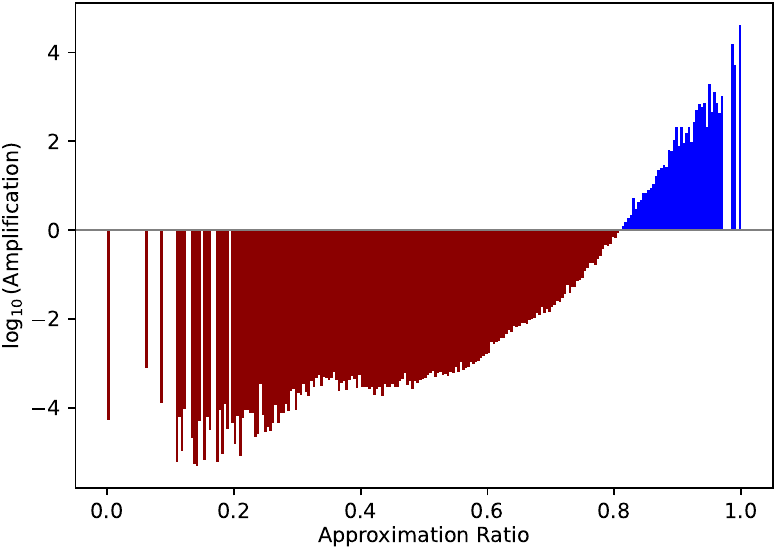}
    \end{subfigure}
    \begin{subfigure}{0.49\columnwidth}
        \centering
        \includegraphics[width=1\columnwidth]{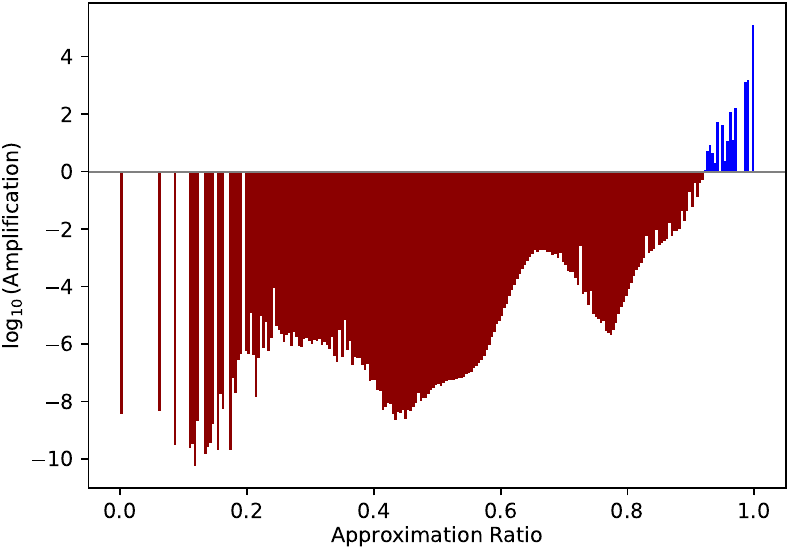}
    \end{subfigure}
    \begin{subfigure}{0.49\columnwidth}
        \centering
        \includegraphics[width=1\columnwidth]{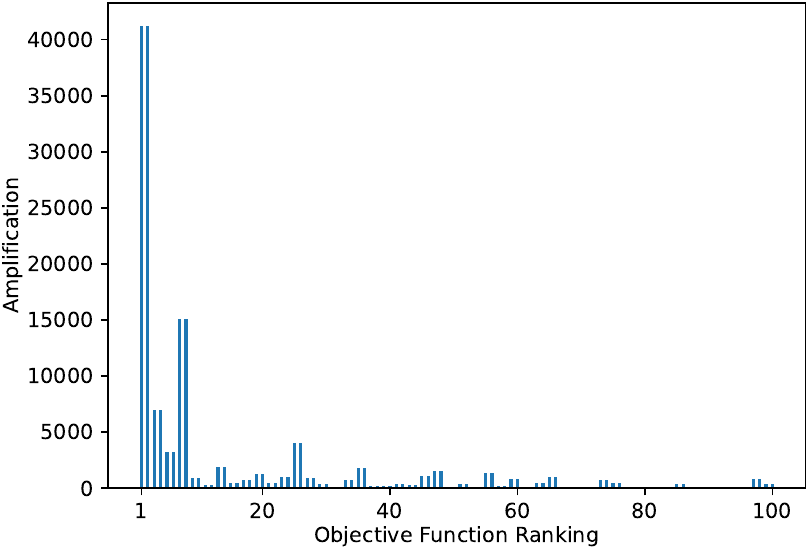}
    \end{subfigure}
    \begin{subfigure}{0.49\columnwidth}
        \centering
        \includegraphics[width=1\columnwidth]{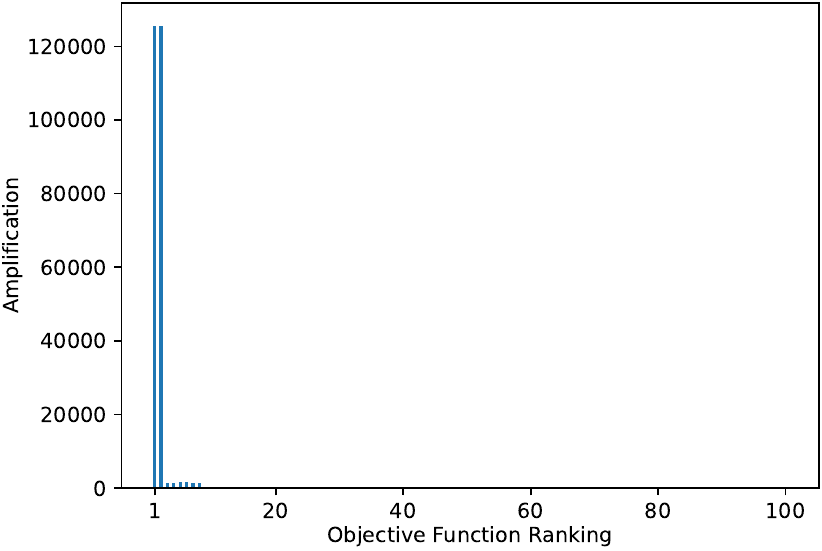}
    \end{subfigure}
    \caption{Simulation results for (left) $p=10$ and (right) $p=100$ non-variational QWOA applied to the $n=18$ weighted maxcut problem. (Top) Probability distributions for approximation ratio as measured from the initial equal superposition state $\ket{s}$ and the amplified state $\ket{\gamma,t,\beta}$. (Middle) Logarithmic scale solution amplification as a function of approximation ratio. (Bottom) Amplification of solutions with the 100 highest objective function values.}
    \label{fig:maxcut_results}
\end{figure}

\newpage
\cref{fig:maxcut_results} includes various figures analysing the final amplified state. From these results it is clear that probability amplitude has in both cases been channelled primarily into the optimal and near-optimal solutions. It is also clear that the globally optimal solution in both cases is maximally amplified, and that the mirror-symmetric degenerate pairs of solutions are identically amplified, as expected. Note that the approximation ratio, defined as,
\begin{equation}
    \text{Approximation Ratio} = \frac{f(\bm{x})}{\text{optimal}\{f(\bm{x}):\bm{x} \in S\}},
\end{equation}
is used for the figures, providing clarity, as regardless of the problem or problem instance, an optimal solution has approximation ratio equal to $1$.

\newpage
\section{Generalising to other problems: k-means clustering}
\label{sec:kmeans}

Cluster analysis involves dividing a set of observations or data-points into clusters based on similarity. Here we will focus on the k-means clustering problem which is to find the partitioning of $n$ multi-dimensional real vectors (data-points) into $k$ clusters which minimises the total squared euclidean distance between each data-point and its respective cluster's centroid.

Consider a particular clustering or solution to be characterised by a vector, ${\bm{x} = (x_1, x_2, ..., x_n)}$, where ${x_j \in \{0,1,...,k-1\}}$ specifies the allocation of data-point $j$ into cluster $x_j$. We define a space of solutions $S$ containing every $\bm{x}$ with cardinality $N = k^n$. $S$ has considerable degeneracy, as any particular clustering can be relabelled in many ways (up to $k!$) and also contains solutions with less than $k$ clusters. Never-the-less, we suggest this approach is highly effective as it permits an efficiently implementable mixer under which degenerate solutions evolve identically, and the objective function naturally penalises solutions with less than $k$ clusters. The objective function (to be minimised) can be expressed as,
\begin{equation}
    f(\bm{x}) = \sum_{i=0}^{k-1} \frac{1}{|C_i(\bm{x})|} \sum_{\bm{a},\bm{b}\in C_i(\bm{x})} {||\bm{a}-\bm{b}||}^2,
\end{equation}
where $C_i(\bm{x}) = \{\bm{v}_j : x_j=i\}$ is cluster $i$ and $||\bm{a}-\bm{b}||$ is the euclidean distance between data points $\bm{a}$ and $\bm{b}$ which are contained in $C_i$.

As discussed in \cref{sec:definitions}, embedding solutions within the computational basis states of the input register can be achieved by assigning an $m$ qubit sub-register to each of the $n$ integer variables. For a binary encoding, $m=\lceil \log_2 k \rceil$, and the valid integer variable values are associated with the first $k$ computational basis states of each of these sub-registers. For a one-hot encoding, $m=k$, and the valid integer variable values are encoded in the computational basis states where a single qubit is ``on", i.e. these states can be expressed with the Kronecker delta as, ${\ket{x_j} = \prod_{i=0}^{k-1} \ket{\delta_{i,x_j}} }$. The equal superposition over all feasible solutions in $S$ can be prepared by acting independently on each sub-register,
\begin{equation}
    \ket{s} = \left(U_k \ket{0}^{\otimes m} \right)^{\otimes n} = \ket{k}^{\otimes n} = \frac{1}{\sqrt{k^n}} \sum_{\bm{x} \in S} \ket{\bm{x}},
\end{equation}
where $\ket{k}$ is the equal superposition over the $k$ utilized computational basis states of a sub-register (those which directly encode each of the possible decision variable values). Each circuit implementation (one-hot and binary) of the unitary $U_k$ which prepares $\ket{k}$ is described in \cref{alg:Uk}.

When defining the mixing graph for this problem, the obvious perturbation of a solution which would generate nearest neighbour solutions is to re-allocate one of the data-points to a different cluster. The mixing graph is therefore defined by the set of $n(k-1)$ possible moves in which a single data-point is re-allocated. The adjacency matrix of the mixing graph is given by,
\begin{equation}
    A = \sum_{j=1}^{n} \id^{\otimes j-1} \otimes K_k \otimes \id^{\otimes n-j},
    \label{eq:Adjacency_integer_mixer}
\end{equation}
where $K_k$ is the adjacency matrix of a complete graph over $k$ vertices. More specifically, $K_k$ connects just the $k$ utilized computational basis states within a sub-register and can be expressed as ${K_k = k \ket{k}\bra{k} - \id}$. This mixing graph, also known as a Hamming graph $H(n,k)$, is the Cartesian product of $n$ complete graphs $K_k$.

A continuous-time quantum walk on this mixing graph for time $t$ can be expressed (with a convenient global phase $e^{-\text{i} n t}$) as,
\numparts
\begin{eqnarray}
    U_M(t) &= e^{-\text{i} n t} e^{-\text{i} t A} \\
           &= e^{-\text{i} n t} \prod_{j=1}^n e^{-\text{i} t K_k} \\
           &= e^{-\text{i} n t} \prod_{j=1}^n e^{-\text{i} t \left(k \ket{k}\bra{k} - \id \right)} \\
           &= e^{-\text{i} n t} \prod_{j=1}^n \left(  e^{-\text{i} \left(k-1\right) t} \ket{k}\bra{k}  + e^{\text{i} t} \left( \id - \ket{k}\bra{k} \right) \right) \\
           &= \prod_{j=1}^n \left(  e^{-\text{i} k t} \ket{k}\bra{k}  + \left( \id - \ket{k}\bra{k} \right) \right).
\end{eqnarray}
\endnumparts

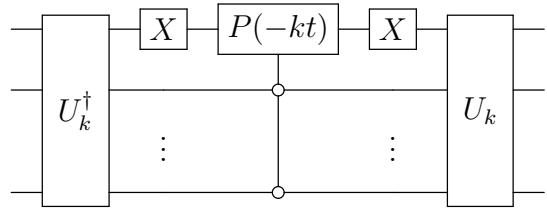
\begin{figure}[htbp]
    \centering
    
    \[ \hspace{1.5cm} \Qcircuit @C=1em  @R=0.7em  {
        & \multigate{3}{U_k^\dagger} & \gate{X} & \gate{P(-k t)} & \gate{X} & \multigate{3}{U_k} & \qw \\
        & \ghost{U_k^\dagger} & \qw & \ctrlo{-1} & \qw & \ghost{U_k} & \qw \\
        & \push{\rule{0em}{1em}} & \colorbox{white}{\vdots} & & \colorbox{white}{\vdots} & \\
        & \ghost{U_k^\dagger} & \qw & \ctrlo{-2} & \qw & \ghost{U_k} & \qw \\
    }\]
    \captionsetup{margin=1cm, font=small}
    \caption{Quantum circuit implementation of the complete-graph continuous-time quantum walk (${e^{-\text{i} t K_k}}$) to be applied within each of the $m$-qubit sub-registers. $P(\phi)$ is the phase gate which maps computational basis states $\ket{0} \mapsto \ket{0}$ and $\ket{1} \mapsto e^{\text{i} \phi}\ket{1}$. The circuit required to implement $U_k$ can be efficiently prepared with the method presented in \cref{alg:Uk}.}
    \label{fig:Mixer_circuit}
\end{figure}

This unitary can be implemented by applying the circuit in \cref{fig:Mixer_circuit} within each of the sub-registers.

Consider an arbitrary solution, ${\bm{u} = (u_1, u_2, ..., u_n)}$, with solution state, ${\ket{\bm{u}}=\prod_{j=1}^n\ket{u_j}}$. The action of the mixer on this state can be expressed,
\numparts
\begin{eqnarray}
    U_M(t) \ket{\bm{u}} &= e^{-\text{i} n t} \left(\prod_{j=1}^n e^{-\text{i} t K_k}\right) \left( \prod_{j=1}^n\ket{u_j} \right) \\
                   &= e^{-\text{i} n t} \prod_{j=1}^n \left( e^{-\text{i} t K_k}\ket{u_j} \right) \\
                   &= \prod_{j=1}^n \left(  \left(  e^{-\text{i} k t} \ket{k}\bra{k}  + \left( \id - \ket{k}\bra{k} \right) \right) \ket{u_j} \right) \\
                   &= \prod_{j=1}^n  \left(  \frac{e^{-\text{i} k t} + k - 1}{k} \ket{u_j}  +  \frac{e^{-\text{i} k t} - 1}{k} \sum_{x_j \neq u_j} \ket{x_j}  \right) \\
                   &= \sum_{h=0}^{n} \left(\frac{e^{-\text{i} k t} + k - 1}{k}\right)^{n-h} \left(\frac{e^{-\text{i} k t} - 1}{k}\right)^h \sum_{\bm{x} \in h_{\bm{u}}} \ket{\bm{x}} \\
                   &= \frac{1}{k^n} \sum_{h=0}^{n} \left(e^{-\text{i} k t} + k - 1\right)^{n-h} \left(e^{-\text{i} k t} - 1\right)^h \sum_{\bm{x} \in h_{\bm{u}}} \ket{\bm{x}}
\end{eqnarray}
\endnumparts
where $x_j \neq u_j$ is restricted just to the valid states from ${\{0,1,...,k-1\}}$, and $h_{\bm{u}}$ as before refers to the subset of solutions which are a distance $h$ from $\bm{u}$ on the mixing graph. Stated more simply, $h_{\bm{u}}$ is composed of the ${n \choose h}(k-1)^h$ solutions which have exactly $h$ data-points allocated to different clusters when compared with $\bm{u}$.

This expression for the action of the mixer on an arbitrary solution state can be shown to satisfy the necessary condition from \cref{eq:mixer_requirement_1}. First, consider the polar forms of each of the complex terms,
\numparts
\begin{eqnarray}
    z_1 &= e^{-\text{i} k t} + k - 1 = r_1 e^{\text{i} \phi_1} \\
    z_2 &= e^{-\text{i} k t} - 1 = r_2 e^{\text{i} \phi_2} \\
    r_1 &= \sqrt{k^2-2\left( k-1 \right) \left( 1 - \cos{k t} \right)} \\
    \phi_1 &= \arctan \left( \frac{-\sin{k t}}{k+\cos{k t} - 1} \right) \\
    r_2 &=  \sqrt{2 \left( 1 - \cos{k t} \right)} \\
    \phi_2 &= \arctan \left( \frac{\sin{k t}}{1 - \cos{k t}} \right)-\pi.
\end{eqnarray}
\endnumparts
Substituting the polar forms, the action of the mixer on $\bm{u}$ simplifies,
\numparts
\begin{eqnarray}
    U_M(t) \ket{\bm{u}} &= \frac{1}{k^n} \sum_{h=0}^{n} \left( r_1 e^{\text{i} \phi_1} \right)^{n-h} \left( r_2 e^{\text{i} \phi_2} \right)^h \sum_{\bm{x} \in h_{\bm{u}}} \ket{\bm{x}} \\
                   &= \frac{e^{\text{i} n \phi_1}}{k^n} \sum_{h=0}^{n} \left( r_1 \right)^{n-h} \left( r_2 \right)^h e^{\text{i} h \left( \phi_2 - \phi_1 \right)} \sum_{\bm{x} \in h_{\bm{u}}} \ket{\bm{x}} \\
                   &= \frac{e^{\text{i} n \phi_1}}{k^n} \sum_{h=0}^{n} r_h(t) e^{-\text{i} h \phi(t)} \sum_{\bm{x} \in h_{\bm{u}}} \ket{\bm{x}},       
\end{eqnarray}
\endnumparts
where $r_h(t) = \left( r_1 \right)^{n-h} \left( r_2 \right)^h$ is a positive real function of $t$ and $\phi(t)$ is a positive real function of $t$ which is constant with respect to distance $h$, expressed as,
\begin{equation}
    \phi(t) = \pi - \arctan \left( \frac{-k \sin{k t}}{(k - 2) (\cos{k t} - 1)} \right).
\end{equation}

Note that the relevant application times for the mixer are over the domain, $(0,\frac{\pi}{k}]$ (as the period of $e^{-\text{i} t K_k}$ is $\frac{2\pi}{k}$), and over this domain, the values for $\phi$ increase monotonically from $\frac{\pi}{2}$ to $\pi$ (for $k>2$). For $k=2$ this mixer becomes functionally equivalent to the hypercube mixer, for which $\phi$ remains constant at $\frac{\pi}{2}$. In any case, the proposed mixing unitary satisfies the first condition set out in \cref{eq:mixer_requirement_1}, or in other words, the mixer distributes probability amplitude globally from an initial solution state with a phase offset that is directly proportional to distance from the initial solution state.

The second necessary condition for effective application of this mixer, is that objective function values as distributed over the mixing graph should satisfy \cref{eq:mixer_requirement_2}. For weighted maxcut, this could be shown analytically, whereas in general, we verify this condition via statistical sampling. Consider the random k-means problem instance generated for $n=12$ data-points (10-dimensional) and $k=3$ clusters, for which the data-points are included in \cref{sec:data-points}. 

\begin{figure}[htbp]
    \centering
    \captionsetup{margin=1cm, font=small}
    \begin{subfigure}{0.50\columnwidth}
        \centering
        \includegraphics[width=1\columnwidth]{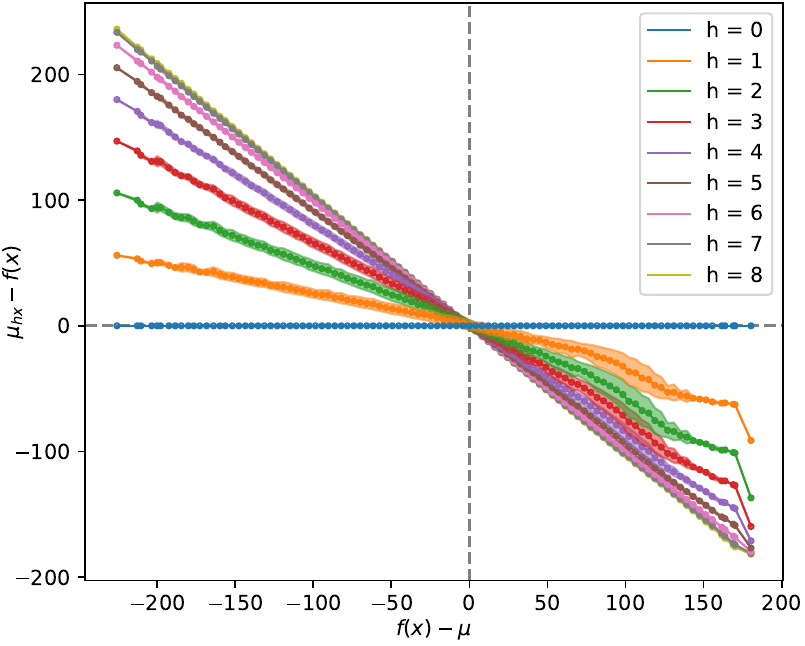}
        \caption{}
        \label{fig:k_means_subset_means}
    \end{subfigure}
    \vspace{-0.25cm}
    \begin{subfigure}{0.49\columnwidth}
        \centering
        \includegraphics[width=1\columnwidth]{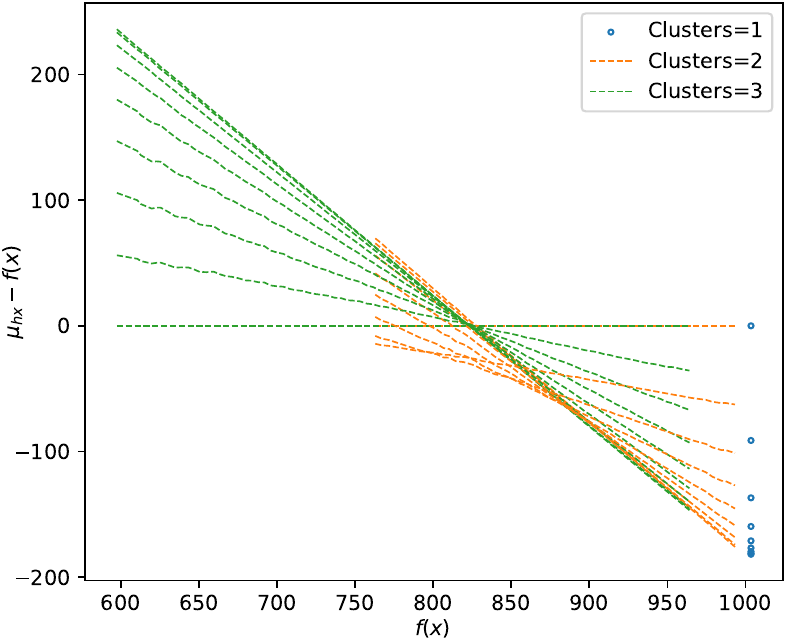}
        \caption{}
        \label{fig:k_means_subset_means_seperated}
    \end{subfigure}
    \caption{Statistical analysis of subset means $\mu_{h\bm{x}}$ for the $n=12$, $k=3$, k-means problem. (a) All solutions. Shading shows variations in the values of $\mu_{h\bm{x}}$ (plus or minus a single standard deviation). (b) Solutions separated by number of clusters.}
\end{figure}

The mean objective function values $\mu_{h\bm{x}}$ are computed for randomly sampled $\bm{x} \in S$ and for $h \in \{ 0,1,2, ..., 8 \}$. Solutions are partitioned into 100 equally spaced bins according to objective function values and $200$ solutions are sampled from each bin (where sufficient solutions are available). The results of this analysis can be seen in \cref{fig:k_means_subset_means}, where it is clear that the condition is approximately satisfied, however there is some disruption at larger values of $f(\bm{x})$. The reason for this becomes clear when solutions are analysed separately based on the number of clusters they contain, as per \cref{fig:k_means_subset_means_seperated}. 

\begin{figure}[htbp]
    \centering
    \captionsetup{margin=1cm, font=small}
    \begin{subfigure}{0.49\columnwidth}
        \centering
        \includegraphics[width=1\columnwidth]{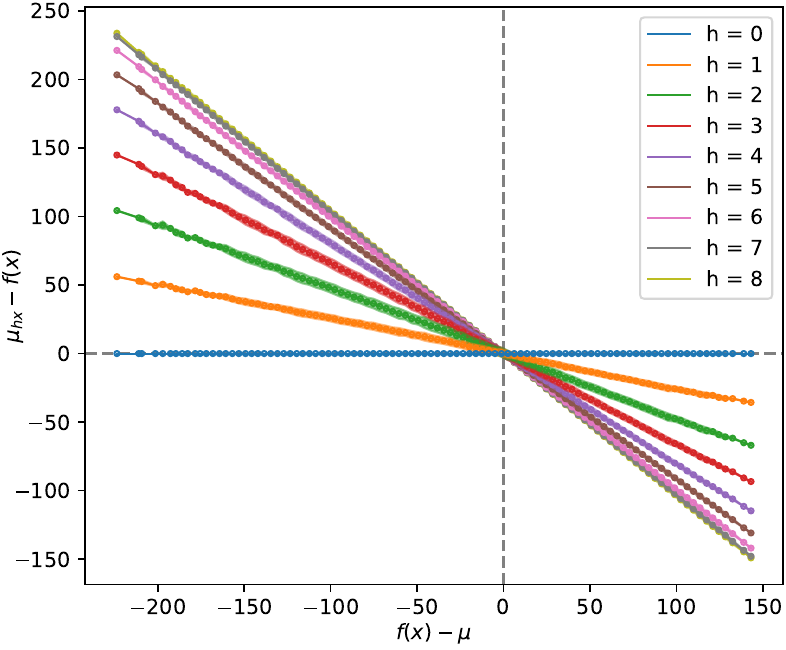}
        \caption{}
    \end{subfigure}
    \vspace{-0.25cm}
    \begin{subfigure}{0.49\columnwidth}
        \centering
        \includegraphics[width=1\columnwidth]{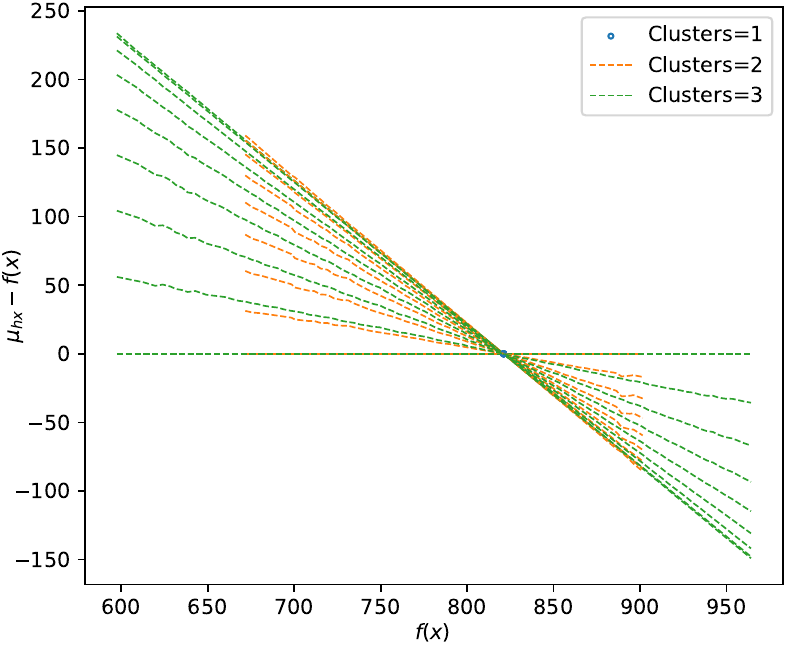}
        \caption{}
    \end{subfigure}
    \caption{Statistical analysis of subset means $\mu_{h\bm{x}}$ for the $n=12$, $k=3$, k-means problem, after transforming the objective function. (a) All solutions. (b) Solutions separated by number of clusters.}
    \label{fig:k_means_subset_means_transformed}
\end{figure}

Adherence to the necessary condition related to subset means can be improved by applying a simple transformation to the objective function,
\begin{equation}
    f(\bm{x}) \longrightarrow f(\bm{x}) - (\mu_{c(\bm{x})}-\mu_k),
\end{equation}
where $c(\bm{x})$ is the number of clusters in solution $\bm{x}$, and $\mu_j$ is defined as the mean objective function value of solutions with $j$ clusters. This acts to align the mean objective function value of solutions containing a different number of clusters. The same analysis is repeated for this transformed objective function, with results shown in \cref{fig:k_means_subset_means_transformed}.

\begin{figure}[htbp]
    \centering
    \captionsetup{margin=1cm, font=small}
    \begin{subfigure}{0.49\columnwidth}
        \centering
        \includegraphics[width=1\columnwidth]{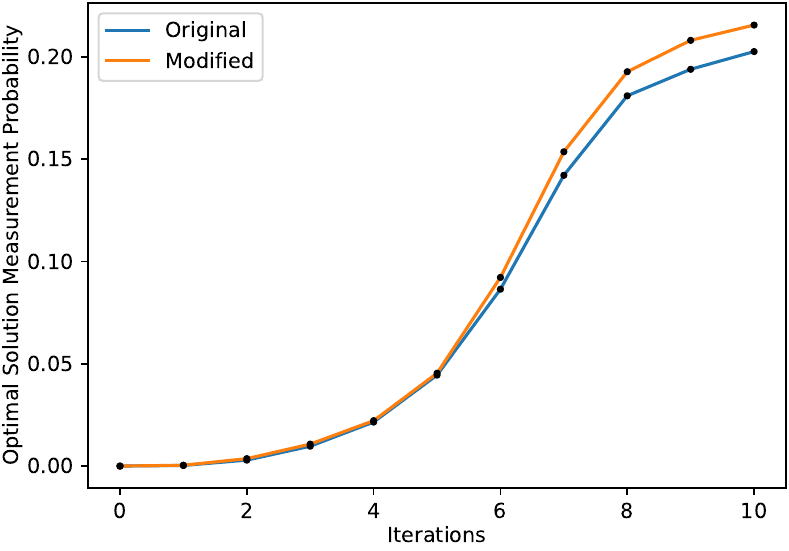}
        \caption{}
    \end{subfigure}
    \vspace{-0.25cm}
    \begin{subfigure}{0.49\columnwidth}
        \centering
        \includegraphics[width=1\columnwidth]{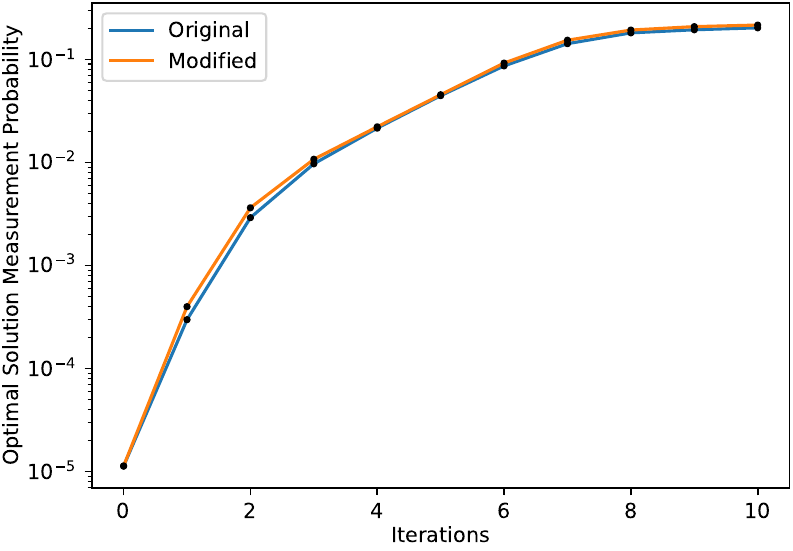}
        \caption{}
    \end{subfigure}
    \caption{Simulation results for $p=10$ non-variational QWOA applied to the k-means problem. Optimal solution measurement probability at each iteration (a) linear scale and (b) logarithmic scale, and shown for both the original and transformed objective function.}
    \label{fig:k_means_optimal_prob}
\end{figure}

We simulate the performance of the non-variational QWOA applied to this problem, both with the regular and transformed objective function, and for $p=10$ iterations. Gradient decent (with respect to expectation value of the objective function) is initialised with $\gamma=1$, $t=0.1$ and $\beta=\frac{1}{p}$. For the unmodified objective function, the amplified state $\ket{\gamma,t,\beta}$ is achieved with $\gamma=1.4960$, $t=0.2346$ and $\beta=0.3370$, whereas for the modified objective function, it is achieved with $\gamma=1.5345$, $t=0.2483$ and $\beta=0.3441$. \cref{fig:k_means_optimal_prob} shows how the globally optimal solution is amplified during the 10 iterations in both cases, with the modified objective function producing slightly better results. 

\begin{figure}[htbp]
    \centering
    \captionsetup{margin=1cm, font=small}
    \begin{subfigure}{0.48\columnwidth}
        \centering
        \includegraphics[width=1\columnwidth]{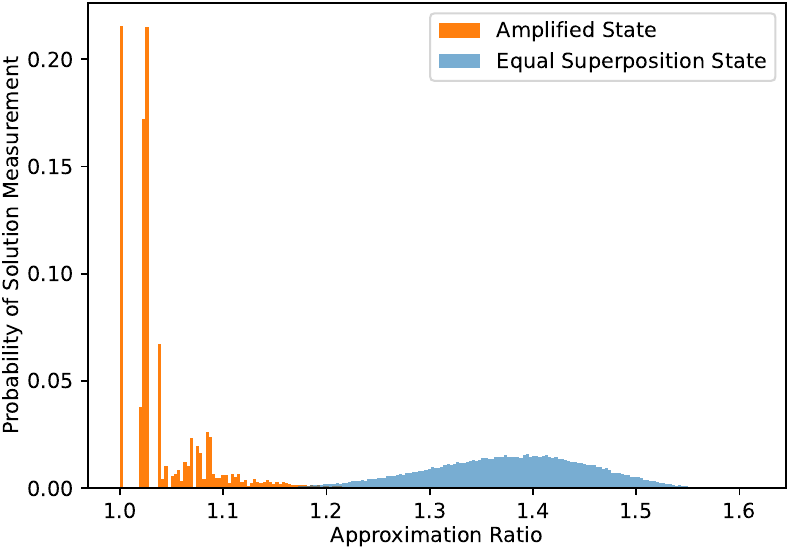}
        \caption{}
    \end{subfigure}
    \vspace{-0.25cm}
    \begin{subfigure}{0.48\columnwidth}
        \centering
        \includegraphics[width=1\columnwidth]{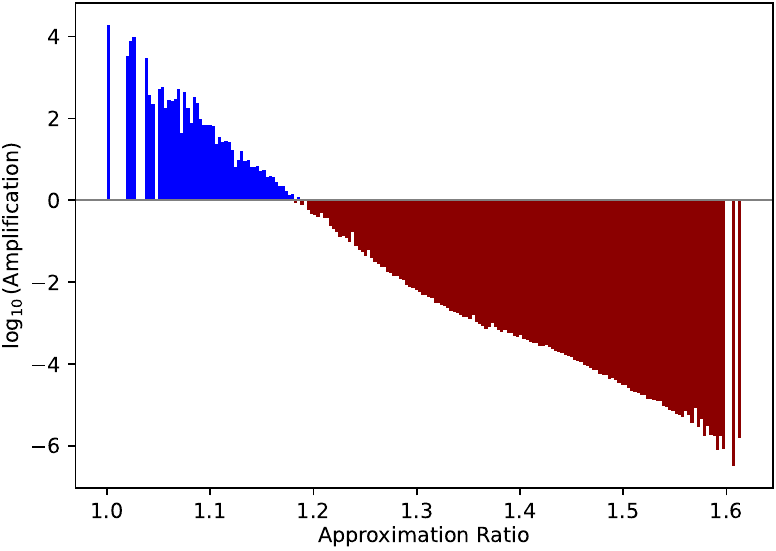}
        \caption{}
    \end{subfigure}
    \caption{Simulation results for $p=10$ non-variational QWOA applied to the k-means problem after transforming the objective function. (a) Probability distributions for approximation ratio as measured from the initial equal superposition state $\ket{s}$ and the amplified state $\ket{\gamma,t,\beta}$. (b) Logarithmic scale solution amplification as a function of approximation ratio.}
    \label{fig:k_means_distributions}
\end{figure}

\begin{figure}[htbp]
    \centering
    \captionsetup{margin=1cm, font=small}
    \includegraphics[width=0.64\columnwidth]{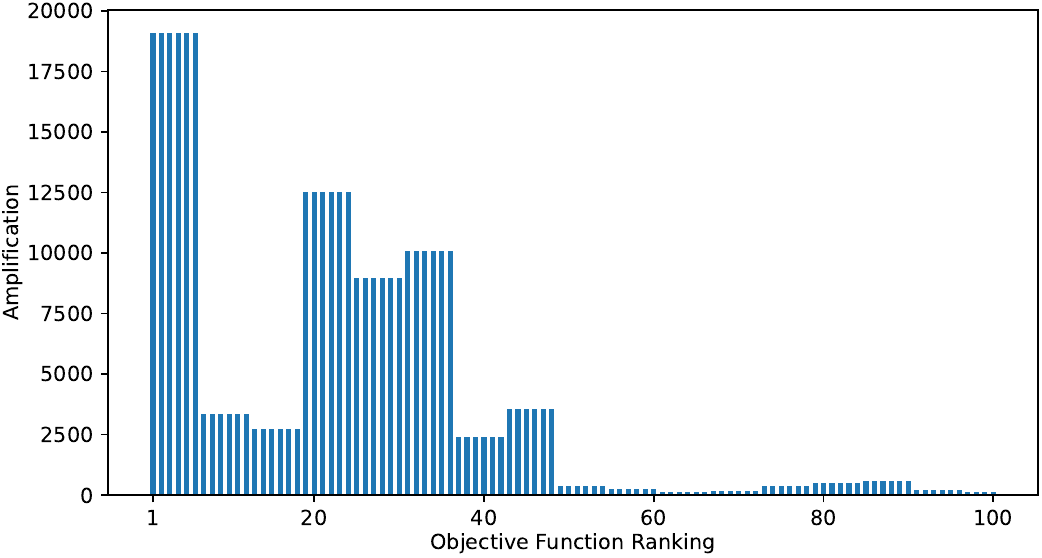}
    \caption{Amplification of the best 100 solutions to the k-means problem within the $p=10$ amplified state $\ket{\gamma,t,\beta}$.}
    \label{fig:k_means_top100}
\end{figure}

The distribution of objective function values before and after amplification, for the modified objective function, is shown in \cref{fig:k_means_distributions}, as well as the distribution of final amplifications. The amplification of the 100 solutions with the lowest objective function values are shown in \cref{fig:k_means_top100}, from which it is clear that the globally optimal solution is selectively amplified. In addition, it is clear that there are $k! = 6$ degenerate copies of each solution and that the degenerate solutions are equally amplified, as expected.

\section{Generalising further: The quadratic assignment problem}
\label{sec:QAP}

The quadratic assignment problem (QAP) is an extremely challenging combinatorial optimisation problem \cite{QAP_hardness_1} which makes the problem an interesting candidate for the study of quantum algorithms. Consider some set of facilities between which materials must be transported. A QAP can be characterised as the problem of assigning each of these facilities to a location, where the number of candidate locations is equal to the number of facilities. Each problem instance can be fully characterised by the distances (or costs associated with transport) between candidate locations and the amount of material flowing between facilities. We define $L_{i,j}$ as the distance or transport-cost between location $i$ and location $j$. Likewise, $F_{i,j}$ quantifies the material flowing between facility $i$ and facility $j$. Consider a solution to a QAP of size $n$ to be characterised by a vector, ${\bm{x} = (x_0, x_1, ..., x_{n-1})}$, where ${x_j \in \{0,1,...,n-1\}}$ specifies the allocation of a facility $j$ to location $x_j$. Only one facility can be assigned to each location, so we define the space of valid solutions $S$ to contain all $N=n!$ permutations, or in other words, all $\bm{x}$ such that there is no repetition in $x_j$. Given a particular solution $\bm{x}$, the objective function (to be minimised) can be expressed as,
\begin{equation}
    f(\bm{x}) = \sum_{i,j} F_{i,j} L_{x_i,x_j}.
    \label{eq:obj_func_QAP}
\end{equation}

As discussed in \cref{sec:definitions}, embedding solutions within the computational basis states of the input register can be achieved by assigning a sub-register to each of the $n$ variables $x_j$ and identifying the $n$ computational basis states of each of these sub-registers which directly encode variable values ${x_j \in \{0,1,...,n-1\}}$. Preparing the equal superposition over all feasible solutions in $S$ is not as simple as for maxcut or k-means clustering, since it cannot be achieved by acting independently on each sub-register. That is, the equal superposition state is not a separable state; the states of the sub-registers depend on each other, due to the non-repetition constraint. We present an efficient method in \cref{sec:permutation_state_preparation} to prepare the equal superposition. The method is recursive in nature and has gate complexity $n^3$. 

There are many ways to define nearest neighbours within the space of permutations. One such choice which is fairly natural, and the one that will be made here, is to consider nearest neighbours as those that can be generated from the set of all possible transpositions (the swapping of two elements of the permutation) of which there are $d=\frac{1}{2} n(n-1)$. This is equivalent to defining nearest neighbour solutions as those with a minimum non-zero Hamming distance between them, as described in \cref{sec:mixing_graph}. This transposition mixing graph has a suitably small diameter, $D=n-1$. If we consider another choice, swapping only adjacent elements of the permutation, this will produce a mixing graph with a diameter which is comparatively large, $D=\frac{1}{2} n(n-1)$. The smaller diameter mixer is preferred because it is better able to partition the solution space into subsets with distinct subset means and hence enables interference effects to reliably occur while mixing probability amplitudes more globally over the space of all solutions (computational paths). The adjacency matrix of the transposition mixing-graph can be expressed as the sum of operators which perform each of the transpositions,
\begin{equation}
    A = \sum_{i=0}^{n-2} \sum_{j=i+1}^{n-1} \text{SWAP}_{i,j},
    \label{eq:Adjacency_permutation_mixer}
\end{equation}
where $\text{SWAP}_{i,j}$ is defined as the permutation matrix associated with swapping the states of the $i^{\text{th}}$ and $j^{\text{th}}$ registers (applying identity to the remaining registers).

Unlike the terms in \cref{eq:Adjacency_binary_mixer} and \cref{eq:Adjacency_integer_mixer}, the individual SWAP operators composing this adjacency matrix do not commute. So the application of this mixer cannot be analysed in the same way since $e^{-\text{i} A t} \neq \prod e^{-\text{i} \text{SWAP} t}$. We can, however, make use of the parity property of permutations in order to study the behaviour of the mixer, using the fact that any single transposition always results in a change of parity. 

In order to demonstrate that a continuous-time quantum walk on this mixing graph satisfies the necessary requirement in \cref{eq:mixer_requirement_1}, we first consider the power series expansion defining the matrix exponential,
\begin{equation}
    U_M(t)\ket{\bm{u}} = e^{-\text{i} A t}\ket{\bm{u}} = \sum_{k=0}^{\infty} (-\text{i})^k \frac{t^k}{k!} A^k \ket{\bm{u}}.
\end{equation}
Due to the parity switching property of transpositions, $A$ acting on solutions in $h_{\bm{u}}$ produces solutions exclusively in either $(h-1)_u$ or $(h+1)_u$. Except of course for the cases where $h=0$ or $h=D=n-1$, where the action of $A$ produces solutions in just $1_u$ or $(n-2)_u$ respectively. This allows for the expression to be modified as follows,
\begin{equation}
    \fl \hspace{1.5cm} U_M(t)\ket{\bm{u}} = \sum_{h=0}^{n-1} \left[ (-\text{i})^h \sum_{\bm{x} \in h_{\bm{u}}} \left( \sum_{k=0}^{\infty} \left(  (-1)^k \frac{t^{h+2k}}{(h+2k)!} \bra{\bm{x}} A^{h+2k} \ket{\bm{u}} \right) \ket{\bm{x}} \right) \right],
\end{equation}
which satisfies the necessary condition in \cref{eq:mixer_requirement_1} so long as,
\begin{equation}
    \sum_{k=0}^{\infty} \left(  (-1)^k \frac{t^{h+2k}}{(h+2k)!} \bra{\bm{x}} A^{h+2k} \ket{\bm{u}} \right),
\end{equation}
always sums to a positive real value. This will occur when the leading term dominates for all cases, that is, for sufficiently small walk times $t$ such that,
\begin{equation}
    \fl \frac{t^{2}}{(h+1)(h+2)} \bra{\bm{x}} A^{h+2} \ket{\bm{u}} < \bra{\bm{x}} A^{h} \ket{\bm{u}} \hspace{0.5cm} \Longrightarrow \hspace{0.5cm} t < \sqrt{\frac{(h+1)(h+2)\bra{\bm{x}} A^{h} \ket{\bm{u}}}{\bra{\bm{x}} A^{h+2} \ket{\bm{u}}}}
\end{equation}
holds for all $\bm{x} \in S$ and for all $h \in \left[0,n-1\right]$. For $h=0$, or $\bm{x} = \bm{u}$, this can be evaluated exactly,
\begin{equation}
     \fl t < \sqrt{\frac{2\bra{\bm{u}} A^{0} \ket{\bm{u}}}{\bra{\bm{u}} A^{2} \ket{\bm{u}}}}   \hspace{0.3cm} \Longrightarrow \hspace{0.3cm}  t < \sqrt{\frac{2}{\bra{\bm{u}} A \sum_{\bm{x} \in 1_u} \ket{\bm{x}}}}  \hspace{0.3cm} \Longrightarrow \hspace{0.5cm}  t < \sqrt{\frac{2}{d}} \hspace{0.3cm} \Longrightarrow \hspace{0.3cm}  t < \frac{2}{n}
\end{equation}

For other cases this is not easily analysed, though an order of magnitude estimation can be made. Values of $(h+1)(h+2)$ are of order $n^2$ and values of $\frac{\bra{\bm{x}} A^{h} \ket{\bm{u}}}{\bra{\bm{x}} A^{h+2} \ket{\bm{u}}}$ should be of order $d^{-2}$ since the action of $A$ increases total probability amplitude by a factor of the graph's degree $d$ ($A$ is not unitary, while the converging series is). Since $d$ is of order $n^2$ we expect positive values for $t \lesssim \frac{1}{n}$. Therefore, for values of $t$ within this range, the action of the transposition mixer can be expressed as,
\begin{equation}
    U_M(t)\ket{\bm{u}} = \sum_{h=0}^{n-1} (-\text{i})^h \sum_{\bm{x} \in h_{\bm{u}}} r_{\bm{x}}(t) \ket{\bm{x}},
\end{equation}
where $r_{\bm{x}}(t)$ are positive-real values for all $\bm{x} \in S$, satisfying \cref{eq:mixer_requirement_1} with $\phi(t)=\frac{\pi}{2}$. 

Efficient implementation of the transposition mixer (continuous-time quantum walk on the transposition graph) can be achieved via Hamiltonian simulation. For example, Berry \emph{et al.} \cite{berry2015simulating} presents a truncated Taylor series method to implement $e^{-\text{i} H t}$ where $H$ is a linear combination of implementable unitary matrices, which is exactly satisfied in this case with $H=A$ and $\text{SWAP}_{i,j}$ playing the role of the individual unitary matrices which are efficiently implemented with a combination of regular 2-qubit SWAP gates. There may be other ways to efficiently approximate the action of this mixer, for example, by Trotterization.

\begin{figure}[htbp]
    \centering
    \captionsetup{margin=1cm, font=small}
    \includegraphics[width=0.5\columnwidth]{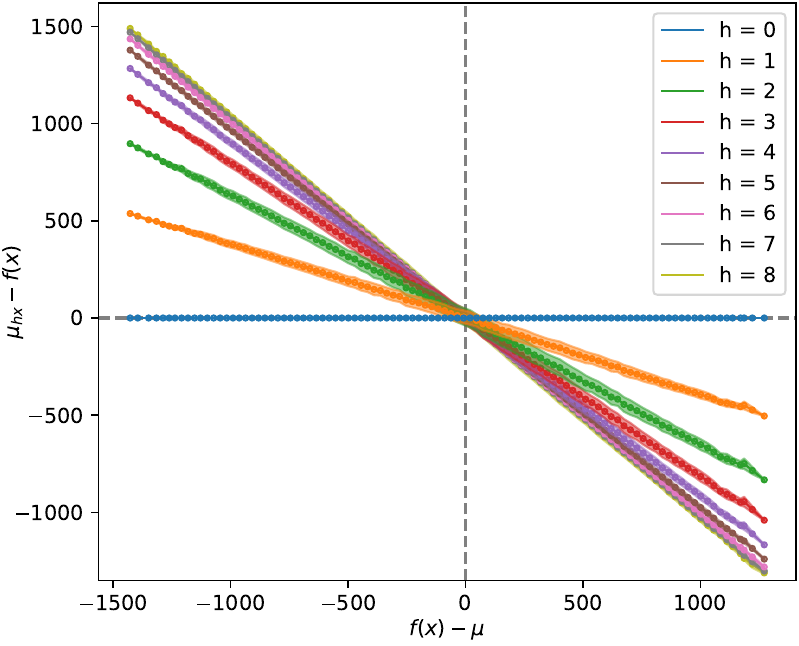}
    \caption{Statistical analysis of subset means $\mu_{h\bm{x}}$ for the $n=9$ quadratic assignment problem. Shading shows variations in the values of $\mu_{h\bm{x}}$ (plus or minus a single standard deviation).}
    \label{fig:QAP_subset_means}
\end{figure}

Now that the transposition mixer has been shown to satisfy the first necessary condition and be efficiently implementable, we demonstrate that objective function values as distributed over the mixing graph satisfy \cref{eq:mixer_requirement_2}. For this purpose and for subsequent simulations, a random QAP instance of size $n=9$ has been generated and included in \cref{sec:random_QAP}. \cref{fig:QAP_subset_means} shows the distribution of subset means in $h_{\bm{x}}$ relative to objective function value $f(\bm{x})$, confirming that the requirement is indeed satisfied.

\begin{figure}[htbp]
    \centering
    \captionsetup{margin=1cm, font=small}
    \begin{subfigure}{0.49\columnwidth}
        \centering
        \includegraphics[width=1\columnwidth]{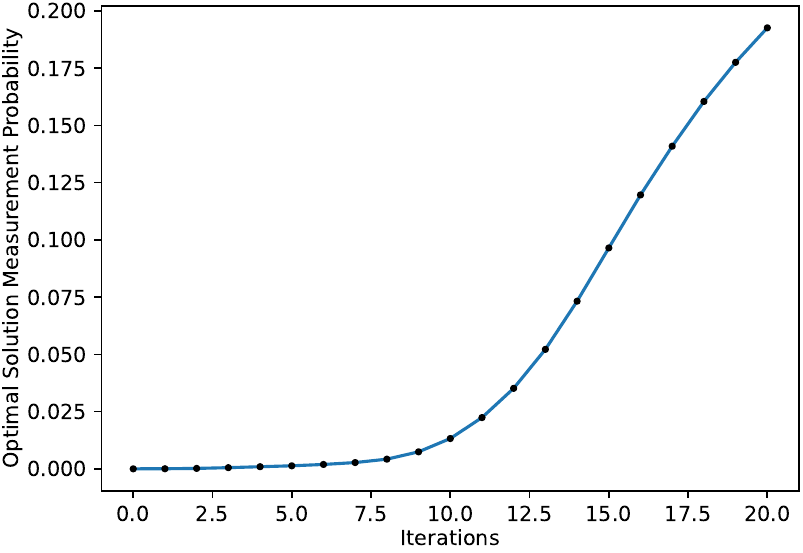}
        \caption{}
    \end{subfigure}
    \begin{subfigure}{0.49\columnwidth}
        \centering
        \includegraphics[width=1\columnwidth]{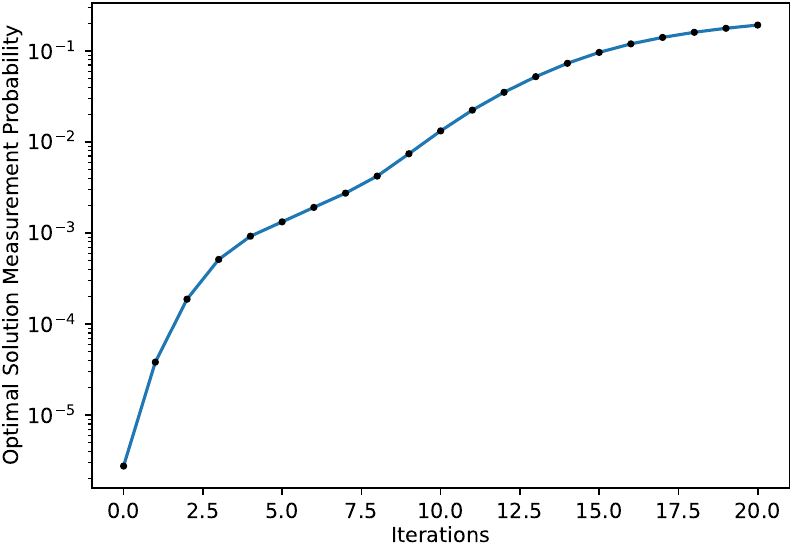}
        \caption{}
    \end{subfigure}
    \vspace{-0.25cm}
    \caption{Simulation results for $p=20$ non-variational QWOA applied to the quadratic assignment problem. Optimal solution measurement probability at each iteration (a) linear scale and (b) logarithmic scale.}
    \label{fig:QAP_optimal_prob}
\end{figure}

\begin{figure}[htbp]
    \centering
    \captionsetup{margin=1cm, font=small}
    \begin{subfigure}{0.48\columnwidth}
        \centering
        \includegraphics[width=1\columnwidth]{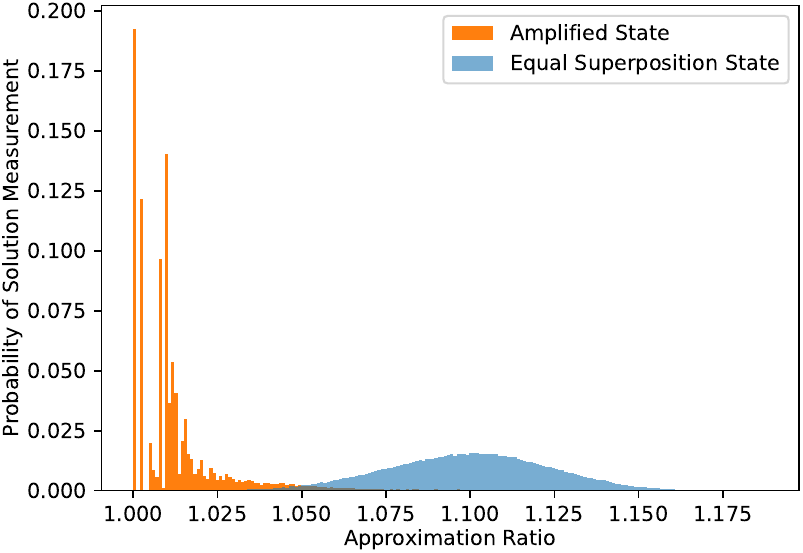}
        \caption{}
    \end{subfigure}
    \vspace{-0.25cm}
    \begin{subfigure}{0.48\columnwidth}
        \centering
        \includegraphics[width=1\columnwidth]{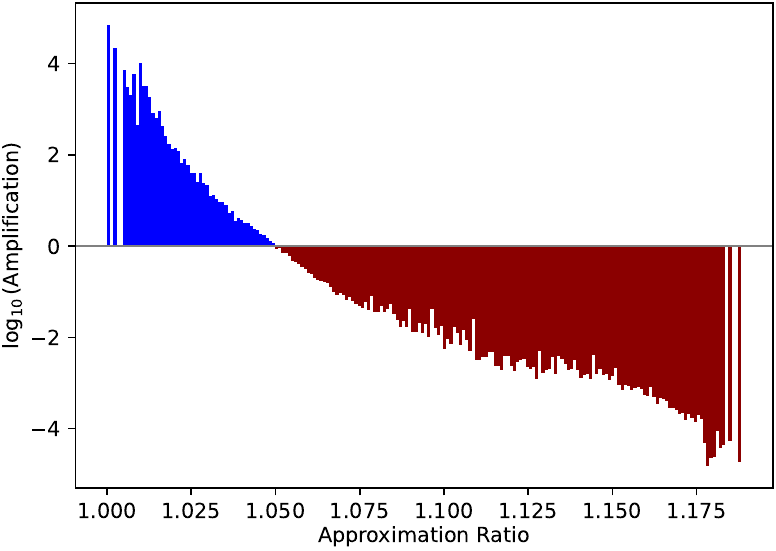}
        \caption{}
    \end{subfigure}
    \caption{Simulation results for $p=20$ non-variational QWOA applied to the quadratic assignment problem. (a) Probability distributions for approximation ratio as measured from the initial equal superposition state $\ket{s}$ and the amplified state $\ket{\gamma,t,\beta}$. (b) Logarithmic scale solution amplification as a function of approximation ratio.}
    \label{fig:QAP_distributions}
\end{figure}

\begin{figure}
    \centering
    \captionsetup{margin=1cm, font=small}
    \includegraphics[width=0.64\columnwidth]{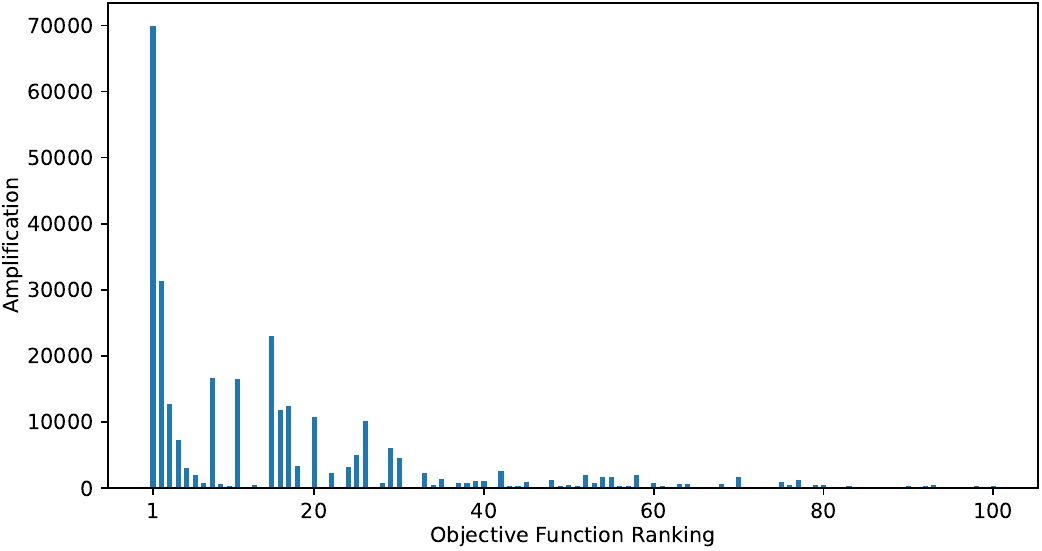}
    \caption{Amplification of the best 100 solutions to the quadratic assignment problem within the $p=20$ amplified state $\ket{\gamma,t,\beta}$.}
    \label{fig:QAP_top100}
\end{figure}

To assess the performance of the non-variational QWOA applied to the randomly generated $n=9$ QAP in \cref{sec:random_QAP}, we simulate the amplified state $\ket{\gamma,t,\beta}$ for $p=20$ iterations. The gradient descent is initiated with $\gamma=1$, $t=\frac{1}{n}$ and $\beta=\frac{1}{p}$ and terminates at optimal values, $\gamma=1.2636$, $t=0.1219$ and $\beta=0.4167$. Evolution of the measurement probability for the optimal solution is shown in \cref{fig:QAP_optimal_prob}, and the final state is analysed in \cref{fig:QAP_distributions} and \cref{fig:QAP_top100}. The results demonstrate a similar performance for the quadratic assignment problem as compared to maxcut and k-means clustering, with the non-variational QWOA solving this particular problem instance. This indicates successful generalisation to a wide range of problem structures, given the significant difference in structure which is introduced with the permutation constraint, compared to the unconstrained binary and non-binary integer problems.

\newpage
\section{Constrained problems: A penalty function approach}
\label{sec:pen_func}
As discussed in \cref{sec:constrained_problems}, if the space of valid solutions is problem instance specific, it is not possible to design a mixing graph which isolates and mixes between only valid solutions. For problems of this kind, the mixing graph is designed to connect a space of feasible solutions which contains, in general, both valid and invalid solutions. In an attempt to enforce or embed the problem constraints, a penalty function method is used in which additional terms are included in the objective function. These terms apply only when a solution is invalid and act to penalise any violation of the problem constraints. 

We introduce three kinds of terms, the first two are designed to ensure that invalid solutions are sufficiently penalised, so that the penalised objective function is optimised by valid solutions. The first of these is a variable term which scales with the extent to which a constraint has been violated. The second is a fixed term which ensures that solutions which violate a constraint by only a small amount remain sufficiently penalised. As demonstrated in detail in \cref{sec:CFLP}, the introduction of these first two penalty terms may adversely effect algorithm performance, due to the introduction of bimodality in the distribution of objective function values across the space of feasible (valid and invalid) solutions. The third penalty term is introduced to correct for this bimodality and improve algorithm performance, while leaving near-optimal invalid solutions adequately penalised.

The influence of each penalty term is controlled by a positive-valued parameter, such that the penalty can be adjusted to improve performance. The penalised objective function is therefore expressed as $f(\bm{x})_{\bm{\lambda}}$, where ${\bm{\lambda} = (\lambda_1, \lambda_2, ...)}$ are the coefficients for each of the terms in the penalty function. Similar to the $\gamma$, $t$ and $\beta$ parameters, appropriate penalty parameters $\bm{\lambda}$ may be known from prior experience. In any case, the penalty parameters can be tuned during the repeated state preparation and measurement process. The general approach is summarised as follows:
\begin{enumerate}
    \item Design the penalty terms and select fixed penalty parameters $\bm{\lambda}_F$ which adequately penalise the invalid solutions, such that the globally optimal values of $f(\bm{x})_{\bm{\lambda}_F}$ belong to valid solutions. 
    \item Using the same penalty terms, define a second objective function, parameterised by tunable penalty parameters $\bm{\lambda}_T$. This objective function $f(\bm{x})_{\bm{\lambda}_T}$ is used within the phase-separation unitary.
    \item Through repeated state preparation and measurement of the amplified state, use gradient ascent/descent to tune the parameters $\{\gamma,t,\beta,\bm{\lambda}_T\}$ (initialised with $\bm{\lambda}_T=\bm{\lambda}_F$) so as to optimise the expectation value of $f(\bm{x})_{\bm{\lambda}_F}$. In other words, optimise $_{\bm{\lambda}_T}\!\bra{\gamma,t,\beta} \hat{f}_{\bm{\lambda}_F} \ket{\gamma,t,\beta}_{\bm{\lambda}_T}$, where $\ket{\gamma,t,\beta}_{\bm{\lambda}_T}$ is the amplified state prepared using $f(\bm{x})_{\bm{\lambda}_T}$.
\end{enumerate}

\section{Constrained problems: Maximum independent set}
\label{sec:MIS}
We'll begin our exploration of constrained problems by studying the (strongly NP-hard \cite{garey1978strong}) maximum independent set problem, as it has a similar structure to maxcut, possessing solutions composed of binary variables. A size $n$ problem instance is defined by an undirected graph $G(V,E)$ with $n$ vertices and an independent set is a subset of the graph's vertices which does not contain any pair of adjacent vertices. Solving the maximum independent set problem involves finding (one of) the independent sets which contains the maximum possible number of vertices. 

As with the maxcut problem, we characterise a solution to the maximum independent set problem as a vector, $\bm{x} = \left(x_1, x_2, ... ,x_n \right)$, where each element of the vector is a binary variable $x_j \in \{0,1\}$ encoding inclusion or exclusion of vertex $j$ in the subset. So the space of solutions $S$ is composed of all possible $\bm{x}$, of which there are $N=2^n$. A sensible perturbation to any particular solution's configuration can be achieved via a single bit-flip, which moves one vertex into or out of the subset. With this choice, the corresponding set of available moves and associated mixing graph is identical to that for maxcut.

In general, many of the solutions in $S$ will be invalid, i.e. many of these solutions will not form an independent set. The independent set constraint will be violated if there is even a single pair of connected vertices included in the subset. We therefore define the penalised objective function (to be maximised) as,
\begin{equation}
    f(\bm{x})_{\bm{\lambda}} = \left( \sum_{j=1}^n x_j \right) - \lambda_1 P_1(\bm{x}) - \lambda_2 P_2(\bm{x}),
\end{equation}
where $P_1(\bm{x})$ counts the number of connected pairs in the subset (quantifies the extent to which the constraint has been violated), 
\begin{equation}
    P_1(\bm{x}) = \sum_{(i,j) \in E} x_i x_j,
\end{equation}
and $P_2(\bm{x})$ flags whether the solution violates the constraint,
\begin{equation}
    P_2(\bm{x}) = \cases{0,\hspace{0.5cm} \mbox{if $ \displaystyle \sum_{(i,j) \in E} x_i x_j = 0$} \\
        1, \hspace{0.5cm} \mbox{otherwise} \\}.
\end{equation}

When selecting the fixed penalty parameters, we aim to ensure that globally optimal solutions are necessarily valid. Consider a globally optimal maximum independent set: If, in an attempt to increase the objective function value, another vertex is added to the subset, this will necessarily introduce at least one pair of adjacent vertices. Therefore, applying a penalty greater than $1$ for each violation (each pair of adjacent vertices) will ensure that the globally optimal solution must be a valid and independent set. As such, we choose $\bm{\lambda}_F = (1.5,0)$.

\begin{figure}[htbp]
    \centering
    \captionsetup{margin=1cm, font=small}
    \begin{subfigure}{0.49\columnwidth}
        \centering
        \includegraphics[width=1\columnwidth]{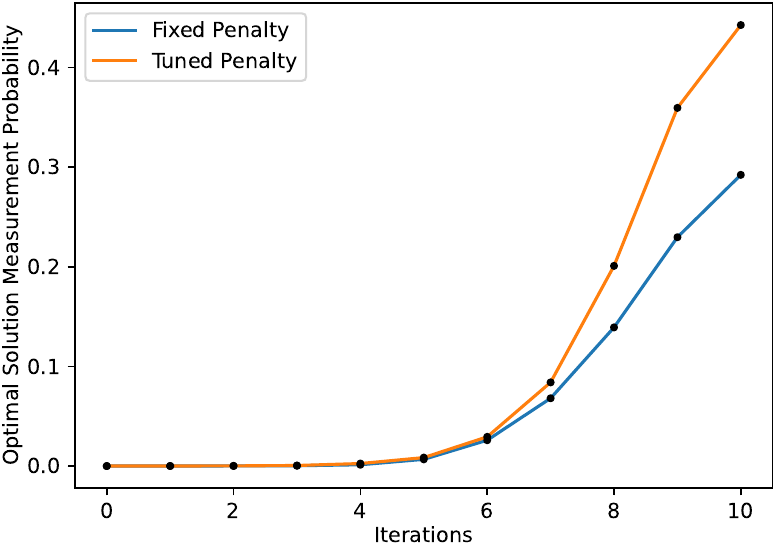}
        \caption{}
    \end{subfigure}
    \vspace{-0.25cm}
    \begin{subfigure}{0.49\columnwidth}
        \centering
        \includegraphics[width=1\columnwidth]{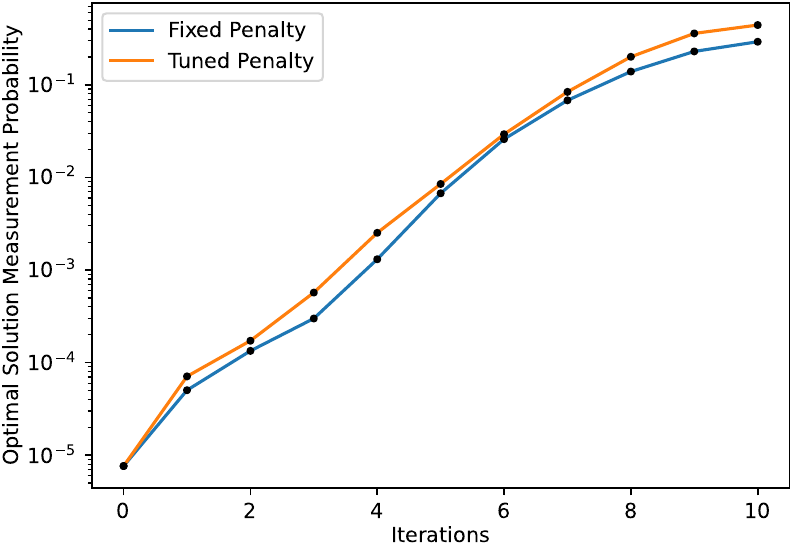}
        \caption{}
    \end{subfigure}
    \caption{Simulation results for $p=10$ non-variational QWOA applied to the maximum independent set problem, for both the fixed and tuned penalty. Optimal solution measurement probability at each iteration (a) linear scale and (b) logarithmic scale.}
    \label{fig:MIS_optimal_prob}
\end{figure}

In order to demonstrate the second necessary condition (\cref{eq:mixer_requirement_2}) and in order to assess the performance of the algorithm, we generate a random $n=18$ graph, as detailed in \cref{sec:random_MIS}. For this problem instance, $1.04\%$ of the solutions are valid, and there are two maximum independent sets, containing 9 vertices each. Selecting $p=10$ iterations, we simulate the non-variational QWOA applied to this randomly generated problem. For comparison, we begin by foregoing the penalty tuning process and simply use the fixed penalty objective function $f(\bm{x})_{\bm{\lambda}_F}$. Beginning with $\gamma=1$, $t=0.1$ and $\beta=\frac{1}{p}$, gradient ascent terminates with the amplified state $\ket{\gamma,t,\beta}$ given by $\gamma=4.0520$, $t=0.5289$ and $\beta=0.1225$. If instead we carry out the penalty tuning process, including initial parameters ${\bm{\lambda}_T=(1.5,0)}$, the amplified state is achieved with $\gamma=3.0098$, $t=0.5724$, $\beta=0.1722$ and ${\bm{\lambda}_T=(1.0370,0.5235)}$. As can be seen in \cref{fig:MIS_optimal_prob} the tuned penalty does perform better, producing greater measurement probability for the two maximum independent sets.  

\begin{figure}[htbp]
    \centering
    \begin{subfigure}{0.49\columnwidth}
        \centering
        \includegraphics[width=1\columnwidth]{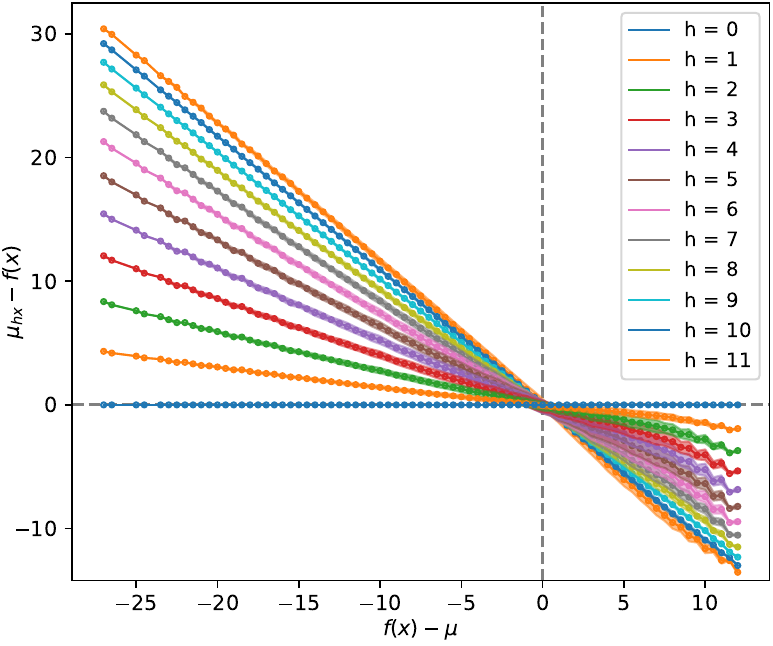}
        \caption{}
    \end{subfigure}
    \vspace{-0.25cm}
    \begin{subfigure}{0.49\columnwidth}
        \centering
        \includegraphics[width=1\columnwidth]{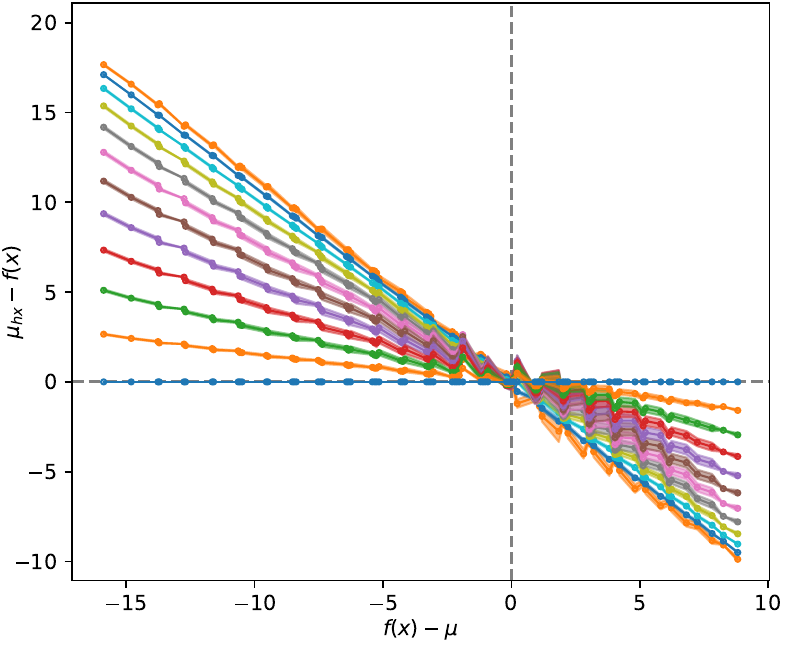}
        \caption{}
    \end{subfigure}
    \caption{Statistical analysis of subset means $\mu_{h\bm{x}}$ for the maximum independent set problem for (a) the fixed penalty $\bm{\lambda}_F = (1.5,0)$ and (b) the tuned penalty $\bm{\lambda}_T = (1.1094,0.4747)$. Shading shows variations in the values of $\mu_{h\bm{x}}$ (plus or minus a single standard deviation).}
    \label{fig:MIS_subset_means}
\end{figure}

Results of an analysis of subset means $\mu_{h\bm{x}}$ for both the fixed and tuned penalty is shown in \cref{fig:MIS_subset_means}, confirming that the second necessary condition is satisfied in both cases. In addition, the analysis shows how the tuned penalty ensures that the optimal solutions have the largest values for $\left(\mu_{h\bm{x}} - f(\bm{x})\right)$, for all $h$, whereas this is not the case for the fixed penalty, perhaps helping to explain, at least in part, how the tuned penalty performs better.

\begin{figure}[htbp]
    \centering
    \captionsetup{margin=1cm, font=small}
    \begin{subfigure}{0.48\columnwidth}
        \centering
        \includegraphics[width=1\columnwidth]{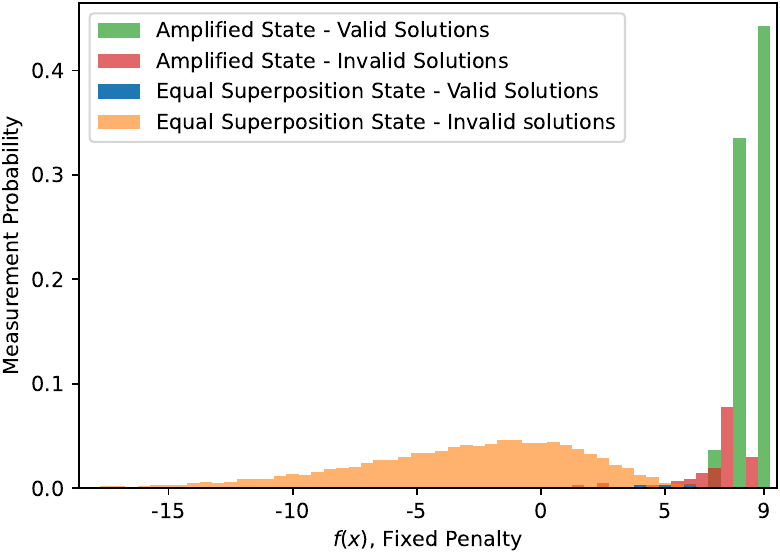}
        \caption{}
    \end{subfigure}
    \vspace{-0.25cm}
    \begin{subfigure}{0.48\columnwidth}
        \centering
        \includegraphics[width=1\columnwidth]{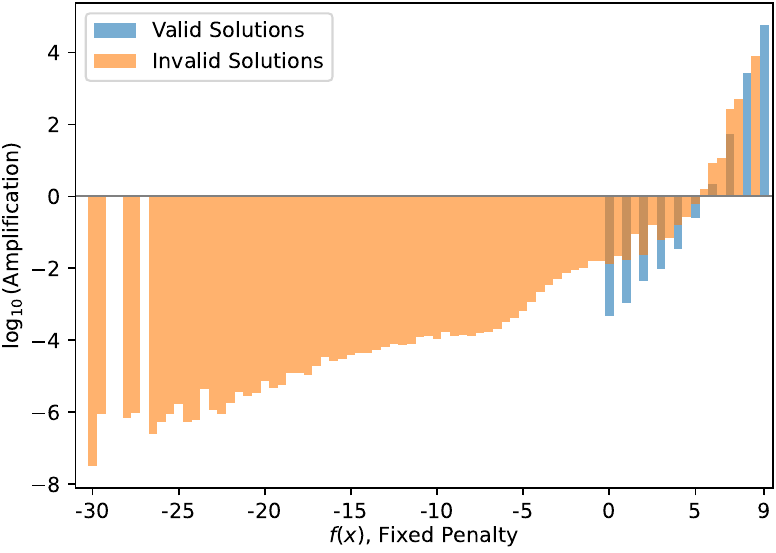}
        \caption{}
    \end{subfigure}
    \caption{Simulation results for $p=10$ non-variational QWOA applied to the maximum independent set problem. (a) Probability distributions for fixed penalty objective function values $f(\bm{x})_{\bm{\lambda}_F}$ as measured from the initial equal superposition state $\ket{s}$ and the amplified state $\ket{\gamma,t,\beta}_{\bm{\lambda}_T}$. (b) Logarithmic scale solution amplification as a function of $f(\bm{x})_{\bm{\lambda}_F}$.}
    \label{fig:MIS_distributions}
\end{figure}

\begin{figure}[htbp]
    \centering
    \captionsetup{margin=1cm, font=small}
    \includegraphics[width=0.64\columnwidth]{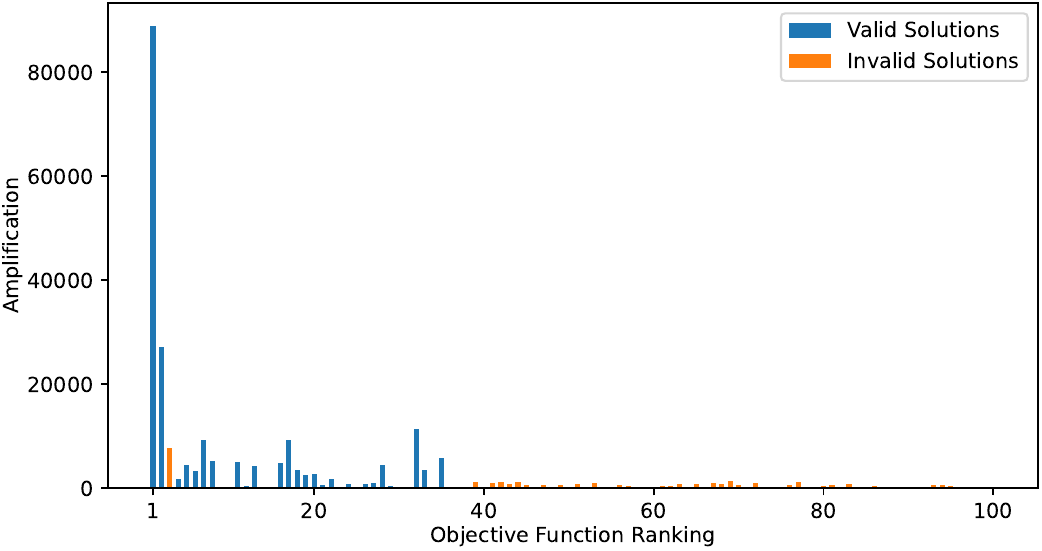}
    \caption{Amplification of the best 100 solutions to the maximum independent set problem within the $p=10$ amplified state $\ket{\gamma,t,\beta}_{\bm{\lambda}_T}$. Solutions are ranked according to $f(\bm{x})_{\bm{\lambda}_T}$.}
    \label{fig:MIS_top100}
\end{figure}

More detailed simulation results (with the tuned penalty) are shown in \cref{fig:MIS_distributions} and \cref{fig:MIS_top100}, indicating that the two maximum independent sets are amplified significantly more than all other solutions, though one of these more than the other. These results, in general, appear very consistent with the positive outcomes demonstrated on previous non-constrained problems, suggesting that the non-variational QWOA may generalise well to constrained problems, or at least, to those problems for which the constraints can be suitably embedded via a penalised objective function.

In a recent study focused on quantum optimisation for constrained problems, Saleem \emph{et al.} \cite{saleem2023approaches} focus specifically on the maximum independent set problem and introduce an alternative variational approach, which was demonstrated on a randomly generated 14 vertex graph. While it is difficult to make direct comparison, we note that the non-variational QWOA solves this particular problem instance, producing a measurement probability for the 8 maximum independent sets of 0.16 for p = 2 iterations, and 0.36 for p = 3 iterations.

\section{Constrained problems: The capacitated facility location problem}
\label{sec:CFLP}
The facility location problem involves the servicing of $n$ customers via some number of facilities located amongst $k$ candidate locations. A solution to the problem is a selection of candidate locations at which to install a facility, and the assignment of each customer to a facility. There are a few parameters which define a particular problem instance. $F_i$ is the cost associated with opening a facility at candidate location $i$, and $L_{j,i}$ measures the transport cost between customer $j$ and facility location $i$. We consider a variant of the problem where each customer is serviced by only a single facility. In addition, we consider the capacitated variant, the capacitated facility location problem (CFLP), where customer $j$ requires a number of resources $R_j$, and a facility at candidate location $i$ has a maximum capacity $C_i$ with respect to total supplied resources. 
 
Consider a particular solution to be characterised by a vector, ${\bm{x} = (x_1, x_2, ..., x_n)}$, where ${x_j \in \{0,1,...,k-1\}}$ specifies the allocation of customer $j$ to a facility at candidate location $x_j$. We define the space of feasible solutions $S$ containing every $\bm{x}$ with cardinality $N = k^n$. For this particular encoding, neighbouring solutions will naturally be those that vary by just a single customer assigned to a different facility location. This encoding of solutions and the neighbourhood associated with it is therefore functionally equivalent to that defined in \cref{sec:kmeans} for k-means clustering, so the same qubit embedding and mixer can be used. 

However, unlike the k-means clustering problem, the CFLP involves a constraint which partitions the set of feasible solutions into valid and invalid subsets. Prior to embedding the capacity constraint via penalty terms, the objective function (to be minimised) is given by,
\begin{equation}
    f(\bm{x}) =  \sum_{j=1}^n R_j L_{j,x_j}  + \sum_{i = 0}^{k-1} \cases{F_i,\hspace{0.5cm} \mbox{if $i \in \bm{x}$} \\
        0, \hspace{0.5cm} \mbox{otherwise} \\ },
    \label{eq:non_penalised}
\end{equation}
where the first sum accounts for the transportation cost of resources and the second sum accounts for the total cost of opening facilities. 

\begin{figure}[htbp]
    \centering
    \captionsetup{margin=1cm,font=small}
    \begin{subfigure}{0.49\columnwidth}
        \centering
        \includegraphics[width=1\columnwidth]{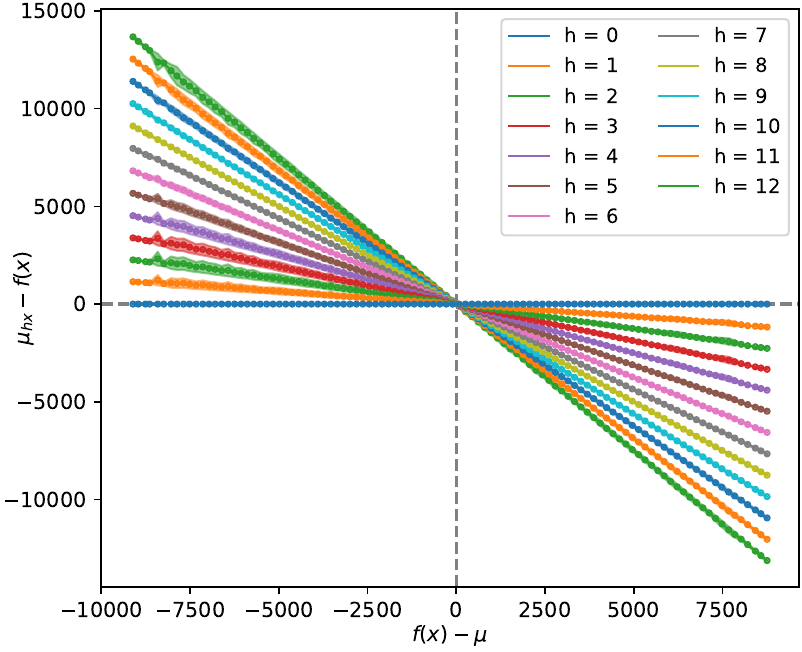}
        \caption{}
        \label{fig:CFLP_unconstrained_subset_means}
    \end{subfigure}
    \vspace{-0.25cm}
    \begin{subfigure}{0.49\columnwidth}
        \centering
        \includegraphics[width=1\columnwidth]{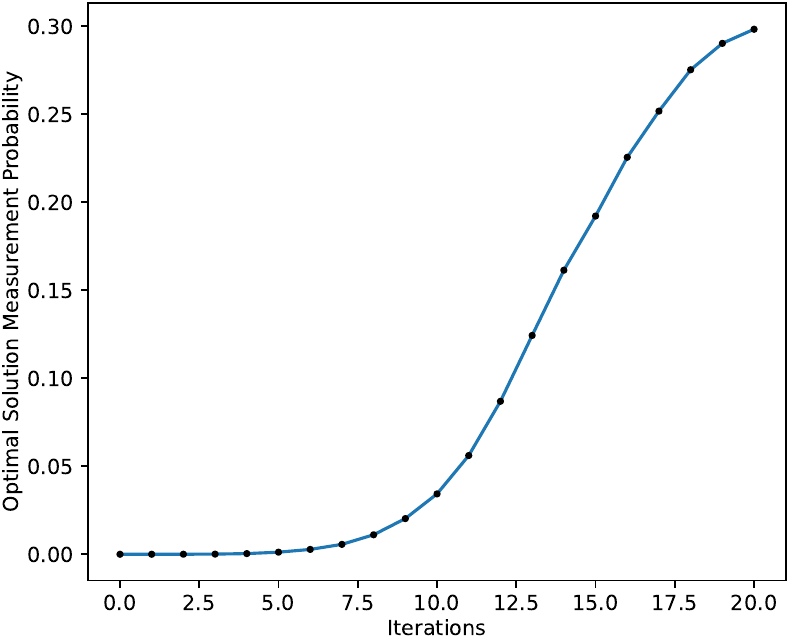}
        \caption{}
        \label{fig:CFLP_unconstrained_meas_prob}
    \end{subfigure}
    \caption{An analysis of the unconstrained variant associated with the $n=12$ CFLP. (a) Subset means satisfying \cref{eq:mixer_requirement_2} and (b) evolution of the optimal solution measurement probability within the amplified state $\ket{\gamma,t,\beta}$.}
\end{figure}

Consider the randomly generated $n=12$, $k=3$, problem instance recorded in \cref{sec:random_CFLP}. To begin with, we demonstrate that the unconstrained variant of the problem is solvable with the non-variational QWOA. An analysis of the subset means $\mu_{h\bm{x}}$ for the unconstrained objective function, distributed over the mixing graph, is shown in \cref{fig:CFLP_unconstrained_subset_means}, demonstrating clearly that the second condition (\cref{eq:mixer_requirement_2}) is satisfied. Using $p=20$ iterations, the amplified state is achieved with $\gamma=2.9258$, $t=0.3147$ and $\beta=0.0353$, for which the globally optimal solution is amplified to a measurement probability of $0.30$, as shown in \cref{fig:CFLP_unconstrained_meas_prob}. 

\begin{figure}[htbp]
    \centering
    \captionsetup{margin=1cm, font=small}
    \includegraphics[width=0.48\columnwidth]{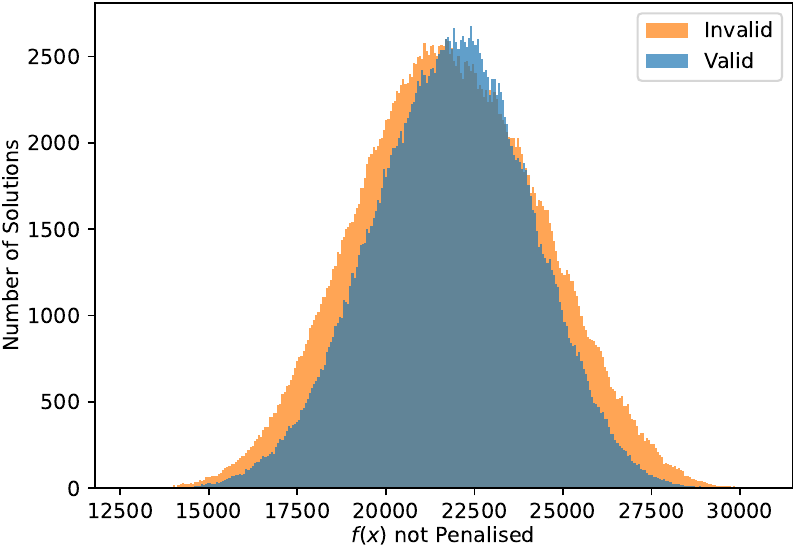}
    \caption{Objective function distributions for the randomly generated CFLP without a penalty function included.}
    \label{fig:CFLP_without_penalty}
\end{figure}

Focusing again on the constrained variant, for the same problem instance, approximately $46\%$ of the solutions are valid. The distribution of objective function values as per \cref{eq:non_penalised} is visualised in \cref{fig:CFLP_without_penalty}, where it is clear that the optimal and nearest-optimal values of $f(\bm{x})$ occur for invalid solutions. 

\newpage
Consider now a trial penalised objective function,
\begin{equation}
    g(\bm{x})_{\bm{\lambda}} = f(\bm{x}) + \sum_{i = 1}^k \cases{0,\hspace{0.5cm} \mbox{if $ \sum_{j: x_j = i} R_j \leq C_i$} \\ 
    \lambda_1 P_{i,1}(\bm{x}) + \lambda_2 P_{i,2}(\bm{x}), \hspace{0.1cm} \mbox{otherwise} \\},
\end{equation}
where $P_{i,1}(\bm{x})$ is a variable term which scales with the average facility-customer distance,
\begin{equation}
    P_{i,1}(\bm{x}) = \left(\frac{1}{n k} \sum_{j=1}^n\sum_{l=1}^k L_{j,l}\right)  \left( \left( \sum_{j: x_j = i} R_j \right) - C_i\right),
\end{equation}
and $P_{i,2}(\bm{x})$ is a fixed term proportional to the average facility opening cost,
\begin{equation}
    P_{i,2}(\bm{x}) = \left(\frac{1}{k} \sum_{j=1}^k F_j\right)  \left\lceil \frac{\left( \sum_{j: x_j = i} R_j \right) - C_i}{C_i} \right\rceil.
\end{equation}

\begin{figure}[htbp]
    \centering
    \captionsetup{margin=1cm,font=small}
    \begin{subfigure}{0.48\columnwidth}
        \centering
        \includegraphics[width=1\columnwidth]{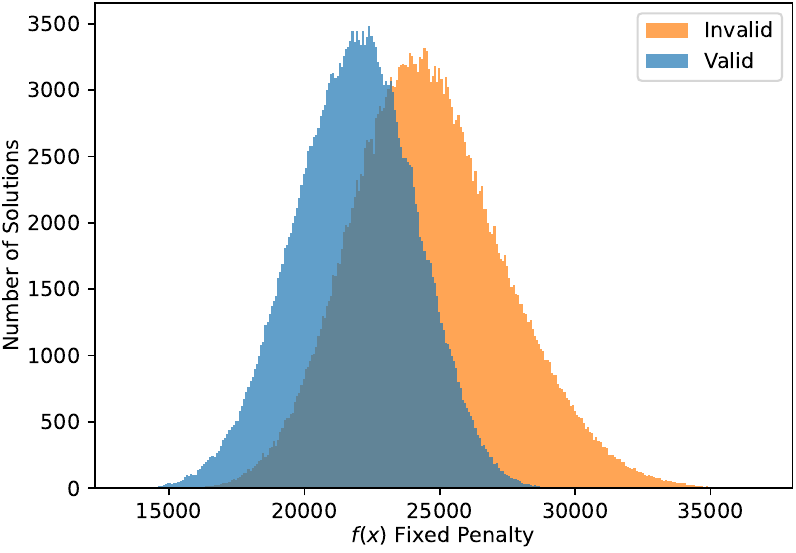}
        \caption{}
        \label{fig:CFLP_with_fixed_penalty}
    \end{subfigure}
    \vspace{-0.25cm}
    \begin{subfigure}{0.48\columnwidth}
        \centering
        \includegraphics[width=1\columnwidth]{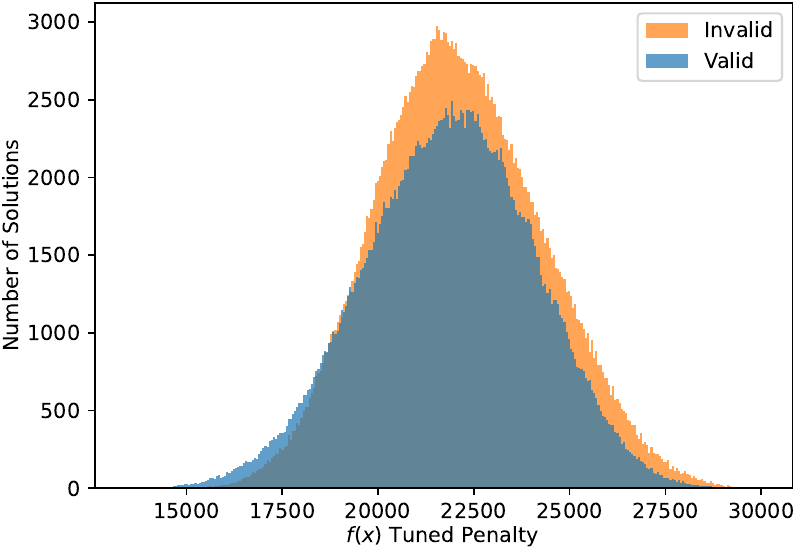}
        \caption{}
        \label{fig:CFLP_with_tuned_penalty}
    \end{subfigure}
    \caption{Objective function distributions for the randomly generated CFLP: (a) with the fixed penalty $\bm{\lambda}_F = (1,1,0)$ and (b) with the tuned penalty ${\bm{\lambda}_T = (0.8966,0.4996,0.1732)}$.}
\end{figure}

\begin{figure}[htbp]
    \centering
    \captionsetup{margin=1cm,font=small}
    \begin{subfigure}{0.49\columnwidth}
        \centering
        \includegraphics[width=1\columnwidth]{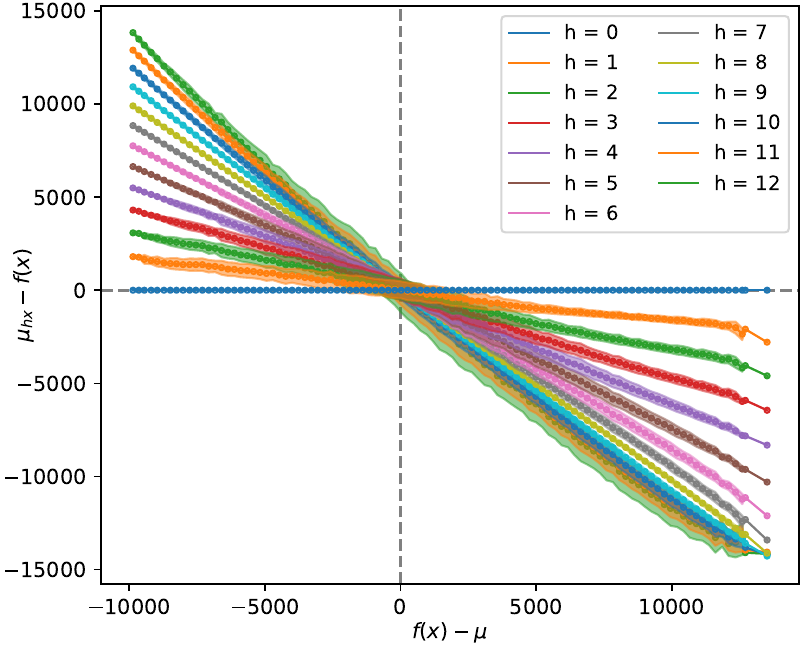}
        \caption{}
    \end{subfigure}
    \vspace{-0.25cm}
    \begin{subfigure}{0.49\columnwidth}
        \centering
        \includegraphics[width=1\columnwidth]{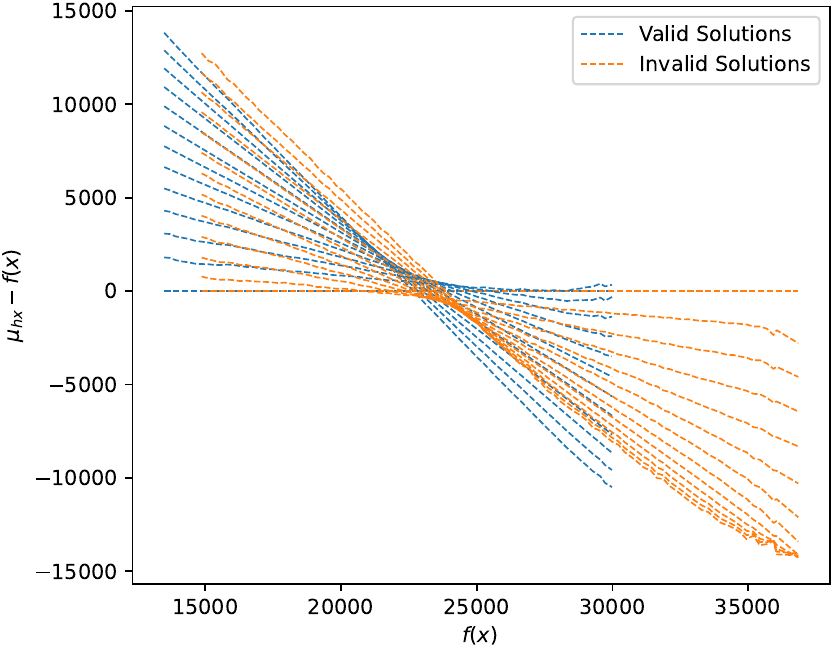}
        \caption{}
    \end{subfigure}
    \caption{An analysis of subset means $\mu_{h\bm{x}}$ for the fixed penalty, $\bm{\lambda}_F = (1,1,0)$. (a) All solutions. Shading shows variations in the values of $\mu_{h\bm{x}}$ (plus or minus a single standard deviation). (b) Solutions separated by validity.}
    \label{fig:CFLP_fixed_pen_subset_means}
\end{figure}

Intuitively, $\lambda_1$ controls a variable penalty, which takes into account the amount by which the capacity of a facility has been exceeded and $\lambda_2$ controls a fixed penalty which applies when the capacity of a facility is exceeded. In order for invalid solutions to be sufficiently penalised, the fixed term covers the average cost to open new facilities, and the variable term covers the average transportation cost per unit resource, ensuring that excessive resources assigned to a single facility are penalised at least as much as any cost savings associated with shorter customer-facility distances. \cref{fig:CFLP_with_fixed_penalty} shows the distribution of penalised objective function values $g(\bm{x})_{\bm{\lambda}}$ with $\bm{\lambda}=(1,1)$, from which it is clear that the invalid solutions are sufficiently penalised. However, penalising the invalid solutions has also introduced a clear bimodality in the distribution. This bimodality disrupts the necessary subset means relationship (\cref{eq:mixer_requirement_2}), as demonstrated in \cref{fig:CFLP_fixed_pen_subset_means}. 

\begin{figure}[htbp]
    \centering
    \captionsetup{margin=1cm,font=small}
    \begin{subfigure}{0.49\columnwidth}
        \centering
        \includegraphics[width=1\columnwidth]{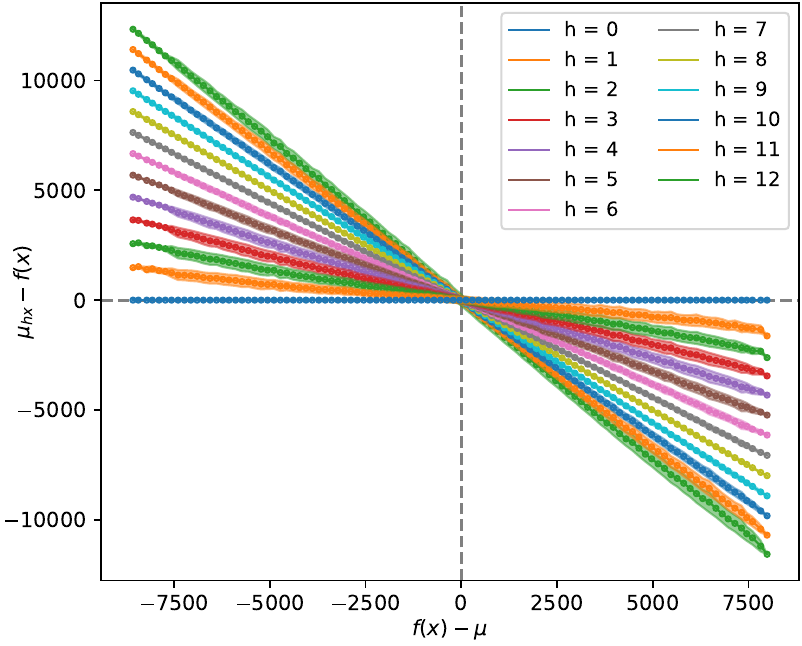}
        \caption{}
    \end{subfigure}
    \vspace{-0.25cm}
    \begin{subfigure}{0.49\columnwidth}
        \centering
        \includegraphics[width=1\columnwidth]{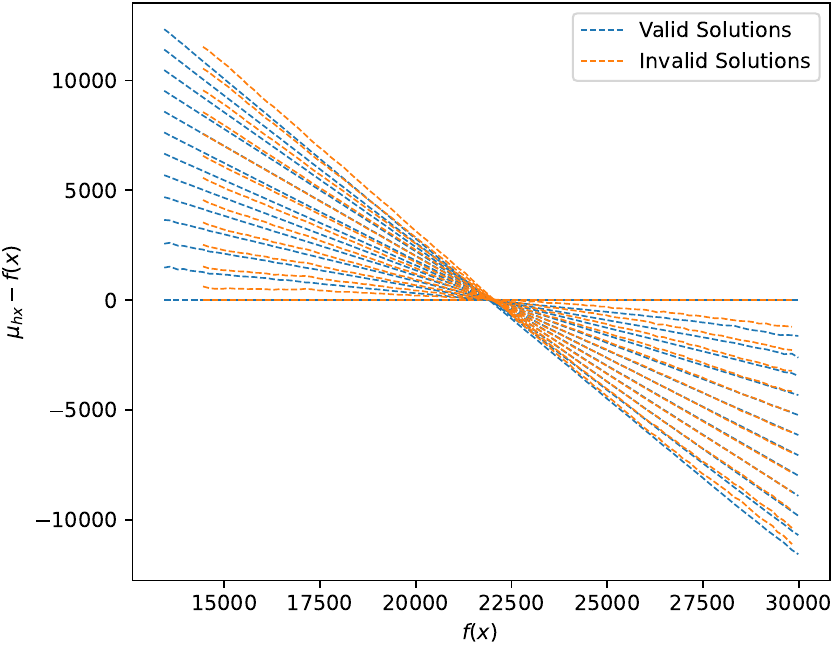}
        \caption{}
    \end{subfigure}
    \caption{An analysis of subset means $\mu_{h\bm{x}}$ for the tuned penalty, ${\bm{\lambda}_T=(0.8966,0.4996,0.1732)}$. (a) All solutions. Shading shows variations in the values of $\mu_{h\bm{x}}$ (plus or minus a single standard deviation). (b) Solutions separated by validity.}
    \label{fig:CFLP_tuned_pen_subset_means}
\end{figure}

\newpage
To correct for this bimodality, we introduce a third penalty term, arriving at the final penalised objective function,
\begin{equation}
    \fl \hspace{1.5cm} f(\bm{x})_{\bm{\lambda}} = g(x,\lambda_1,\lambda_2) - \cases{0,\hspace{0.5cm} \mbox{if $\bm{x}$ is valid} \\
        \lambda_3 \left[ g(x,\lambda_1,\lambda_2) - g(y,\lambda_1,\lambda_2) \right], \hspace{0.5cm} \mbox{otherwise} \\}
\end{equation}
where $\bm{y}$ is the solution (or at least an approximate solution) to the unconstrained variant. \cref{fig:CFLP_with_tuned_penalty} shows that after tuning, the addition of the third penalty term enables correction of the bimodality, and \cref{fig:CFLP_tuned_pen_subset_means} demonstrates how this improves adherence to the necessary condition in \cref{eq:mixer_requirement_2}.

\begin{figure}[htbp]
    \centering
    \captionsetup{margin=1cm,font=small}
    \begin{subfigure}{0.49\columnwidth}
        \centering
        \includegraphics[width=1\columnwidth]{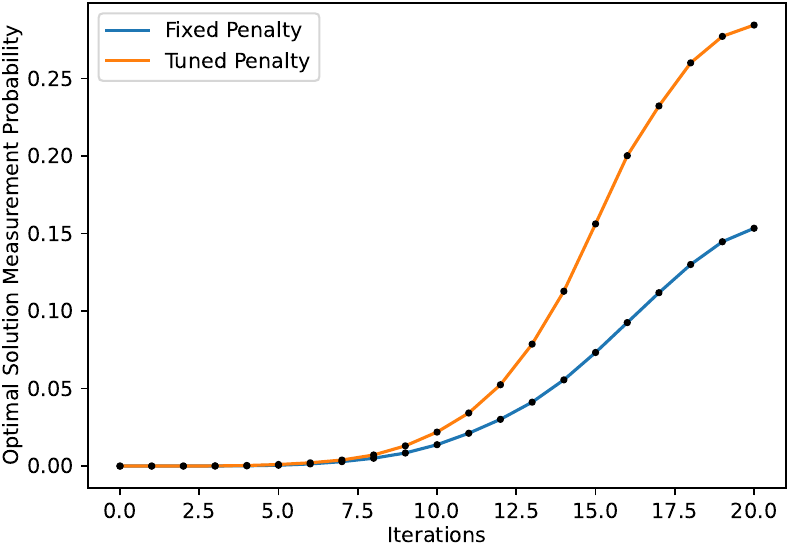}
        \caption{}
    \end{subfigure}
    \vspace{-0.25cm}
    \begin{subfigure}{0.49\columnwidth}
        \centering
        \includegraphics[width=1\columnwidth]{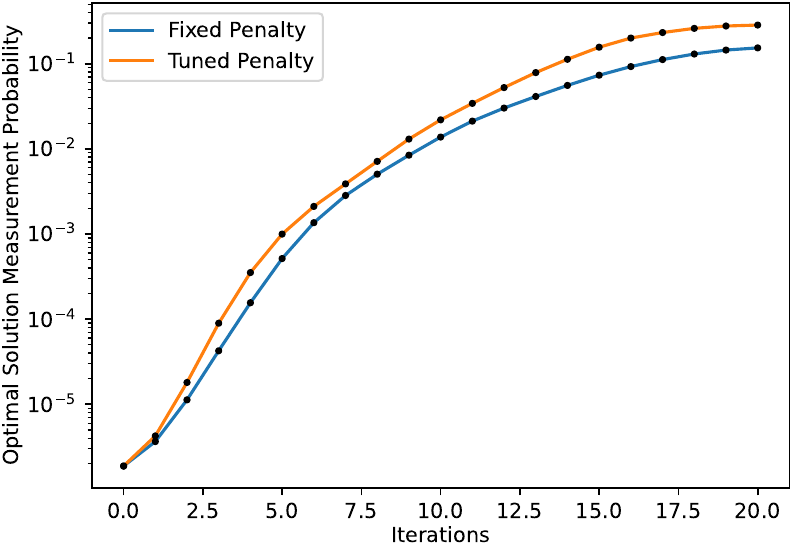}
        \caption{}
    \end{subfigure}
    \caption{Simulation results for $p=20$ non-variational QWOA applied to the CFLP for both the fixed and tuned penalty. Optimal solution measurement probability at each iteration (a) linear scale and (b) logarithmic scale.}
    \label{fig:CFLP_optimal_probs}
\end{figure}

\newpage
Selecting $p=20$ iterations, we simulate the non-variational QWOA applied to this randomly generated CFLP. As with the maximum independent set problem, we begin by foregoing the penalty tuning process and simply use the fixed penalty ${\bm{\lambda}_F = (1,1,0)}$ for the objective function. Beginning with $\gamma=1$, $t=0.1$ and $\beta=\frac{1}{p}$, gradient descent terminates with the amplified state $\ket{\gamma,t,\beta}$ given by $\gamma=2.0697$, $t=0.2185$ and $\beta=0.0780$. If instead we carry out the penalty tuning process, including initial parameters ${\bm{\lambda}_T=\bm{\lambda}_F=(1,1,0)}$, the amplified state is achieved with $\gamma=2.5732$, $t=0.2756$, $\beta=0.0593$ and ${\bm{\lambda}_T=(0.8966,0.4996,0.1732)}$. As can be seen in \cref{fig:CFLP_optimal_probs} the tuned penalty performs significantly better, producing greater measurement probability for the globally optimal valid solution.  

\begin{figure}[htbp]
    \centering
    \captionsetup{margin=1cm, font=small}
    \begin{subfigure}{0.48\columnwidth}
        \centering
        \includegraphics[width=1\columnwidth]{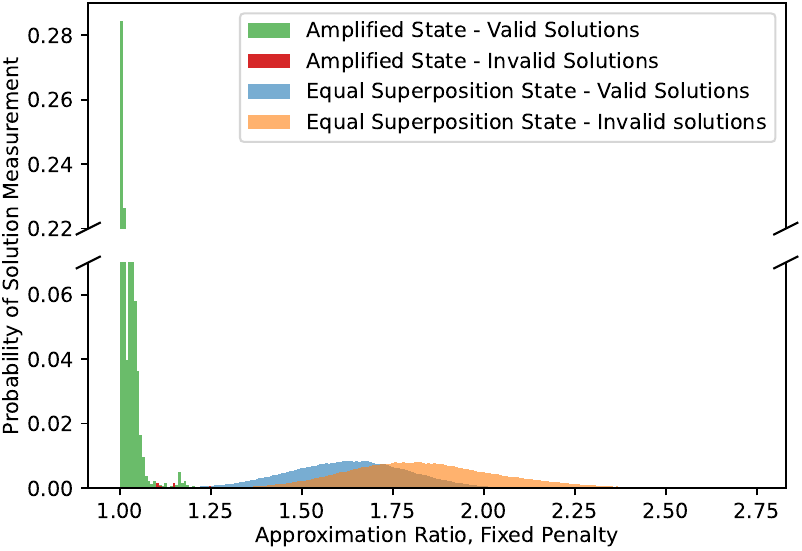}
        \caption{}
    \end{subfigure}
    \vspace{-0.25cm}
    \begin{subfigure}{0.48\columnwidth}
        \centering
        \includegraphics[width=1\columnwidth]{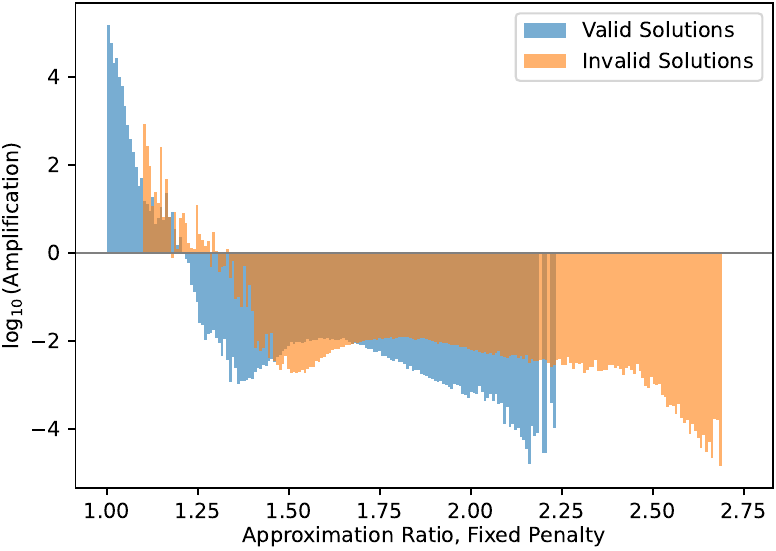}
        \caption{}
    \end{subfigure}
    \caption{Simulation results for $p=20$ non-variational QWOA applied to the CFLP. (a) Probability distributions for fixed penalty objective function values $f(\bm{x})_{\bm{\lambda}_F}$ as measured from the initial equal superposition state $\ket{s}$ and the amplified state $\ket{\gamma,t,\beta}_{\bm{\lambda}_T}$. (b) Logarithmic scale solution amplification as a function of $f(\bm{x})_{\bm{\lambda}_F}$.}
    \label{fig:CFLP_distributions}
\end{figure}

\begin{figure}[htbp]
    \centering
    \captionsetup{margin=1cm, font=small}
    \includegraphics[width=0.64\columnwidth]{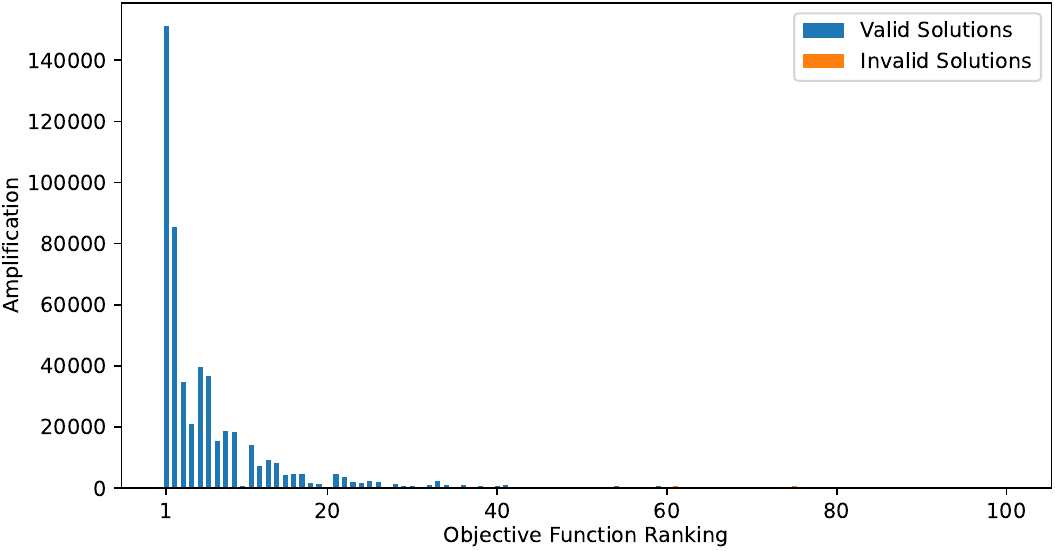}
    \caption{Amplification of the best 100 solutions to the CFLP within the $p=20$ amplified state $\ket{\gamma,t,\beta}_{\bm{\lambda}_T}$. Solutions are ranked according to $f(\bm{x})_{\bm{\lambda}_T}$.}
    \label{fig:CFLP_top100}
\end{figure}

More detailed simulation results (with the tuned penalty) are shown in \cref{fig:CFLP_distributions} and \cref{fig:CFLP_top100}, indicating that the globally optimal solution is amplified significantly more than all other solutions, and that the amplification process has focused probability amplitude preferentially into optimal and near-optimal valid solutions. The non-variational QWOA therefore appears to be capable of solving at least in some instances, more complicated constrained combinatorial optimisation problems, via careful design of the penalty function. 

\section{Conclusions}

This paper explores in depth a quantum algorithm for combinatorial optimisation, which is motivated from first principles. We design the mixing unitary as a continuous-time quantum walk over a mixing graph for which distance acts as a measure of solution similarity. As the mixing unitary distributes probability amplitude from one solution state to another, it imparts a phase-shift proportional to the distance between respective vertices on the mixing graph. Analysing the statistics of objective function values as distributed over the mixing graph, reveals how the nett effect of individual phase-shifts produced by the phase-separation unitary is able to offset the distance dependent phase shifts induced by the mixing unitary, resulting in controllable interference effects.

Insights into the interference process, produced by the alternating application of these two unitaries, enable intelligent control of mixing times and phase-shift magnitudes, to produce an essentially non-variational algorithm. For an arbitrary number of iterations, preparation of the amplified state is fully determined by three user-defined parameters, $\gamma$, $t$ and $\beta$, combined with the standard deviation of objective function values (efficiently approximated via random sampling). Due to the reliable interference process at the heart of the algorithm, a wide range of values for $\gamma$, $t$ and $\beta$ produce significant amplification of optimal and near-optimal solutions. In addition, optimal parameters can be found via a fixed complexity 3-dimensional local optimisation. 

Appropriate mixing graphs can be designed for a variety of significant combinatorial problems with very different structures, as illustrated in the current paper. For example, we provide efficiently implementable mixing unitaries for problems with binary variables, non-binary integer variables and for problems whose solutions are characterised by permutations. Ideally an appropriately designed mixing graph connects only valid solutions to a constrained problem. Where this is not possible, a carefully designed and tunable penalty function is used.

Numerical simulations show that the non-variational QWOA solves random problem instances for several problems, where various problem structures, as well as both unconstrained and constrained problems, are considered. A key component of future work will involve large scale and detailed benchmarking (via both simulations and physical implementation) in order to demonstrate, or at least provide strong evidence for, a quantum advantage relative to classical heuristics (e.g. local search methods). It may also be possible to quantify average case performance, in which case we suspect the algorithm may be capable of finding globally optimal solutions to some classically intractable (NP-hard) problems, more frequently than classical heuristics, with a number of iterations which is polynomial in problem size.

Future work is also likely to involve the application of this algorithm to more complicated and practically relevant constrained problems such as the capacitated vehicle routing and multiple traveling salesman problems. We also suggest that a study of worst case performance would be beneficial. Within the theoretical framework we have provided, it may be possible to identify when and why poor performance would occur, and it may even be possible to correct for this in order to improve the algorithm's robustness.

\appendix
\section{Establishing a useful approximation}
\label{sec:deriving_the_approximation}

Consider the sum of approximately coherent (in-phase) complex numbers,
\begin{equation}
    \sum_m r_m e^{\text{i} \phi_m}.
\end{equation}
The goal is to find an approximation for the resultant complex number, and in particular, we would like to be able to estimate the resultant's phase. We begin by making a number of provisional assumptions: the $\phi_m$ are normally distributed ($\phi_m \sim N(\mu,\sigma)$), the $r_m$ values are all equal, and there are a large number of terms in the sum. We also make the substitution, $\phi_m = \mu + z_m \sigma$, where now $z_m \sim N(0,1)$. Given these assumptions, it is possible to make an accurate approximation by integrating over a probability distribution for $z_m$ which is distributed as per the standard normal distribution,
\begin{equation}
    \fl \hspace{2cm} \sum_m r_m e^{\text{i} \phi_m}  = e^{\text{i}\mu} \sum_m r_m e^{\text{i} \sigma z_m} \approx \left( \sum_m r_m \right) e^{\text{i}\mu} \int_{-\infty}^{\infty} e^{\text{i} \sigma z_m} \frac{1}{\sqrt{2 \pi}} e^{-\frac{1}{2} z_m^2} \,d z_m.
\end{equation}
There are a number of ways to evaluate this integral, one of these is recognising its form as that of the Fourier transform for a Gaussian function and making use of the established result, 
\begin{equation}
    \frac{1}{\sqrt{2\pi}} \int_{-\infty}^{\infty} e^{-\alpha t^2} e^{-\text{i} \omega t}  \,d t = \frac{1}{\sqrt{2\alpha}}e^{\frac{-\omega^2}{4\alpha}}
\end{equation}
from which we can evaluate the approximation by substituting $\omega = -\sigma$ and $\alpha=\frac{1}{2}$,
\begin{equation}
    \sum_m r_m e^{\text{i} \phi_m} \approx \left( \sum_m r_m \right) e^{\frac{-\sigma^2}{2}} e^{\text{i} \mu}.
\end{equation}

While this approximation is accurate under the assumptions, we must generalise to cases where $r_m$ are not equal, and where $\phi_m$ need not be normally distributed. Since we suggest that normally distributed $\phi_m$ is not an entirely unreasonable assumption, rather than abandoning the above approximation altogether, suppose we trial a function of similar form, substituting the weighted mean of $\phi_m$ for $\mu$, defined as, 
\begin{equation}
    E = \frac{\sum_m r_m \phi_m}{\sum_m r_m} = \sum_m p_m \phi_m, \hspace{1cm} p_m = \frac{r_m}{\sum_m r_m},
\end{equation}
and the weighted variance of $\phi_m$ for $\sigma^2$, defined as,
\begin{equation}
    V = \frac{\sum_m r_m \left( \phi_m - E \right)^2}{\sum_m r_m} = \sum_m p_m \delta_m^2, \hspace{1cm}  \delta_m = \phi_m - E,
\end{equation}
such that the proposed estimation is given by,
\begin{equation}
    \label{eq:approximation_appendix}
    \sum_m r_m e^{\text{i} \phi_m} \approx \left( \sum_m r_m \right) e^{-\frac{V}{2}} e^{\text{i} E}.
\end{equation}

In order to demonstrate the approximation's effectiveness under the relaxed assumptions, we'll show that both sides of the expression are equal when substituting Taylor series expansions around residuals $\delta_m=0$, discarding terms of third order and above. Beginning with the left hand side, and setting the weighted mean as a global phase, each term is now expressed with a relative phase equal to the residual,
\begin{equation}
    \sum_m r_m e^{\text{i} \phi_m} = e^{\text{i} E} \sum_m r_m e^{\text{i} \delta_m} = \left( \sum_m r_m \right) e^{\text{i} E} \sum_m p_m e^{\text{i} \delta_m}.
\end{equation}
Substituting the Taylor series expansion of $e^{\text{i} \delta_m}$ around $\delta_m=0$ gives the following approximation (discarding terms 3rd order and above).
\begin{equation}
    \fl \hspace{1cm} \sum_m r_m e^{\text{i} \phi_m} \approx \left( \sum_m r_m \right) e^{\text{i} E} \left[  \left( \sum_m p_m \right) + \text{i} \left( \sum_m p_m \delta_m \right) - \frac{1}{2} \left( \sum_m p_m \delta_m^2 \right)\right].
\end{equation}
By definition, $\sum_m p_m=1$ and $\sum_m p_m \delta_m=0$, so the expression simplifies as follows,
\begin{equation}
    \sum_m r_m e^{\text{i} \phi_m} \approx \left( \sum_m r_m \right) e^{\text{i} E} \left[ 1 - \frac{1}{2} \left( \sum_m p_m \delta_m^2 \right)\right].
\end{equation}
To show that the right hand side is equivalent, we just need to analyse the exponential of $\frac{-V}{2}$,
\begin{equation}
    e^{-\frac{V}{2}} = e^{-\frac{\sum_m p_m \delta_m^2}{2}} = \prod_m e^{-\frac{p_m \delta_m^2}{2}}
\end{equation}
Substituting the Taylor series expansion around $\delta_m=0$ and discarding terms of fourth order and above, the expression simplifies (as required),
\begin{equation}
    e^{-\frac{V}{2}} = \prod_m e^{-\frac{p_m \delta_m^2}{2}} \approx \prod_m \left( 1 - \frac{p_m \delta_m^2}{2} \right) \approx 1 - \frac{1}{2} \left( \sum_m p_m \delta_m^2 \right).
\end{equation}

Given the equivalence under Taylor series expansion, we take \cref{eq:approximation_appendix} as the final approximation for the sum of approximately coherent complex terms. Each term in the expression can be understood intuitively, $e^{\text{i} E}$ corresponds with the fact that the resultant has a phase centred within that of the individual contributions, and $e^{\frac{-V}{2}}$ corresponds with the tendency of the resultant to shrink as the phases of individual contributions become more dispersed. The result is significant, as it allows us to analyse the sum of an arbitrarily large number of complex terms, simply by studying their statistics, given of course that the terms are approximately in-phase. For the purposes of this study, we explicitly assume the relevant complex terms remain approximately in-phase.

\begin{figure}[htbp]
    \centering
    \captionsetup{margin=1cm, font=small}
    \begin{subfigure}{0.48\columnwidth}
        \centering
        \includegraphics[width=1\columnwidth]{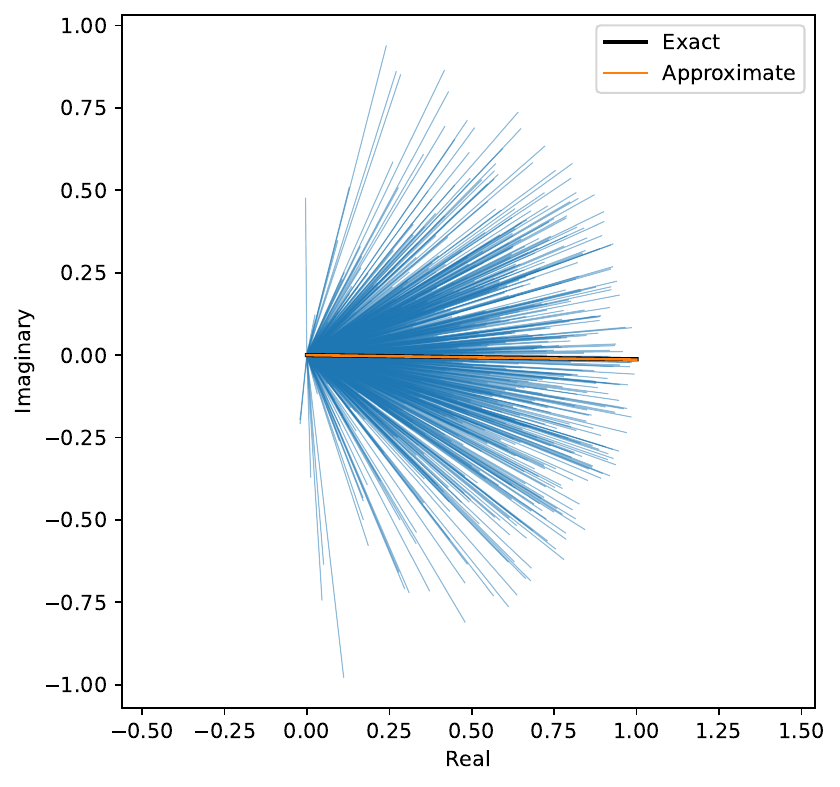}
    \end{subfigure}
    \begin{subfigure}{0.47\columnwidth}
        \centering
        \includegraphics[width=1\columnwidth]{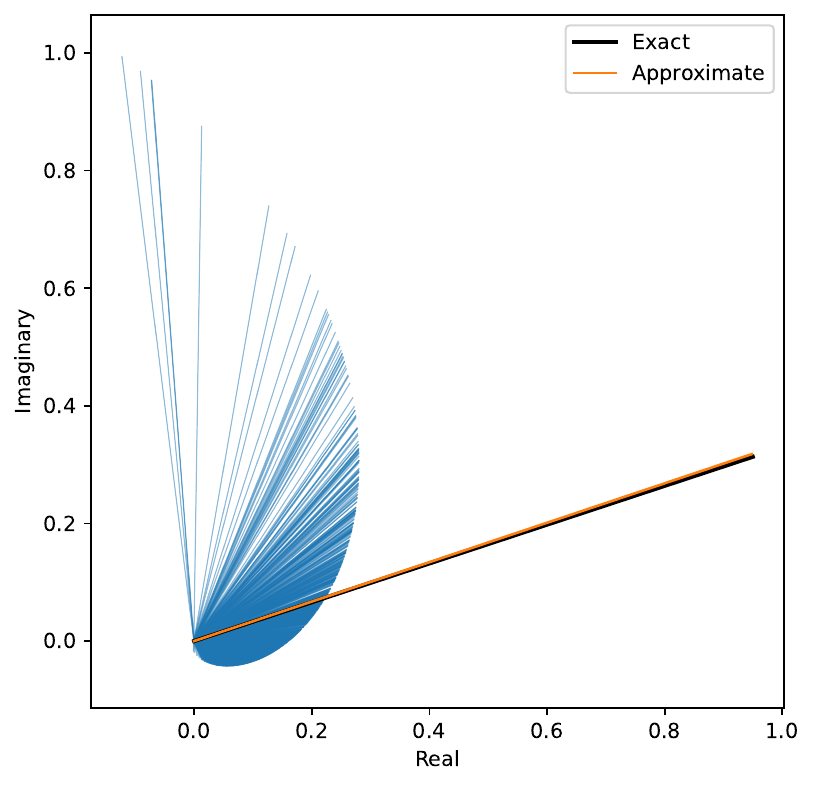}
    \end{subfigure} \\
    \begin{subfigure}{0.48\columnwidth}
        \centering
        \includegraphics[width=1\columnwidth]{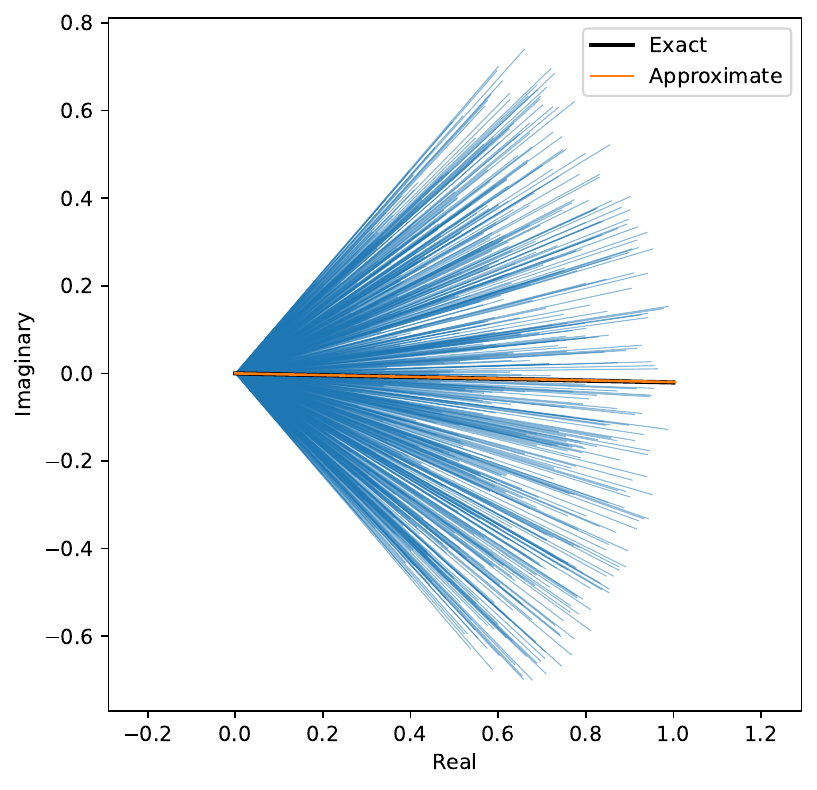}
    \end{subfigure}
    \begin{subfigure}{0.47\columnwidth}
        \centering
        \includegraphics[width=1\columnwidth]{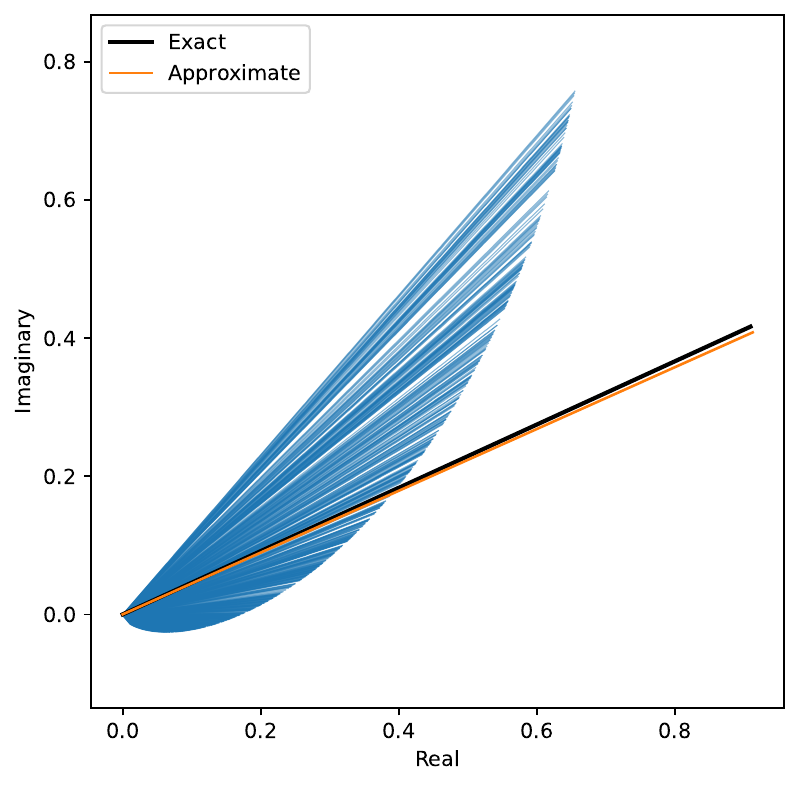}
    \end{subfigure}
    \caption{Vector sum depictions of the sum of complex numbers as per \cref{eq:approximation_appendix}. The resultant is shown with a black line, and the approximated resultant with an orange line. The individually contributing terms are shown with blue lines. The resultant is shown with length 1 for clarity. The top row shows normally distributed $\phi_m$ and the bottom row uniformly distributed $\phi_m$. The left column shows uniformly distributed $r_m$ and the right column shows $r_m$ increasing with $\phi_m$.}
    \label{fig:vector_sums}
\end{figure}

In order to verify that the approximation is appropriately accurate for the purposes of this analysis, we study its dependence on coherence and on the nature of the distributions in $r_m$ and $\phi_m$. In order to do so, we randomly generate distributions for both $\phi_m$ and $r_m$. In every instance, $1000$ terms are summed. Four different cases are considered from the combination of the following: The phases $\phi_m$ are drawn from either a normal distribution, or from a uniform distribution, in either case with a specified standard deviation. The weights $r_m$ are either drawn from a uniform distribution over the interval $\left[0,1\right)$ or are given by a monotonically increasing function of $\phi_m$. Examples of each of these cases are shown in \cref{fig:vector_sums}, where in all cases the standard deviation is $0.5$ radians.
\newpage

\begin{figure}[htbp]
    \centering
    \captionsetup{margin=1cm, font=small}
    \begin{subfigure}{0.49\columnwidth}
        \centering
        \includegraphics[width=1\columnwidth]{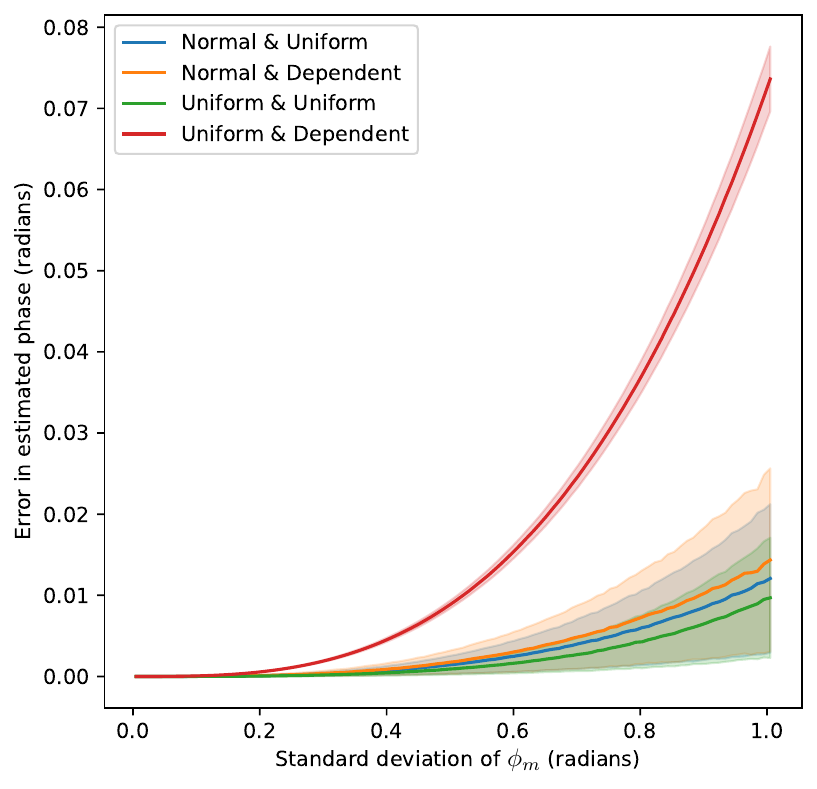}
        \caption{}
    \end{subfigure}
    \begin{subfigure}{0.49\columnwidth}
        \centering
        \includegraphics[width=1\columnwidth]{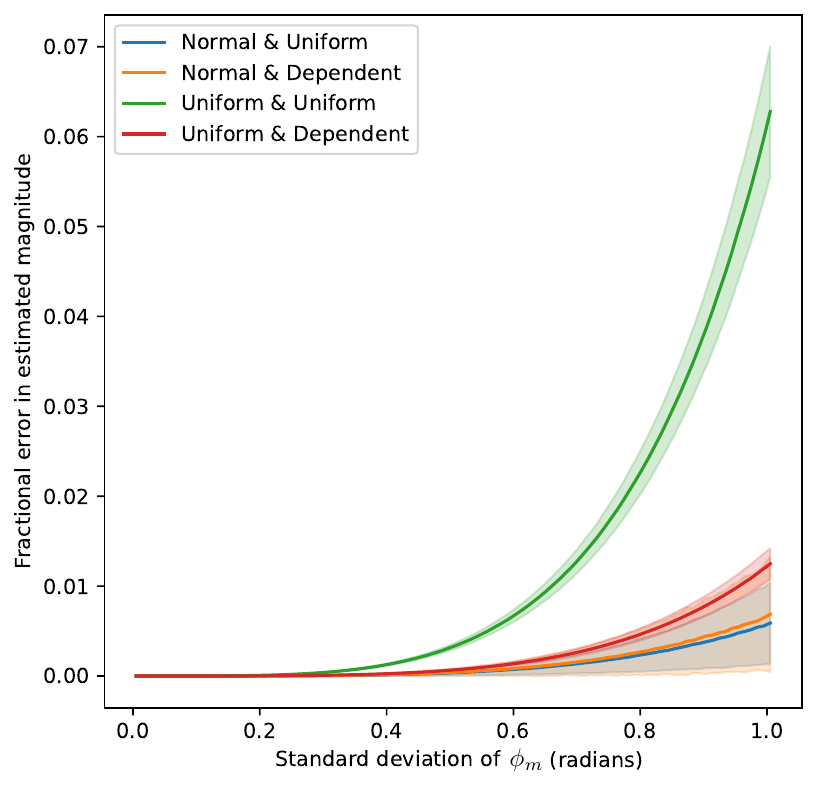}
        \caption{}
    \end{subfigure}
    \vspace{-0.25cm}
    \caption{These figures show the scaling in error associated with the approximation from \cref{eq:approximation_appendix} as a function of increasing standard deviation of $\phi_m$. (a) error in phase and (b) error in magnitude. Shaded regions show plus/minus a single standard deviation.}
    \label{fig:error_scaling}
\end{figure}

\newpage
Despite the fact that these complex terms do not appear to be approximately coherent, but rather their phases vary substantially, the approximated resultant appears to be very accurate, particularly for the normally distributed $\phi_m$. \cref{fig:error_scaling} shows the result of one million such instances for each of these four cases, with standard deviations of $\phi_m$ distributed between $0$ and $1$ radians (accounting for a wide range of coherence in complex phase). These plots show that the estimation remains quite accurate, even moving beyond the regime within which the phases could be reasonable described as coherent.

\newpage
\section{Random weighted graph for maxcut}
\label{sec:maxcut_graph}
Generating the following random weighted graph begins by generating a random, undirected graph, via the Erdos-Renyi model, in which each possible edge is included with some probability (in this case $\text{probability}=0.5$). After this, each edge is assigned a weight from the uniform distribution $(0,1]$.

The randomly generated maxcut graph ($n=18$) is defined by the following edge weights, where $w(u,v)$ denotes the weight of an edge connecting vertex $\bm{u}$ and $\bm{v}$:

\begin{center}
\begin{tabular}{ l l l l }
${w(0,1) = 0.989287}$, & ${w(0,2) = 0.032138}$, & ${w(0,3) = 0.351590}$, & ${w(0,6) = 0.036441}$, \\
${w(0,7) = 0.584669}$, & ${w(0,9) = 0.731521}$, & ${w(0,11) = 0.141659}$, & ${w(0,12) = 0.371674}$, \\ ${w(0,14) = 0.387308}$, & ${w(0,16) = 0.360192}$, & ${w(1,2) = 0.690304}$, & ${w(1,3) = 0.974171}$, \\
${w(1,5) = 0.909797}$, & ${w(1,6) = 0.424171}$, & ${w(1,8) = 0.580351}$, & ${w(1,14) = 0.121489}$, \\
${w(1,16) = 0.205978}$, & ${w(1,17) = 0.086657}$, & ${w(2,3) = 0.017430}$, & ${w(2,7) = 0.790093}$, \\
${w(2,8) = 0.839485}$, & ${w(2,9) = 0.296687}$, & ${w(2,11) = 0.379847}$, & ${w(2,12) = 0.338815}$, \\
${w(2,13) = 0.285064}$, & ${w(2,15) = 0.130135}$, & ${w(3,10) = 0.009759}$, & ${w(3,12) = 0.963173}$, \\
${w(3,13) = 0.099826}$, & ${w(3,15) = 0.714542}$, & ${w(4,6) = 0.985471}$, & ${w(4,8) = 0.808318}$, \\
${w(4,9) = 0.685892}$, & ${w(4,12) = 0.173569}$, & ${w(4,17) = 0.571766}$, & ${w(5,6) = 0.242078}$, \\
${w(5,7) = 0.183907}$, & ${w(5,11) = 0.137946}$, & ${w(5,12) = 0.757602}$, & ${w(5,14) = 0.385144}$, \\
${w(5,15) = 0.519802}$, & ${w(5,16) = 0.693374}$, & ${w(6,7) = 0.670602}$, & ${w(6,9) = 0.696379}$, \\
${w(6,10) = 0.111345}$, & ${w(6,12) = 0.514704}$, & ${w(6,14) = 0.854230}$, & ${w(7,8) = 0.599075}$, \\
${w(7,9) = 0.825837}$, & ${w(7,10) = 0.617606}$, & ${w(7,13) = 0.605946}$, & ${w(7,16) = 0.983943}$, \\
${w(7,17) = 0.085275}$, & ${w(8,11) = 0.349141}$, & ${w(8,13) = 0.268013}$, & ${w(8,16) = 0.142334}$, \\
${w(8,17) = 0.731623}$, & ${w(9,13) = 0.808777}$, & ${w(9,14) = 0.205945}$, & ${w(9,16) = 0.554085}$, \\
${w(9,17) = 0.304275}$, & ${w(10,11) = 0.974365}$, & ${w(10,13) = 0.393311}$, & ${w(10,16) = 0.274597}$, \\
${w(11,12) = 0.586437}$, & ${w(11,13) = 0.928648}$, & ${w(11,17) = 0.512909}$, & ${w(12,13) = 0.795810}$, \\
${w(12,14) = 0.581403}$, & ${w(12,16) = 0.041086}$, & ${w(12,17) = 0.759730}$, & ${w(13,14) = 0.336961}$, \\
${w(13,15) = 0.198718}$, & ${w(13,17) = 0.191082}$, & ${w(14,15) = 0.186035}$, & ${w(14,17) = 0.761156}$. \\
\end{tabular}
\end{center}

\newpage
\section{Constructing $U_k$}
\label{alg:Uk}

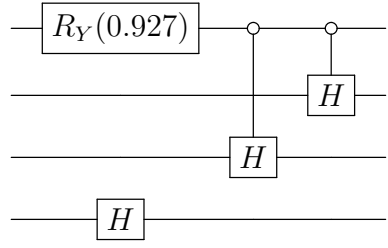
\begin{figure}[htbp]
    \centering
    \[ \hspace{1.5cm} \Qcircuit @C=1em  @R=0.7em  {
        & \gate{R_Y(0.927)} & \ctrlo{2}  & \ctrlo{1}  & \qw \\
        &       \qw         &    \qw     &  \gate{H}  & \qw \\
        &       \qw         &  \gate{H}  &     \qw    & \qw \\
        &     \gate{H}      &    \qw     &     \qw    & \qw \\
    }\]
    \captionsetup{margin=1cm, font=small}
    \caption{Binary encoding quantum circuit implementation of the unitary $U_k$ for $k=10$, prepared as per \cref{alg:Uk}.}
    \label{fig:U_k_binary}
\end{figure}

\emph{Binary encoding:} This algorithm/pseudocode constructs a quantum circuit implementation of $U_k$ such that
\begin{equation}
    U_k \ket{0}^{\otimes m} = \frac{1}{\sqrt{k}} \sum_{x=0}^{k-1} \ket{\bm{x}}.
\end{equation}
The circuit reduces to the trivial case of $H^{\otimes m}$ when $k=2^m$. The method can be derived using a binary tree which distributes probability, as required, between the $\ket{0}$ and $\ket{1}$ states for each qubit. An example circuit for $k=10$ is illustrated in \cref{fig:U_k_binary}.

\begin{algorithm}[htbp]
    \caption{Constructing $U_k$ (binary)}
    \setstretch{1.1}   
    \begin{algorithmic}
        \State $m \gets \lceil \text{log}_2(k) \rceil$
        \State $r \gets m - 1$
        \State Label qubits $q_0, q_1, ..., q_r$ where $q_0$ is the least significant qubit
        \State Apply $\text{R}_{\text{y}}\left(2 \arccos{\sqrt{\frac{2^r}{k}}}\right)$ to $q_r$
        \State $R \gets k - 2^r$
        \State $j \gets \text{log}_2(R)$
        \While {$j \text{ mod } 1 \neq 0$}
            \State $j \gets \lfloor j \rfloor$
            \State Apply $\text{R}_{\text{y}}\left(2 \arccos{\sqrt{\frac{2^j}{R}}}\right)$ to $q_j$ controlled on $q_r$    
            \State $R \gets R - 2^j$
            \State $r \gets j$
            \State $j \gets \text{log}_2(R)$
        \EndWhile
        \For{$\forall i: i<j$}
            \State Apply H to $q_i$
        \EndFor
        \While{$j \neq m - 1$}
            \State $r \gets j + 1$
            \While{$\lfloor \frac{k}{2^r} \rfloor \text{ mod } 2 \neq 1$}
                \State $r \gets r + 1$
            \EndWhile
            \For{$\forall i: j \leq i < r$}
                \State Apply H to $q_i$ controlled on zero at $q_r$
            \EndFor
            \State $j \gets r$
        \EndWhile
    \end{algorithmic}
\end{algorithm}

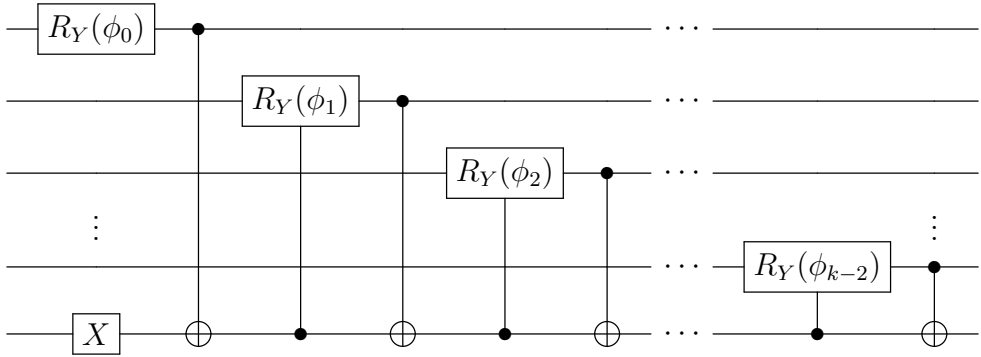
\begin{figure}[htbp]
    \centering
    \[ \fl \hspace{1.5cm} \Qcircuit @C=1em  @R=0.7em  {
        & \gate{R_Y(\phi_0)} & \ctrl{5}  & \qw & \qw & \qw & \qw & \qw & \cdots & & \qw & \qw & \qw \\
        & \qw & \qw  & \gate{R_Y(\phi_1)} & \ctrl{4} & \qw & \qw & \qw & \cdots & & \qw & \qw & \qw \\
        & \qw & \qw  & \qw & \qw & \gate{R_Y(\phi_2)} & \ctrl{3} & \qw & \cdots & & \qw & \qw & \qw \\
        & \colorbox{white}{\vdots} & & & & & & & & & & \colorbox{white}{\vdots} & & \\
        & \qw & \qw  & \qw & \qw & \qw & \qw & \qw & \cdots & & \gate{R_Y(\phi_{k-2})} & \ctrl{1} & \qw \\
        & \gate{X} & \targ & \ctrl{-4} & \targ & \ctrl{-3} & \targ & \qw & \cdots & & \ctrl{-1} & \targ & \qw \\
    }\]
    \captionsetup{margin=1cm, font=small}
    \caption{One-hot encoding quantum circuit implementation of the unitary $U_k$ for arbitrary $k$.}
    \label{fig:U_k_onehot}
\end{figure}

\noindent \emph{One-hot encoding:} \cref{fig:U_k_onehot} depicts a quantum circuit implementation of $U_k$ such that
\begin{equation}
    U_k \ket{0}^{\otimes k} = \frac{1}{\sqrt{k}} \sum_{x=0}^{k-1} \prod_{i=0}^{k-1} \ket{\delta_{i,x}}.
\end{equation}

\noindent The rotation angles $\phi_i$ are given by the expression,
\begin{equation}
    \phi_i = 2 \arcsin{\left(\frac{1}{\sqrt{k-i}}\right)}.
\end{equation}

\newpage
\section{Random data-points for k-means}
\label{sec:data-points}

Each value is randomly generated from the uniform distribution $(0,10]$.

\vspace{0.5cm} 
$\fl \hspace{1cm} \begin{array}{cccccccccc} [ 9.11002 & 0.06106 & 4.82484 & 6.19904 & 1.73007 & 9.69564 & 3.27416 & 9.04641 & 8.89987 & 3.90831 ] \end{array} $ 

\vspace{0.2cm} 
 $\fl \hspace{1cm} \begin{array}{cccccccccc} [ 0.67598 & 2.91592 & 3.81196 & 3.54562 & 4.70264 & 0.45163 & 6.26537 & 2.48306 & 7.78254 & 8.71130 ] \end{array} $ 

\vspace{0.2cm} 
 $\fl \hspace{1cm} \begin{array}{cccccccccc} [ 8.94023 & 7.73040 & 0.47080 & 4.68225 & 9.95584 & 9.96049 & 5.93501 & 7.75738 & 8.67906 & 1.48174 ] \end{array} $ 

\vspace{0.2cm} 
 $\fl \hspace{1cm} \begin{array}{cccccccccc} [ 4.25387 & 5.45516 & 7.78103 & 6.79578 & 5.39876 & 9.90842 & 0.81172 & 9.65141 & 1.23032 & 7.56302 ] \end{array} $ 

\vspace{0.2cm} 
 $\fl \hspace{1cm} \begin{array}{cccccccccc} [ 2.55319 & 5.53473 & 8.11386 & 5.84780 & 3.16515 & 5.31585 & 8.12552 & 3.41561 & 1.60965 & 5.27394 ] \end{array} $ 

\vspace{0.2cm} 
 $\fl \hspace{1cm} \begin{array}{cccccccccc} [ 3.24497 & 2.64679 & 2.25402 & 9.37613 & 3.91658 & 4.33125 & 2.92193 & 2.21017 & 1.68782 & 6.17135 ] \end{array} $ 

\vspace{0.2cm} 
 $\fl \hspace{1cm} \begin{array}{cccccccccc} [ 5.31082 & 5.54856 & 1.87199 & 9.74049 & 1.62758 & 9.21121 & 0.60003 & 6.19258 & 0.56318 & 4.51460 ] \end{array} $ 

\vspace{0.2cm} 
 $\fl \hspace{1cm} \begin{array}{cccccccccc} [ 6.63098 & 5.86293 & 6.14683 & 6.41343 & 9.15442 & 5.80235 & 9.20749 & 2.70775 & 3.74269 & 2.54521 ] \end{array} $ 

\vspace{0.2cm} 
 $\fl \hspace{1cm} \begin{array}{cccccccccc} [ 7.10254 & 4.16486 & 0.13647 & 3.91372 & 8.46266 & 6.47524 & 1.74747 & 8.57903 & 0.35589 & 9.84442 ] \end{array} $ 

\vspace{0.2cm} 
 $\fl \hspace{1cm} \begin{array}{cccccccccc} [ 4.41710 & 8.58625 & 2.09502 & 9.78096 & 4.32488 & 2.70198 & 4.54662 & 6.53021 & 2.63868 & 7.14099 ] \end{array} $ 

\vspace{0.2cm} 
 $\fl \hspace{1cm} \begin{array}{cccccccccc} [ 6.86481 & 2.01457 & 9.41148 & 3.74956 & 0.87813 & 0.85384 & 0.43303 & 0.94263 & 6.65593 & 2.07344 ] \end{array} $ 

\vspace{0.2cm} 
 $\fl \hspace{1cm} \begin{array}{cccccccccc} [ 3.84940 & 8.90650 & 8.47315 & 8.64937 & 8.17979 & 7.40941 & 3.91822 & 0.52376 & 8.66642 & 2.44087 ] \end{array} $ 

\newpage
\section{Preparing an equal superposition over permutations}
\label{sec:permutation_state_preparation}

Here we discuss an efficient method for preparing an equal superposition over all solution states $\bm{x}$ directly encoding permutations of length $n$, of which there are $N=n!$. The method has gate complexity $O(n^3)$. Importantly, the input register remains separable from any ancilla qubits, which is not always the case with other methods as in \cite{marsh2020combinatorial} and \cite{chiew2019graph}. If the input register were entangled with any ancilla register, the mixing unitary would not result in interference between probability amplitudes distributed between solution states.

The method is recursive in nature. Given the equal superposition of permutations of size $n-1$, the equal superposition of permutations of size $n$ can be efficiently prepared by introducing a new sub-register prepared in an equal superposition of the $n$ states encoding integer variable values ${x_j \in \{0,1,...,n-1\}}$. Each existing sub-register has an additional qubit appended in the $\ket{0}$ state, (only if necessary for the binary encoding). Each of the pre-existing sub-registers is then entangled with the new register by conditionally incrementing the value stored in the pre-existing sub-register. The value stored in the pre-existing sub-register is incremented only if it has an integer value greater than or equal to the integer value stored in the new sub-register.

\begin{figure}[htbp]
    \centering
    \[ \fl \Qcircuit @C=0.2em  @R=0.3em  {
             & \gate{X} & \qw & \targ & \qw & \qw & \ctrl{1} & \targ & \qw & \qw & \qw & \qw & \qw & \qw & \qw & \qw & \qw & \qw & \ctrl{1} & \targ & \qw & \qw & \qw & \qw & \qw & \qw & \qw & \qw & \qw & \qw & \qw & \qw & \qw & \qw & \qw & \qw & \qw \\
             & \qw & \targ & \ctrl{-1} & \ctrl{1} & \targ & \targ & \ctrl{-1} & \qw & \qw & \qw & \qw & \qw & \qw & \ctrl{1} & \targ & \ctrl{1} & \targ & \targ & \ctrl{-1} & \qw & \qw & \qw & \qw & \qw & \qw & \qw & \qw & \qw & \qw & \qw & \qw & \qw & \qw & \qw & \qw & \qw \\
             & \qw & \qw & \qw & \targ & \ctrl{-1} & \qw & \qw & \qw & \qw & \qw & \qw & \ctrl{1} & \targ & \targ & \ctrl{-1} & \targ & \ctrl{-1} & \qw & \qw & \qw & \qw & \qw & \qw & \qw & \qw & \qw & \qw & \qw & \qw & \qw & \qw & \qw & \qw & \qw & \qw & \qw \\
             & \qw & \qw & \qw & \qw & \qw & \qw & \qw & \qw & \qw & \qw & \qw & \targ & \ctrl{-1} & \qw & \qw & \qw & \qw & \qw & \qw & \qw & \qw & \qw & \qw & \qw & \qw & \qw & \qw & \qw & \qw & \qw & \qw & \qw & \qw & \qw & \qw & \qw \\
             & & & & & & & & & & & & & & & & & & & & & & & & & & & & & & & & & & & & & \\
             & \multigate{1}{U_k^{(k=2)}} & \ctrl{-4} & \qw & \qw & \qw & \qw & \qw & \qw & \qw & \ctrl{1} & \targ & \qw & \qw & \qw & \qw & \qw & \qw & \qw & \qw & \qw & \qw & \qw & \qw & \qw & \qw & \ctrl{1} & \targ & \qw & \qw & \qw & \qw & \qw & \qw & \qw & \qw & \qw \\
             & \ghost{U_k^{(k=2)}} & \qw  & \qw & \qw & \qw & \qw & \qw & \ctrl{1} & \targ & \targ & \ctrl{-1} & \qw & \qw & \qw & \qw & \qw & \qw & \qw & \qw & \qw & \qw & \ctrl{1} & \targ & \ctrl{1} & \targ & \targ & \ctrl{-1} & \qw & \qw & \qw & \qw & \qw & \qw & \qw & \qw & \qw \\
             & \qw & \qw & \qw & \qw & \qw & \qw & \qw & \targ & \ctrl{-1} & \qw & \qw & \qw & \qw & \qw & \qw & \qw & \qw & \qw & \qw & \ctrl{1} & \targ & \targ & \ctrl{-1} & \targ & \ctrl{-1} & \qw & \qw & \qw & \qw & \qw & \qw & \qw & \qw & \qw & \qw & \qw \\
             & \qw & \qw & \qw & \qw & \qw & \qw & \qw & \qw & \qw & \qw & \qw & \qw & \qw & \qw & \qw & \qw & \qw & \qw & \qw & \targ & \ctrl{-1} & \qw & \qw & \qw & \qw & \qw & \qw & \qw & \qw & \qw & \qw & \qw & \qw & \qw & \qw & \qw \\
             & & & & & & & & & & & & & & & & & & & & & & & & & & & & & & & & & & & & & \\
             & \multigate{2}{U_k^{(k=3)}} & \qw & \qw & \qw & \qw & \ctrl{-9} & \ctrl{-10} & \qw & \qw & \ctrl{-4} & \ctrl{-5} & \qw & \qw & \qw & \qw & \qw & \qw & \qw & \qw & \qw & \qw & \qw & \qw & \qw & \qw & \qw & \qw & \qw & \qw & \qw & \qw & \qw & \qw & \ctrl{1} & \targ & \qw \\
             & \ghost{U_k^{(k=3)}} & \qw & \qw & \qw & \qw & \qw & \qw   & \qw & \qw & \qw & \qw & \qw & \qw & \qw & \qw & \qw & \qw & \qw & \qw & \qw & \qw & \qw & \qw & \qw & \qw & \qw & \qw & \qw & \qw & \ctrl{1} & \targ & \ctrl{1} & \targ & \targ & \ctrl{-1} & \qw\\
             & \ghost{U_k^{(k=3)}} & \qw & \qw & \ctrlo{-10} & \qw & \qw & \qw  & \ctrlo{-5} & \qw & \qw & \qw & \qw & \qw & \qw & \qw & \qw & \qw & \qw & \qw & \qw & \qw & \qw & \qw & \qw & \qw & \qw & \qw & \ctrl{1} & \targ & \targ & \ctrl{-1} & \targ & \ctrl{-1} & \qw & \qw & \qw\\
             & \qw & \qw & \qw & \qw & \qw & \qw & \qw  & \qw & \qw & \qw & \qw & \qw & \qw & \qw & \qw & \qw & \qw & \qw & \qw & \qw & \qw & \qw & \qw & \qw & \qw & \qw & \qw & \targ & \ctrl{-1} & \qw & \qw & \qw & \qw & \qw & \qw & \qw\\
             & & & & & & & & & & & & & & & & & & & & & & & & & & & & & & & & & & & & &\\
             & \multigate{3}{U_k^{(k=4)}} & \qw & \qw & \qw & \qw & \qw & \qw & \qw & \qw & \qw & \qw & \qw & \qw & \qw & \qw & \ctrl{-13} & \ctrl{-14} & \ctrl{-14} & \ctrl{-15} & \qw & \qw & \qw & \qw & \ctrl{-8} & \ctrl{-9} & \ctrl{-9} & \ctrl{-10} & \qw & \qw & \qw & \qw & \ctrl{-3} & \ctrl{-4} & \ctrl{-4} & \ctrl{-5} & \qw\\
             & \ghost{U_k^{(k=4)}} & \qw & \qw & \qw & \qw & \qw & \qw & \qw & \qw & \qw & \qw & \qw & \qw & \ctrl{-14} & \ctrl{-15} & \qw & \qw & \qw & \qw & \qw & \qw & \ctrl{-9} & \ctrl{-10} & \qw & \qw & \qw & \qw & \qw & \qw & \ctrl{-4} & \ctrl{-5} & \qw & \qw & \qw & \qw & \qw\\
             & \ghost{U_k^{(k=4)}} & \qw & \qw & \qw & \qw & \qw & \qw & \qw & \qw & \qw & \qw & \qw & \qw & \qw & \qw & \qw & \qw & \qw & \qw & \qw & \qw & \qw & \qw & \qw & \qw & \qw & \qw & \qw & \qw & \qw & \qw & \qw & \qw & \qw & \qw & \qw\\
             & \ghost{U_k^{(k=4)}} & \qw & \qw & \qw & \qw & \qw & \qw & \qw & \qw & \qw & \qw & \ctrlo{-15} & \qw & \qw & \qw & \qw & \qw & \qw & \qw & \ctrlo{-10} & \qw & \qw & \qw & \qw & \qw & \qw & \qw & \ctrlo{-5} & \qw & \qw & \qw & \qw & \qw & \qw & \qw & \qw\\     
    }\]
    \captionsetup{margin=1cm, font=small}
    \caption{Example one-hot encoding quantum circuit implementation of the unitary $U_s$ for preparing an equal superposition over all solution states $\bm{x}$ directly encoding permutations of length $n=4$.}
    \label{fig:U_perm}
\end{figure}
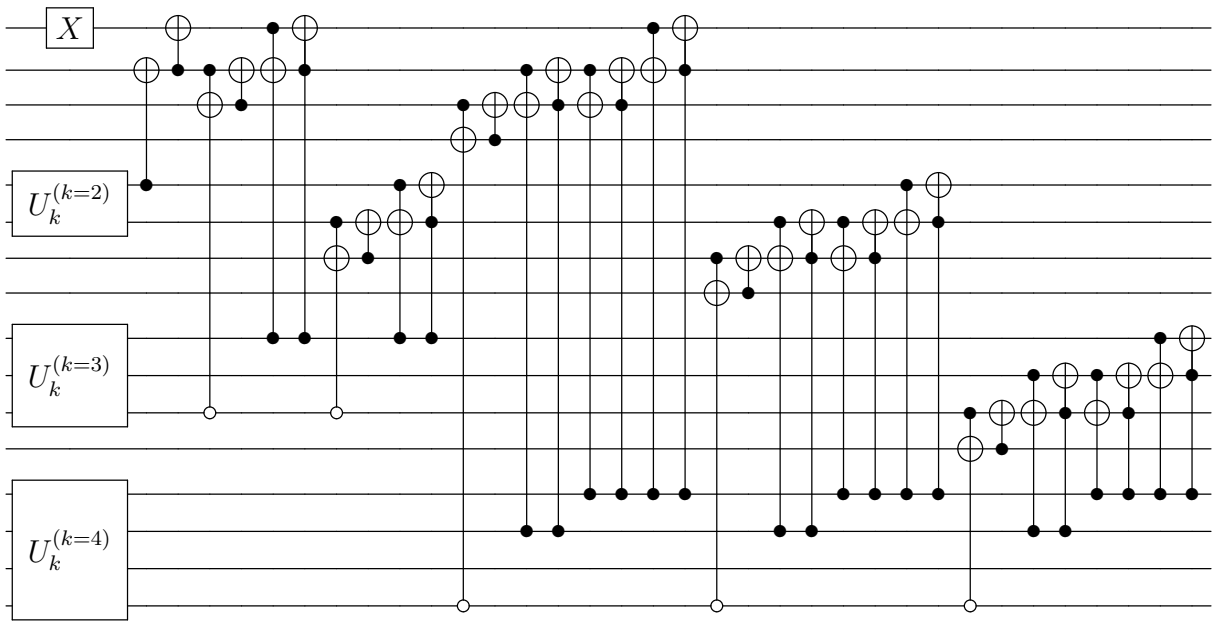

An example circuit is included in \cref{fig:U_perm} specifically for the case where $n=4$ and for a one-hot encoding. A different method, applicable to the one-hot encoding, is also presented in \cite{Grover_mixer}, which also has an $n^3$ gate complexity, though it has a very different structure.

\newpage
\section{Random quadratic assignment problem}
\label{sec:random_QAP}

A list of candidate facility locations in a 2D grid is populated by random numbers drawn from the uniform distribution over $\left[0,30\right)$:
\[ 
f = \left(
\begin{array}{cc}
    3.301559 & 3.301559 \\ 
    26.321011 & 26.321011 \\ 
    12.484666 & 12.484666 \\ 
    27.655557 & 27.655557 \\ 
    0.826060 & 0.826060 \\ 
    24.856629 & 24.856629 \\ 
    12.733276 & 12.733276 \\ 
    7.914894 & 7.914894 \\ 
    4.433159 & 4.433159
\end{array} \right)
\]

The distance matrix is populated by taking euclidean distances between each pair of locations:
\[
\fl L = \left(
\begin{array}{ccccccccc}
    0.0000 &24.6975 &14.4967 &29.5593 &4.9043 &21.5563 &9.4326 &4.6191 &14.7543 \\ 
    24.6975 &0.0000 &24.4557 &7.9174 &25.9272 &9.2966 &16.1992 &20.3657 &22.6338 \\ 
    14.4967 &24.4557 &0.0000 &31.8189 &19.3559 &16.5448 &11.3483 &12.3271 &27.1493 \\ 
    29.5593 &7.9174 &31.8189 &0.0000 &29.6063 &17.2137 &22.3389 &25.7415 &23.3119 \\ 
    4.9043 &25.9272 &19.3559 &29.6063 &0.0000 &24.4421 &12.5951 &8.1405 &11.0807 \\ 
    21.5563 &9.2966 &16.5448 &17.2137 &24.4421 &0.0000 &12.1287 &16.9481 &25.3065 \\ 
    9.4326 &16.1992 &11.3483 &22.3389 &12.5951 &12.1287 &0.0000 &4.8195 &16.7791 \\ 
    4.6191 &20.3657 &12.3271 &25.7415 &8.1405 &16.9481 &4.8195 &0.0000 &14.8919 \\ 
    14.7543 &22.6338 &27.1493 &23.3119 &11.0807 &25.3065 &16.7791 &14.8919 &0.0000 \\ 
\end{array} \right)
\]

The inter-facility flows are drawn randomly from a uniform distribution over $\left[0,20\right)$:
\[
\fl F = \left(
\begin{array}{ccccccccc}
    0.0000 &12.8188 &8.0291 &18.0712 &10.7207 &15.8060 &13.1888 &6.9539 &16.4688 \\ 
    11.7805 &0.0000 &16.1912 &1.0064 &17.8705 &17.2467 &15.0627 &3.9854 &12.8095 \\ 
    1.9235 &19.5465 &0.0000 &19.1171 &14.0134 &8.8951 &13.4869 &2.9430 &1.1696 \\ 
    14.6809 &19.2269 &2.8263 &0.0000 &10.8089 &10.9314 &18.6236 &15.2934 &9.2809 \\ 
    10.0583 &7.0718 &6.5861 &14.6330 &0.0000 &0.3073 &19.4842 &18.7635 &14.8562 \\ 
    16.6853 &7.4937 &17.3569 &18.0753 &1.2478 &0.0000 &12.4494 &18.8339 &16.2280 \\ 
    16.5752 &18.0927 &5.1124 &12.1212 &18.4499 &13.6680 &0.0000 &16.0441 &18.8221 \\ 
    19.4013 &6.7816 &14.9683 &14.6019 &12.8320 &17.1668 &10.4133 &0.0000 &3.0366 \\ 
    12.1731 &4.1699 &15.3266 &13.4975 &18.0310 &6.4509 &11.7254 &15.0682 &0.0000 \\ 
\end{array}  \right)
\]

\newpage
\section{Random graph for maximum independent set}
\label{sec:random_MIS}
The following random graph, is generated via the Erdos-Renyi model, in which each possible edge is included with some probability (in this case $\text{probability}=0.2$). One exception, is that the process was repeated as necessary to produce a connected graph. 

\begin{eqnarray*}
\fl E = \{&(0,5), (0,6), (0,9), (0,12), (0,13), (0,16), (1,16), (2,3), (2,6), (2,8), (2,17), (3,13), \\
\fl & (3,15), (4,15), (4,16), (5,7), (5,10), (5,13), (5,15), (6,8), (6,14), (6,17), (7,11), (7,14), \\
\fl &(10,16), (11,13), (12,16), (13,14), (8,10), (8,11), (8,12), (9,16)\}    
\end{eqnarray*}

\section{Random capacitated facility location problem}
\label{sec:random_CFLP}

Required resources, integers drawn uniformly from $\left[200,800\right)$,

\[
R = \left(
\begin{array}{cccccccccccc}
    232 & 520 & 465 & 765 & 229 & 277 & 540 & 324 & 395 & 428 & 351 & 381
\end{array} \right)
\]

Capacities (this value was selected in order to produce a good balance of valid and invalid solutions),
\[
C = \left(
\begin{array}{ccc}
    2290 & 2290 & 2290
\end{array} \right)
\]

Candidate facility locations, random locations in a 2D grid, drawn uniformly from $\left[0,8\right)$,
\[ 
f = \left(
\begin{array}{cc} 
    7.038715 & 7.380405 \\
    5.415957 & 1.366392 \\
    3.990352 & 4.843390 
\end{array} \right)
\]

Customer locations, random locations in a 2D grid, drawn uniformly from $\left[0,8\right)$,
\[ 
c = \left(
\begin{array}{cc}
    1.957538 & 5.164732 \\
    1.883436 & 3.333173 \\
    2.993788 & 6.839264 \\
    3.714122 & 3.777548 \\
    6.795077 & 4.626900 \\
    5.407863 & 5.090097 \\
    5.493991 & 2.284864 \\
    6.828993 & 6.443164 \\
    7.101305 & 6.207911 \\
    0.196783 & 5.268291 \\
    5.804895 & 5.965286 \\
    6.682594 & 0.684806
\end{array} \right)
\]

Distances between customers and each candidate facility location (Euclidean distances),
\[ 
L = \left(
\begin{array}{ccc}
    5.543245 & 5.136929 & 2.058055 \\
    6.554159 & 4.043134 & 2.592268 \\
    4.080965 & 5.984916 & 2.230841 \\
    4.902397 & 2.951256 & 1.101055 \\
    2.764263 & 3.540181 & 2.813068 \\
    2.811617 & 3.723713 & 1.438820 \\
    5.324539 & 0.921780 & 2.967657 \\
    0.960419 & 5.269752 & 3.258399 \\
    1.174163 & 5.126471 & 3.397050 \\
    7.160522 & 6.516486 & 3.817291 \\
    1.877464 & 4.615311 & 2.133359 \\
    6.705062 & 1.438378 & 4.953987
\end{array}  \right)
\]

Facility opening costs, drawn uniformly from $\left[1000,2000\right)$,
\[
F = \left(
\begin{array}{ccc}
    1070.126610 & 1746.937054 & 1014.019952
\end{array} \right)
\]

\section*{References}
\bibliographystyle{iopart-num}
\bibliography{refs}

\end{document}